\documentclass{emulateapj}

\newcommand{\myemail}{catherine.walsh@qub.ac.uk}

\usepackage{graphics}
\usepackage{url}
\usepackage{subfigure}
\usepackage{amsmath}
\usepackage{textcomp}

\shorttitle{Photochemistry in Protoplanetary Disks}
\shortauthors{Walsh et al.}

\begin{document}

\title{Chemical Processes in Protoplanetary Disks. II.\\
On the Importance of Photochemistry and X-ray Ionization}

\author{Catherine Walsh\altaffilmark{1}, Hideko Nomura\altaffilmark{2}, 
T. J. Millar\altaffilmark{1} and Yuri Aikawa\altaffilmark{3}}
\altaffiltext{1}{Astrophysics Research Centre, School of Mathematics and Physics, Queen's University
Belfast, University Road, Belfast, Northern Ireland, UK, BT7 1NN}
\altaffiltext{2}{Department of Astronomy, Graduate School of Science, Kyoto University, Kyoto
606-8502, Japan}
\altaffiltext{3}{Department of Earth and Planetary Sciences, Kobe University, 1-1 Rokkodai-cho, Nada,
Kobe 657-8501, Japan}

\email{\myemail}

\begin{abstract}
We investigate the impact of photochemistry and X-ray ionization on the molecular composition of,
and ionization fraction in, a protoplanetary disk surrounding a typical T~Tauri star.  
We use a sophisticated physical model, which includes a robust treatment 
of the radiative transfer of UV and X-ray radiation, and calculate the time-dependent 
chemical structure using a comprehensive chemical network.  
In previous work, we approximated the photochemistry and X-ray ionization, here,
we recalculate the photoreaction rates using the explicit UV wavelength spectrum 
and wavelength-dependent reaction cross sections.  
We recalculate the X-ray ionization rate using our explicit elemental composition and X-ray 
energy spectrum.  
We find photochemistry has a larger influence on the molecular composition than 
X-ray ionization.  
Observable molecules sensitive to the photorates include OH, HCO$^+$, N$_2$H$^+$, 
H$_2$O, CO$_2$ and CH$_3$OH.  
The only molecule significantly affected by the X-ray ionization is N$_2$H$^+$ indicating 
it is safe to adopt existing approximations of the X-ray ionization 
rate in typical T~Tauri star-disk systems.  
The recalculation of the photorates increases the abundances of neutral molecules 
in the outer disk, highlighting the importance of 
taking into account the shape of the UV spectrum in protoplanetary disks.
A recalculation of the photoreaction rates also affects the gas-phase chemistry due to the 
adjustment of the H/H$_2$ and C$^+$/C ratios.  
The disk ionization fraction is not significantly affected by the methods adopted to calculate the 
photochemistry and X-ray ionization.
We determine there is a probable `dead zone' where accretion is suppressed, 
present in a layer, $Z/R$ $\lesssim$ 0.1 - 0.2, in the disk midplane, 
within $R$~$\approx$~200 AU.  
\end{abstract}


\section{INTRODUCTION}
\label{introduction}

Protoplanetary disks have several vital functions in star and planet formation: 
they (1) aid the dissipation 
of angular momentum away from the protostellar system, 
(2) allow the efficient accretion of matter from the constituent cloud 
material onto the central star, and 
(3) contain the material which may eventually form an accompanying planetary system.  

Protoplanetary disks are physically and thus, chemically, complex objects 
\citep[see e.g.,][]{bergin07}.  
They are heavily irradiated by UV radiation from their parent T~Tauri star and are permeated by 
X-rays and excess UV photons thought to arise from an
accretion shock generated as disk material impinges upon the stellar surface 
\citep{herbig86,kastner97}.  
Beyond a radius $r$~$\gtrsim$~100~AU, the UV radiation field originating from the 
parent star decreases in strength 
due to a combination of absorption of UV photons by the intervening disk material and 
geometrical dilution.  
Here, irradiation by UV photons originating from the interstellar radiation field (ISRF) 
increases in importance and, as a result, the wavelength dependence (or shape) of the radiation field 
varies as a function of disk radius and height \citep{aikawa99,willacy00}.   
As a result, in modern disk models, wavelength-dependent radiative transfer is preferred
\citep{vanzadelhoff03} as is the inclusion of UV excess radiation 
\citep{bergin03,nomura05} and X-rays \citep{glassgold97,aikawa99}.  
Along with cosmic-rays, the X-ray and UV radiation controls the ionization fraction in the disk which has
consequences on the disk accretion rate and the location and extent of `dead zones', 
regions where accretion is potentially suppressed 
\citep{balbus91,gammie96}.  
The varying radiation field will also have a direct effect on the disk chemical structure, controlling
the abundance and distribution of atoms, ions and molecules through photochemistry and influencing the
molecular composition of the icy grain mantle via non-thermal desorption 
and this has been demonstrated in many works
\citep[see e.g.,][]{vanzadelhoff03,aikawa06,walsh10,kamp10,vasyunin11}.     
For these reasons, the treatment of photo processes in protoplanetary disks should be 
thoroughly investigated, in order to aid the interpretation of observational data, especially 
with the impending completion of the Atacama Large Millimeter Array (ALMA) which, for the first time,
will enable the observation of molecular emission from nearby ($\sim$ 140~pc) protoplanetary disks 
on around sub-milli-arcsecond scales with unprecedented spectral resolution.  

A plethora of molecules have been detected in protoplanetary disks via line emission in the 
(sub)mm and infrared regions of the electromagnetic spectrum.  
Early observations at (sub)mm wavelengths were made using the James Clerk Maxwell Telescope 
(JCMT) and IRAM 30~m radio telescope \citep[e.g.,][]{kastner97,dutrey97,vanzadelhoff01,thi04}, 
with more recent detections using the Submillimeter Array (SMA) \citep[e.g.,][]{qi06,qi08,oberg10}.  
Most molecules observed in this spectral region are small, simple, abundant molecules, molecular ions 
and radicals (e.g., CO, CN, CS, HCO$^+$, N$_2$H$^+$, HCN) and associated isotopologues 
(e.g., $^{13}$CO, DCO$^{+}$, and C$^{34}$S).  
A recent survey of disks around T~Tauri and Herbig~Ae stars has led to 
the first successful detection of SO in a circumstellar disk \citep{fuente10}, 
with the authors also reporting a tentative detection of H$_2$S.  
Due to the limitations of existing telescopes and the small angular size of disks, 
the most complex molecule observed to date is formaldehyde, H$_2$CO \citep{dutrey97,aikawa03}.  

Use of the Infrared Spectrograph (IRS) on the Spitzer Space Telescope 
increased the inventory of gas-phase 
molecules detected in disks to include OH, H$_2$O, CO$_2$ and C$_2$H$_2$ 
\citep{lahuis06,carr08,salyk08,pontoppidan10}. 
Existing (sub)mm observations probe the colder, outer disk material whereas infrared 
observations probe the warmer gas in the inner disk surface in the so-called `planet-forming' region.  
There have also been detections of water ice absorption features 
in `edge-on' T~Tauri systems \citep{creech-eakman02,terada07,schegerer10}. 
More recently, \citet{hogerheijde11} report the first detection of the ground-state rotational 
emission lines of both spin isomer states of water in a protoplanetary disk using the 
Heterodyne Instrument for the Far-Infrared (HIFI) mounted on the Herschel Space Observatory.
These sets of observations, of both gas and ice, give us a reasonably sufficient 
benchmark with which we can compare our results.  
 
In \citet{walsh10}, henceforth referred to as Paper I, we used the physical disk 
model described in \citet{nomura05} and \citet{nomura07} to compute the chemical 
structure of a typical protoplanetary 
disk on small scales (sub-milli-arcsecond in the inner disk for an object at 
the distance of Taurus, $\sim$~140~pc), investigating the effects of the addition 
of non-thermal desorption mechanisms (cosmic-ray-induced desorption, 
photodesorption and X-ray desorption) and grain-surface chemistry on the 
disk chemical structure.  
In that work, we presented results from models in which we approximated the 
photoreaction rates by scaling the rates  
from the UMIST Database for Astrochemistry or UDfA \citep{woodall07}, 
which assume the interstellar radiation field (ISRF), by the wavelength 
integrated UV flux at each point in the disk 
(see Section~\ref{photochemistry} for further details).  
Here, we report results from models in which we explicitly calculate the 
photodissociation and photoionization rates taking into consideration the 
 UV spectrum at each point and the wavelength-dependent 
absorption cross section for each photoreaction.  
In addition, we recalculate the X-ray ionization rate everywhere in the disk accounting for the 
elemental composition of the gas and include the direct X-ray ionization of elements, 
in both cases, using the X-ray energy spectrum at each point.  

In Section~\ref{physicalmodel}, we give a brief overview of our physical model.  
In Section~\ref{chemicalmodel} we describe our chemical network and 
processes we include in our calculation of the chemical structure with a thorough 
description of the methods used to compute the photochemical and X-ray ionization rates  
(Sections~\ref{photochemistry} and \ref{xrayionization}, respectively).   
In Section~\ref{ionizationfraction}, we describe the theory behind the identification of regions 
of our disk in which angular momentum transport and thus, accretion, may be suppressed.  
The results of our calculations are presented in Section~\ref{results} with a summary given 
in Section~\ref{summary}.

\section{PROTOPLANETARY DISK MODEL}

\subsection{Physical Model}
\label{physicalmodel}

The physical model of a protoplanetary disk we use is from 
\citet{nomura05} with the addition of X-ray heating as described in 
\citet{nomura07}.  
The degree of ionization in the disk depends on the disk parameters 
adopted and the resulting surface density distribution.  
The theoretical foundation of our model comes from the 
\emph{standard accretion disk model} of \citet{lynden74} and \citet{pringle81} which 
defines a surface density distribution for the disk given the central star's 
mass and radius and a disk accretion rate, $\dot{M}$.  
The kinematic viscosity in the disk is parameterised according to the 
work of \citet{shakura73}, the so-called, \emph{$\alpha$-prescription}.  
We consider an axisymmetric disk surrounding a typical T~Tauri star with mass, 
$M_\ast$~=~0.5~$M_\odot$, radius, $R_\ast$~=~2~$R_\odot$ 
and effective temperature, $T_\ast$~=~4000~K.  
We adopt a viscous parameter, 
$\alpha$~=~0.01 and a mass accretion rate, $\dot{M}$~=~10$^{-8}$~$M_\odot$~yr$^{-1}$.   

\begin{figure*}
\subfigure{\includegraphics[width=0.5\textwidth]{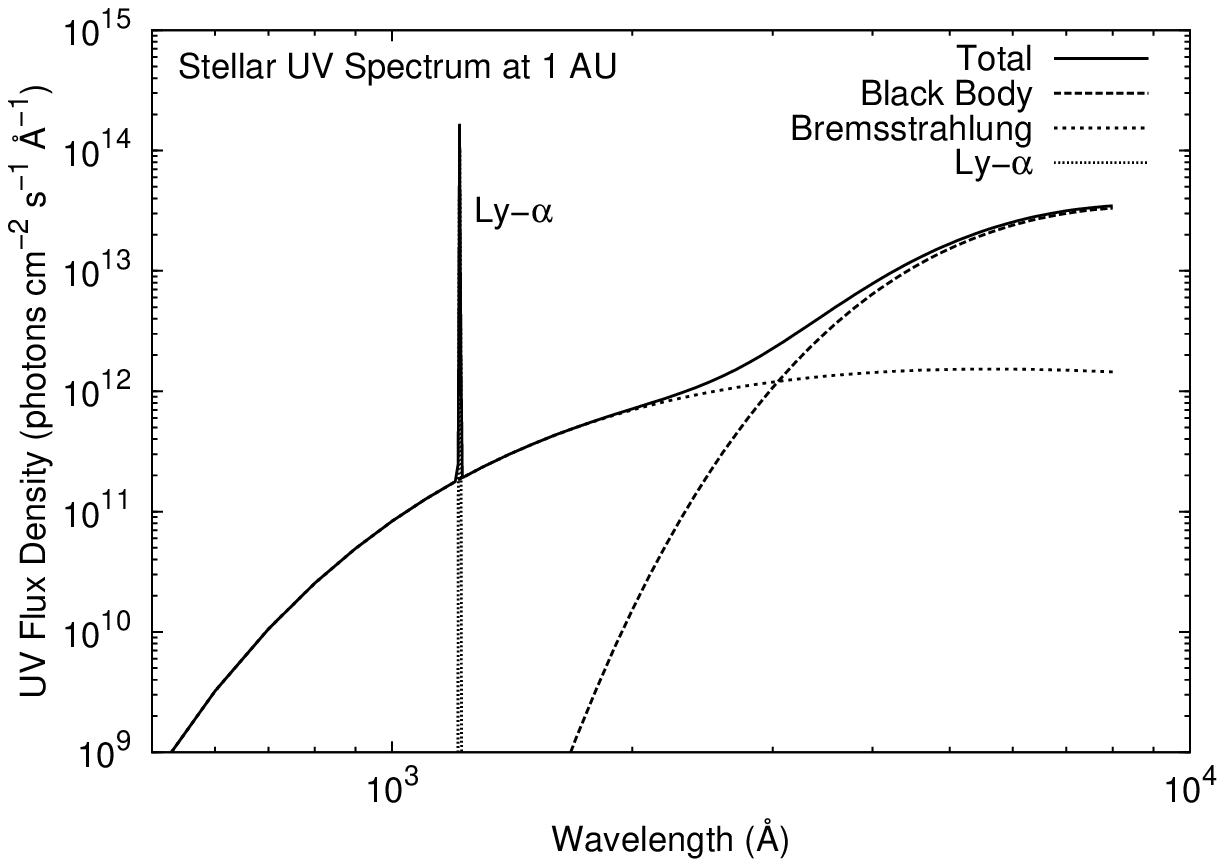}}
\subfigure{\includegraphics[width=0.5\textwidth]{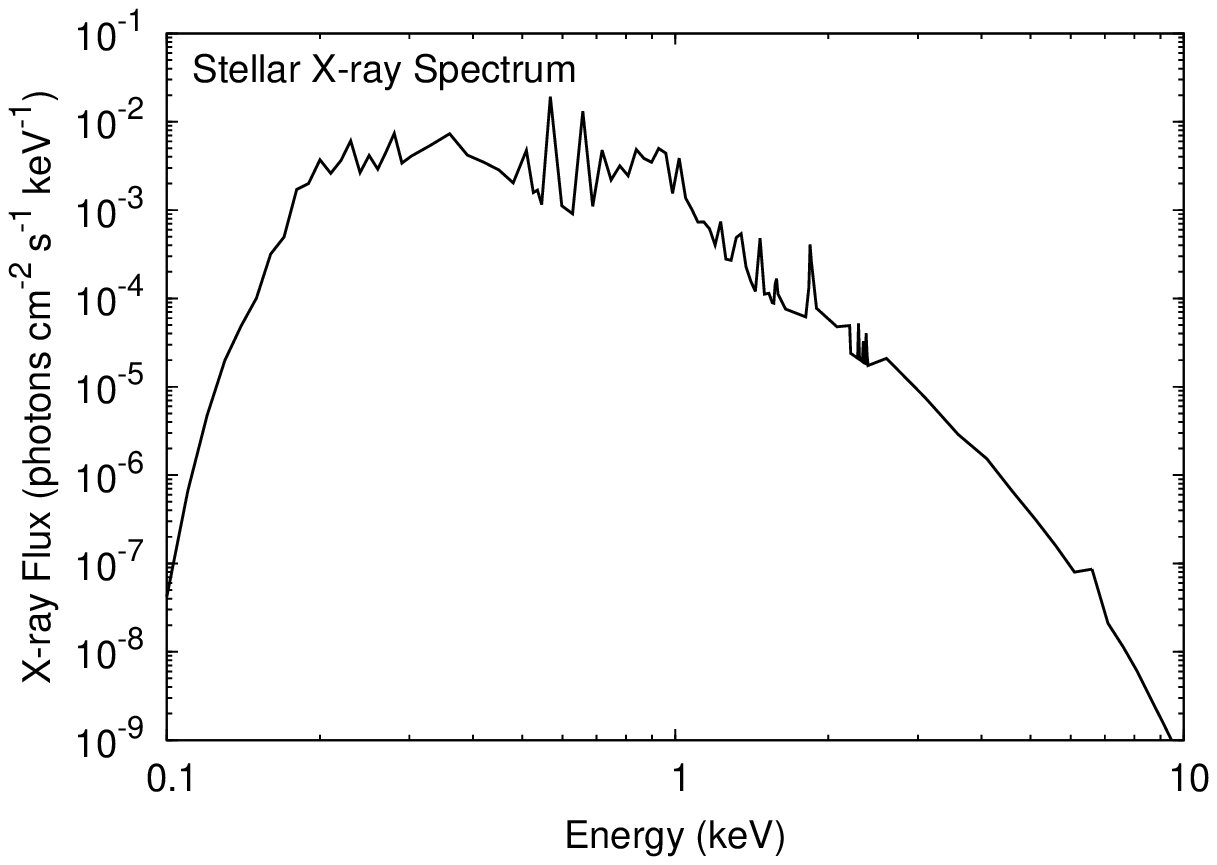}}
\caption{Stellar UV flux density in the disk surface at 1~AU in 
photons cm$^{-2}$ s$^{-1}$ \AA$^{-1}$ (left) 
and binned X-ray flux density in photons cm$^{-2}$ s$^{-1}$ keV$^{-1}$ 
(right).  The latter assumes a distance to source of 56~pc.}
\label{figure1}
\end{figure*}

We use a model X-ray spectrum created by fitting the observed XMM-Newton spectrum of the classical 
T~Tauri star, TW Hydrae \citep{kastner02} with a two-temperature thin thermal plasma model 
(MEKAL model see e.g., \citet{liedahl95}).  
The derived best-fit parameters for the plasma temperatures are $kT_1$~=~0.8~keV and $kT_2$~=~0.2~keV 
and for the foreground interstellar hydrogen column density, $N_\mathrm{H}$~=~2.7~$\times$~10$^{20}$~cm$^{-2}$.  
For the X-ray extinction, we include attenuation due to ionization of all elements and Compton 
scattering by hydrogen.  
The X-ray luminosity is $L_\mathrm{X}$~$\sim$~$10^{30}$~erg~s$^{-1}$ 
and the resulting high-resolution X-ray spectrum is given in Figure~1 of \citet{nomura07}
 assuming a distance to source of 56~pc,  
and is reproduced in binned form here in the right-hand panel of Figure~\ref{figure1}.

The UV radiation field in disks has two sources, the star and the interstellar medium.  
In this disk model, the radiation field due 
to the T~Tauri star has three components: black-body emission at the star's effective temperature, 
optically thin hydrogenic bremsstrahlung emission and strong Lyman-$\alpha$ line emission.  
For the UV extinction, we include absorption and scattering by dust grains.  
We assume the dust and gas in the disk is well mixed and adopt 
a dust-size distribution model which reproduces the observational extinction 
curve of dense clouds \citep{weingartner01}.  
 The calculation of the dust opacity in the disk is described in 
Appendix~D of \citet{nomura05} with the 
resulting wavelength-dependent 
absorption coefficient shown in Figure~D.1. 
The total FUV luminosity  in our model is $L_\mathrm{UV}$~$\sim$~$10^{31}$~erg~s$^{-1}$ 
with the calculation of the radiation field in the disk described in detail in 
Appendix~C of \citet{nomura05}.  
We display the resulting stellar flux density in the disk surface at a radius of 1~AU, 
including each individual component, 
in the left-hand panel of Figure~\ref{figure1}.  
The main source of UV photons shortward of 2000~$\AA$ is due to Brehmsstrahlung and Lyman-$\alpha$ 
radiation with the Lyman-$\alpha$ line contributing around 10$^{3}$ times the UV continuum 
photon flux at $\approx$~1216~$\AA$  over an assumed FWHM of $\approx$~2~$\AA$  \citep[see e.g.,][]{herczeg02}.

\begin{figure*}
\subfigure{\includegraphics[width=0.5\textwidth]{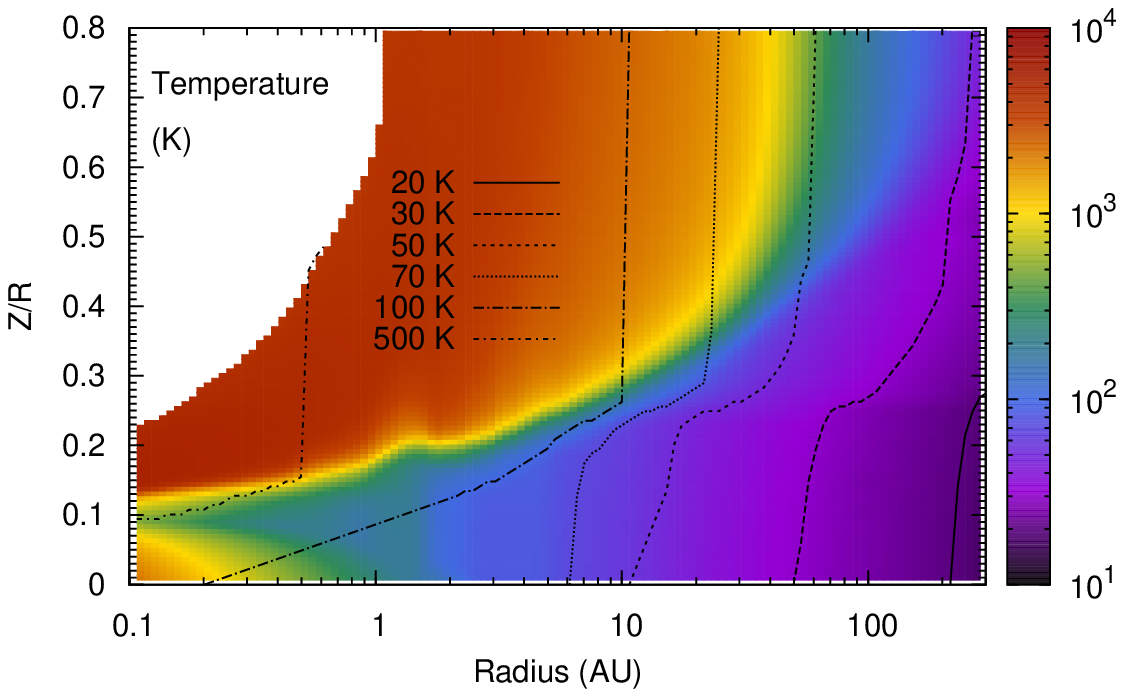}}
\subfigure{\includegraphics[width=0.5\textwidth]{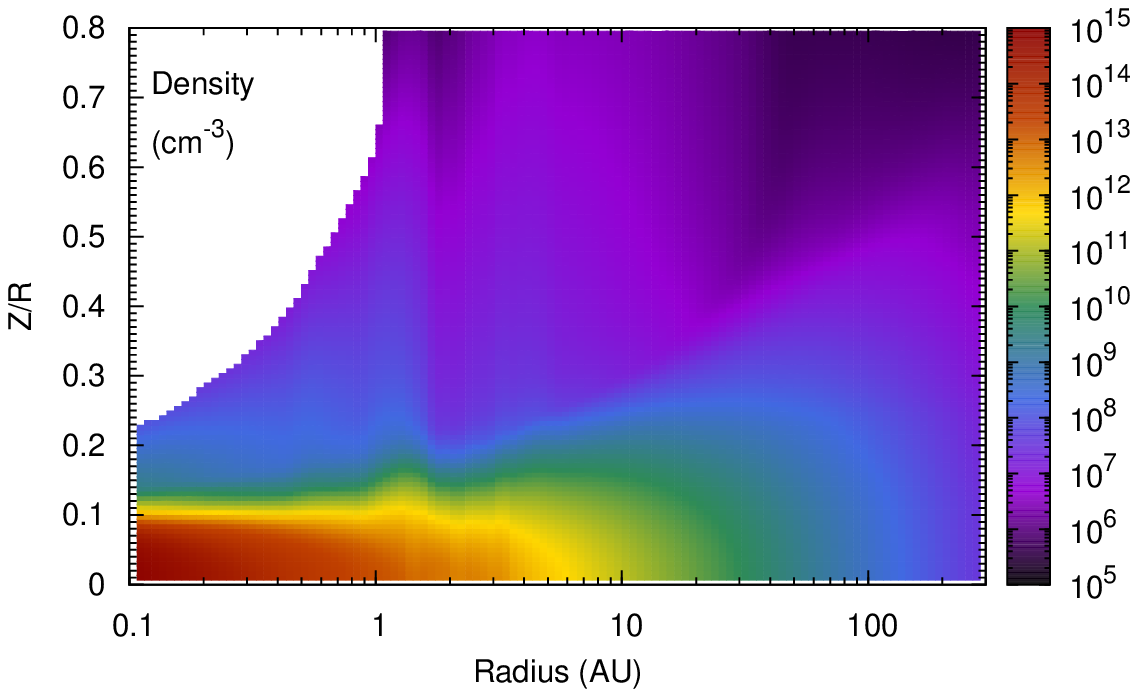}}
\subfigure{\includegraphics[width=0.5\textwidth]{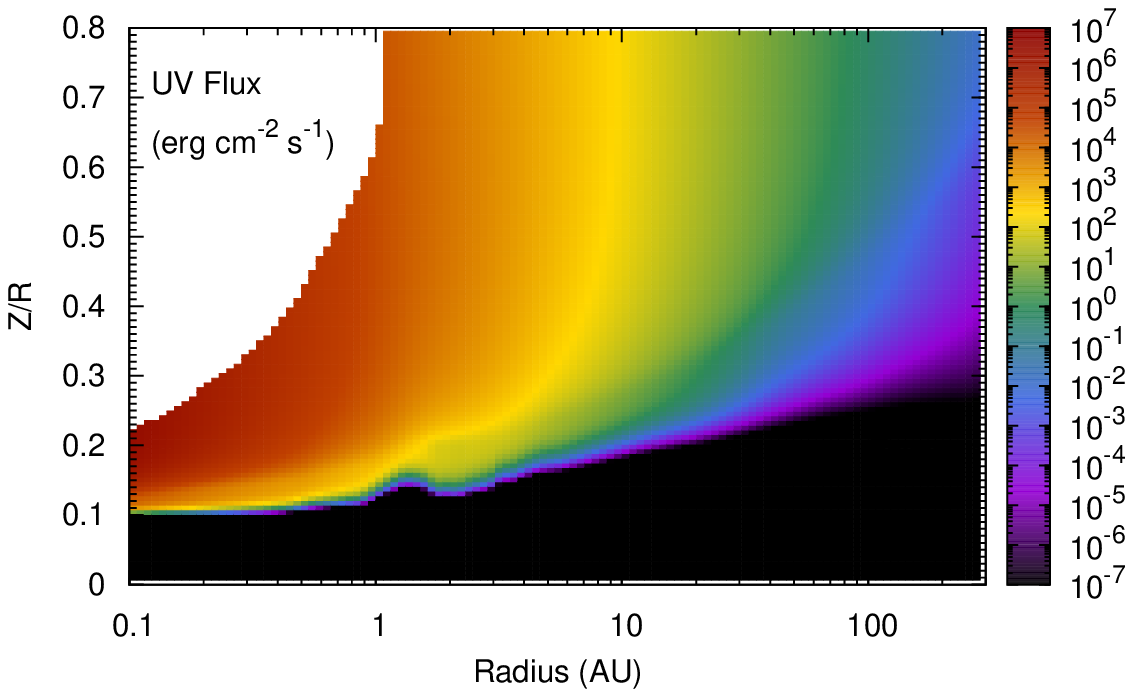}}
\subfigure{\includegraphics[width=0.5\textwidth]{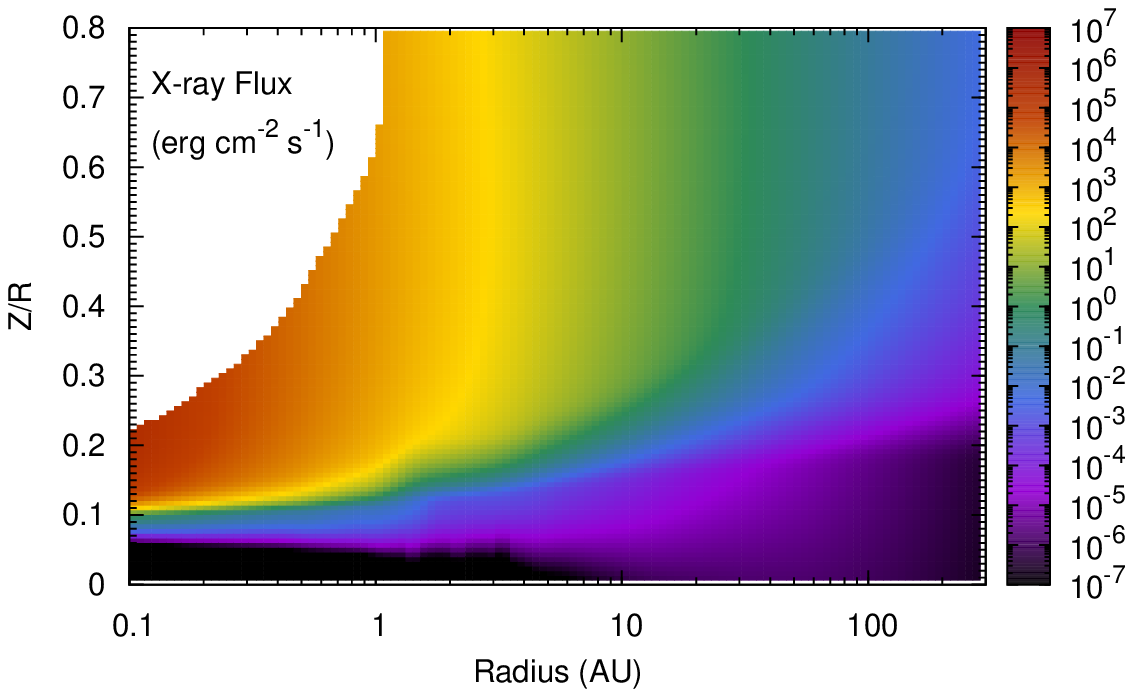}}
\caption{Temperature (top left), number density (top right), UV flux (bottom left) 
and X-ray flux (bottom right) as a function of disk radius, $R$, 
and height (scaled by the radius i.e., $Z/R$). 
On the temperature panel, the colour map represents the gas temperature 
whereas the contours represent the dust temperature.}
\label{figure2}
\end{figure*}

The resulting disk physical structure is given in Figures~1 and 13 in Paper I and we 
refer readers to the Appendix of that publication for a thorough discussion.  
In Figure~\ref{figure2}, we display the gas and dust temperatures in K (top right), the gas number 
density in cm$^{-3}$ (top left), the 
wavelength-integrated UV flux (bottom left) and X-ray flux (bottom right) 
both in units of erg~cm$^{-2}$~s$^{-1}$, as a function of disk radius and height (scaled by the radius).  
 In the temperature plot, the colour map represents the gas temperature whereas the contours represent the 
dust temperature.  As expected, the disk surface closest to the parent star is subjected to the largest flux of both 
UV and X-ray radiation.  The disk midplane is effectively completely shielded from UV photons  
over the radial extent of our disk model.  
The higher energy X-ray photons, although resulting in a lower flux in the disk surface, are 
able to penetrate the disk more effectively  
leading to a small, yet appreciable, X-ray flux in the disk midplane beyond $\approx$~10~AU. 

\subsection{Chemical Model}
\label{chemicalmodel}

As in Paper I, our gas-phase chemistry is
extracted from the latest release of the UMIST Database for Astrochemistry  or 
`Rate06', available at \url{http://www.udfa.net} \citep{woodall07}.  
We use almost the entire Rate06 network removing only those reactions involving 
fluorine- and phosphorus-containing species.  
Here, we also include the subset of three-body reactions from Rate06, since 
these may play a significant role in the densest regions of our disk model  
($n$~$\sim$~10$^{15}$~cm$^{-3}$).  

We allow gas-grain interactions i.e., the accretion of gas-phase species onto dust grains 
with removal of the grain mantle via thermal desorption, 
cosmic-ray-induced desorption 
\citep{leger85,hasegawa93} and photodesorption \citep{westley95,oberg07,willacy07}.  
In Paper I we provided a thorough description of the methods used to determine 
our accretion and desorption rates. 
We also include the grain-surface network from \citet{hasegawa92} and \citet{hasegawa93}.

Here, we calculate our photodesorption rates taking into 
account the molecular composition of the grain mantle.  
Experiments into the photodesorption of UV irradiated ices conducted by \citet{oberg09a,oberg09b} 
suggest that the photodesorption yields are dependent on the ice composition with pure
CO, N$_2$, CO$_2$ and H$_2$O ices giving different desorption yields 
(listed in Table~\ref{table1}).  The picture is further complicated by evidence of 
codesorption in mixed ices, with the photodesorption yield of N$_2$ increasing when 
present in a 1:1 N$_2$/CO ice mixture \citep{oberg07}.  
More recent work suggests that in organic ices, irradiation by UV photons initiates chemistry 
with photodesorption products detected other than those present in the original ice mixture 
\citep{oberg09c}. 
These latter effects are not included in this model for simplicity, however, 
they should be investigated in future models.  

In Paper I, we used a constant photodesorption yield of $3 \times 10^{-3}$  
photon$^{-1}$ as determined for pure water ice by \citet{westley95}.  
To take into consideration the composition of the grain mantle we now determine the 
photodesorption rate for a specific species according to
\begin{equation}
k_\mathrm{pd}^i = G_\mathrm{UV}Y_\mathrm{UV}^i\sigma_\mathrm{d} x_\mathrm{d} x_s^i \;  \mbox{s}^{-1},
\end{equation}
where $G_\mathrm{UV}$ is the radiation field in units of photons cm$^{-2}$ s$^{-1}$, 
$Y_\mathrm{UV}^i$ photon$^{-1}$ is the specific desorption yield of species $i$, as listed in 
Table~\ref{table1}, $x_\mathrm{s}^i = n_\mathrm{s}^i/n_\mathrm{s}^\mathrm{tot}$ is the fractional abundance
of species $i$ on the dust grains and $\sigma_\mathrm{d}$ and $x_\mathrm{d}$ are the dust-grain 
geometrical cross section and fractional abundance, respectively.  
We have also reviewed our set of molecular desorption energies in light 
of recent experimental results and in Table~\ref{table2} we list those species for 
which the binding energies, $E_\mathrm{d}$, have been updated.  

Our initial fractional abundances are the result of a dark cloud model 
run with typical molecular cloud parameters 
i.e., $T$~=~10~K, $n$(H$_2$)~=~10$^{4}$~cm$^{-3}$ and 
$A_{v}$~=~10~mag using the set of oxygen-rich low-metallicity elemental abundances from 
\citet{graedel82} as listed in Table~8 of \citet{woodall07}.  
In the generation of our initial abundances we allow for freeze out and 
thermal desorption and we extract abundances at a time of $10^{5}$~years 
which is thought to be the age of dark clouds on the brink of star formation.   

\begin{deluxetable}{lcc}
\tablecaption{Photodesorption Yields \label{table1}}
\tablewidth{0pt}
\tablehead{\colhead{Species} & \colhead{Yield (molecules photon$^{-1}$)} & \colhead{Reference}}
\startdata
CO                & 2.7 $\times$ 10$^{-3}$ & 1 \\
N$_2$             & 1.8 $\times$ 10$^{-4}$ & 1 \\
CO$_2$            & 2.3 $\times$ 10$^{-3}$ & 2 \\
H$_2$O            & 1.3 $\times$ 10$^{-3}$ & 2 \\
All other species & 3.0 $\times$ 10$^{-3}$ & 3
\enddata
\tablerefs{(1) \citet{oberg09a}, (2) \citet{oberg09b}, (3) \citet{westley95}}
\end{deluxetable}
 
\begin{deluxetable}{lcc}
\tablecaption{Molecular Binding Energies \label{table2}}
\tablewidth{0pt}
\tablehead{\colhead{Species} & \colhead{Binding Energy (K)} & \colhead{Reference}}
\startdata
CO            & 855  & 1 \\
N$_2$         & 790  & 1 \\
CO$_2$        & 2990 & 2 \\
C$_2$S        & 5320 & 2 \\
H$_2$O        & 4800 & 3 \\
SO$_2$        & 5330 & 2 \\
NH$_3$        & 2790 & 3 \\
CH$_4$        & 1090 & 4 \\
HCOOH         & 5000 & 5 \\
CH$_3$OH      & 4930 & 3 \\
CH$_3$CHO     & 3800 & 5 \\
C$_2$H$_6$    & 2300 & 5 \\
HCOOCH$_3$    & 4000 & 5 \\
CH$_3$OCH$_3$ & 3300 & 5 \\
C$_2$H$_5$OH  & 5200 & 5 
\enddata
\tablerefs{(1) \citet{oberg05}, (2) \citet{edridge10}, (3) \citet{brown07}, 
(4) \citet{herrero10}, (5) \citet{oberg09c}}
\end{deluxetable}
 
\subsubsection{Photochemistry}
\label{photochemistry}

The photoreaction rates in Rate06 are calculated assuming the UV radiation field is given 
by the Draine field \citep{draine78}, an adequate assumption for the unshielded ISRF.  
For use in chemical models, these rates are subsequently parameterised according to 
optical depth or $A_{v}$.  
In Paper I, we approximated our photoreaction rates in the disk, $k_\mathrm{ph}$, by 
scaling the rates from Rate06 using the wavelength-integrated UV flux at each point, 
$G_\mathrm{UV}(r,z) = \int_{912\;\AA}^{2000\;\AA} G_\lambda(r,z) \; \mbox{d}\lambda$ i.e.,
\begin{equation}
k_\mathrm{ph} = \frac{G_\mathrm{UV}}{G_0} k_{0} \; \mbox{s}^{-1}
\label{photoapprox}
\end{equation}
where $G_0$ is the unshielded interstellar UV flux and $k_0$ is the rate calculated for the 
unshielded interstellar medium.  
Note, $G_\lambda(r,z)$ includes both the stellar and interstellar components 
of the radiation field.  
This approximation is unsuitable for use in protoplanetary disks as these objects 
are irradiated by UV radiation with three components: the central star, UV excess due 
to accretion and the interstellar radiation field.  
Hence, the UV spectrum at each point in the disk will not only vary with disk radius and height, 
but will also bear no resemblance to the interstellar radiation field 
(see Figure 4 in \citet{nomura05}).  
For these reasons, we have investigated a recalculation of the photorates in the 
disk taking into consideration the UV spectrum at each point \citep{vandishoeck87,vandishoeck06}.   

The photodissociation rate due to continuous absorption, 
$k_\mathrm{ph}^\mathrm{c}$, is calculated using 
\begin{equation}
k_\mathrm{ph}^\mathrm{c} = \int_{\lambda_\mathrm{min}}^{\lambda_\mathrm{max}} 
\sigma(\lambda)I(\lambda)\;\mathrm{d}\lambda \;
\mathrm{s}^{-1},
\label{photorates1}
\end{equation}
where, $\lambda$ is the wavelength, $\sigma(\lambda)$ is the cross section 
and $I(\lambda)$ is the mean 
intensity of UV radiation.  
The rate for photoionization, $k_\mathrm{pi}$, is calculated using the same equation.  
For indirect photodissociation via absorption into a bound upper state, $u$, from a lower state, 
$l$, the rate is 
\begin{equation}
k_\mathrm{ph}^\mathrm{l} = \frac{\pi e^2}{mc^2}\lambda_\mathrm{ul}^2 f_\mathrm{ul}\mu_{u}I(\lambda) \; \mathrm{s}^{-1},
\label{photorates2}
\end{equation}
where, $\lambda_\mathrm{ul}$ is the wavelength of the line transition, 
$f_\mathrm{ul}$ is the oscillator strength of the transition and $\mu_\mathrm{ul}$ is an efficiency factor.  
The parameters, $e$, $c$ and $m$ are the electron charge, the speed of light and 
the atomic or molecular mass, respectively.  
The total photodissociation rate is found by summing over all possible channels.  
The photoreaction cross sections, $\sigma(\lambda)$, are those adopted in 
\citet{vanzadelhoff03}
which originate from calculations by \citet{vandishoeck88}, updated by \citet{jansen95a,jansen95b} and 
\citet{vandishoeck06}.  
The photo cross sections are downloadable from \url{http://www.strw.leidenuniv.nl/}$\sim$\url{ewine/photo/}.   
For species which do not have a calculated cross section, we use the rate for a similar type of molecule.  

We should note here we include Lyman-$\alpha$ radiation in the calculation of the 
wavelength-integrated UV flux (Equation~\ref{photoapprox}) 
which accounts for approximately 85\% of the  total flux \citep[see e.g.,][]{bergin03}, however, 
we ignore it in the calculation of the wavelength-dependent UV radiation field and subsequent photorates 
(see Equations~\ref{photorates1} and \ref{photorates2}).  
 The exclusion of Lyman-$\alpha$ is primarily due to the difficulties in treating the scattering 
of Lyman-$\alpha$ photons from the surface into the disk \citep[see e.g.,][]{bergin03}.  
Lyman-$\alpha$ scattering differs from that of the background 
UV photons since the scattering occurs   
predominantly by H atoms rather than dust grains. 
Indeed, Lyman-$\alpha$ radiation has been historically neglected in protoplanetary 
disk models and has only very recently been addressed in the work by \citet{bethell11}.  
 However, the photorates calculated according to Equations~\ref{photorates1} and \ref{photorates2} are more accurate 
than those determined using Equation~\ref{photoapprox} since the shape of the background 
radiation field is included in the calculation. 
 The photorates calculated using Equation~\ref{photoapprox} include the 
total flux of UV photons (background plus Lyman-$\alpha$) heavily overestimating the strength of the UV field
at wavelengths other than the Lyman-$\alpha$ wavelength ($\approx$~1216~$\AA$). 
The effect of including the wavelength-dependence in the photorates is 
apparent in differences in the relative abundances of molecules (see Table~\ref{table4}) e.g., 
 the column density ratio of 
CO$_2$/H$_2$O changes from $\approx$~0.045 in model UV-old to $\approx$~1.90 in model UV-new at 100~AU.  
We discuss the possible effects of including Lyman-$\alpha$ radiation in the 
calculation of the photorates in Section~\ref{lyman-alpha}.

We include the self-shielding of H$_2$ using the approximation from 
\citet{federman79} in the generation of our physical model giving us the initial conditions in 
our disk, however,
we do not explicitly include the self- and mutual shielding of H$_2$ and CO 
in our calculation of the subsequent chemical structure.  
We find the dominant component of the radiation field in the disk surface is the radial component which 
is the direct stellar radiation (the vertical component consists of both the contribution from the 
interstellar radiation field and scattered stellar radiation).  
In addition, throughout the majority of the disk, the stellar radiation dominates over the interstellar 
radiation.   
Hence, we argue against the validity of adopting the usual plane-parallel approximation 
for the calculation of the self-shielding factors and the application of shielding factors computed 
for irradiated interstellar clouds, to protoplanetary disks.  
To correctly include the effects of self- and mutual shielding in disks, a self-consistent two-dimensional 
treatment is needed which takes into consideration the time-varying H$_2$ and CO abundances throughout the disk, 
the column densities in the radial \emph{and} vertical direction and the two-dimensional physical structure 
of the disk which will effect the line widths and line strengths and hence, shielding factors.     
We discuss this issue further in Section~\ref{selfshielding}.

\begin{deluxetable}{lcccc}
\tablecaption{Column Densities\label{table4}}
\tablewidth{0pt}
\tablehead{\colhead{Species} & \colhead{UV-old} & \colhead{XR+UV-old} & \colhead{UV-new} & \colhead{XR+UV-new}}
\startdata
\cutinhead{0.1 AU}
H           & 1.4(21) & 1.4(21) & 4.5(20) & 4.2(20) \\
H$_2$       & 1.1(26) & 1.1(26) & 1.1(26) & 1.1(26) \\
CO          & 9.7(21) & 9.5(21) & 9.5(21) & 9.6(21) \\
\textbf{HCO$^+$}     & \textbf{2.1(14)} & \textbf{2.5(14)} & \textbf{3.2(12)} & \textbf{1.7(12)} \\
HCN         & 5.2(19) & 5.0(19) & 5.4(19) & 4.9(19) \\
CN          & 8.2(13) & 6.9(13) & 4.3(13) & 6.5(13) \\
CS          & 1.3(17) & 1.5(17) & 1.5(17) & 1.5(17) \\
C$_2$H      & 1.0(15) & 9.9(14) & 9.6(14) & 9.8(14) \\
H$_2$CO     & 6.9(15) & 8.2(15) & 6.5(15) & 8.3(15) \\
\textbf{N$_2$H$^+$}  & \textbf{3.7(09)} & \textbf{8.7(08)} & \textbf{2.0(11)} & \textbf{9.2(10)} \\
\textbf{OH}          & \textbf{6.0(16)} & \textbf{6.0(16)} & \textbf{6.1(15)} & \textbf{5.1(15)} \\
H$_2$O      & 2.8(22) & 2.8(22) & 2.8(22) & 2.8(22) \\
CO$_2$      & 3.8(18) & 5.4(18) & 3.7(18) & 5.5(18) \\
C$_2$H$_2$  & 6.1(18) & 9.6(18) & 9.3(18) & 9.1(18) \\
CH$_3$OH    & 2.3(18) & 2.4(18) & 2.3(18) & 2.4(18) \\
\cutinhead{1 AU}
H           & 1.1(21) & 1.1(21) & 8.8(20) & 8.3(20) \\
H$_2$       & 1.9(25) & 1.9(25) & 1.9(25) & 1.9(25) \\
CO          & 1.9(21) & 1.9(21) & 1.9(21) & 1.9(21) \\
HCO$^+$     & 1.7(14) & 1.9(14) & 9.4(13) & 6.5(13) \\
HCN         & 2.0(18) & 2.0(18) & 2.0(18) & 2.0(18) \\
CN          & 2.7(14) & 2.7(14) & 1.0(14) & 1.3(14) \\
CS          & 8.7(12) & 8.4(12) & 6.2(12) & 6.3(12) \\
C$_2$H      & 4.1(14) & 4.4(14) & 9.1(13) & 1.5(14) \\
H$_2$CO     & 3.1(13) & 3.1(13) & 3.2(13) & 3.1(13) \\
\textbf{N$_2$H$^+$}  & \textbf{5.2(10)} & \textbf{9.3(09)} & \textbf{2.7(12)} & \textbf{5.1(11)} \\
OH          & 1.7(16) & 1.8(16) & 2.4(16) & 2.2(16) \\
H$_2$O      & 2.2(21) & 2.2(21) & 2.2(21) & 2.2(21) \\
CO$_2$      & 7.7(20) & 7.7(20) & 7.7(20) & 7.7(20) \\
C$_2$H$_2$  & 1.6(14) & 1.9(14) & 1.5(14) & 1.8(14) \\ 
CH$_3$OH    & 1.0(15) & 1.0(15) & 1.0(15) & 1.0(15) \\
\cutinhead{10 AU}
H           & 7.3(20  & 7.2(20) & 6.1(20) & 5.7(20) \\
H$_2$       & 2.6(24) & 2.6(24) & 2.6(24) & 2.6(24) \\
CO          & 2.0(20) & 2.0(20) & 2.0(20) & 2.0(20) \\
HCO$^+$     & 4.8(13) & 4.5(13) & 1.4(14) & 1.6(14) \\
HCN         & 8.0(14) & 2.1(15) & 4.5(14) & 6.9(14) \\
CN          & 7.3(13) & 8.1(13) & 3.0(14) & 3.4(14) \\
CS          & 4.7(13) & 4.4(13) & 4.3(13) & 5.5(13) \\
C$_2$H      & 5.4(13) & 6.7(13) & 1.1(14) & 1.6(14) \\
H$_2$CO     & 2.0(12) & 2.0(12) & 1.2(12) & 8.1(11) \\
\textbf{N$_2$H$^+$}  & \textbf{2.0(10)} & \textbf{5.7(09)} & \textbf{2.8(11)} & \textbf{9.2(10)} \\
OH          & 8.9(15) & 9.4(15) & 2.1(16) & 2.3(16) \\
\textbf{H$_2$O}      & \textbf{5.4(15)} & \textbf{5.9(15)} & \textbf{8.5(16)} & \textbf{9.2(16)} \\
\textbf{CO$_2$}      & \textbf{2.8(16)} & \textbf{3.7(16)} & \textbf{2.7(17)} & \textbf{6.5(17)} \\
C$_2$H$_2$  & 7.5(13) & 3.2(13) & 7.9(13) & 3.7(13) \\ 
\textbf{CH$_3$OH}    & \textbf{1.0(08)} & \textbf{9.1(07)} & \textbf{5.3(09)} & \textbf{5.2(09)} \\
\cutinhead{100 AU}
H           & 3.3(19) & 1.7(19) & 3.2(19) & 1.5(19) \\
H$_2$       & 2.0(23) & 2.0(23) & 2.0(23) & 2.0(23) \\
CO          & 9.1(18) & 9.7(18) & 8.9(18) & 9.3(18) \\
HCO$^+$     & 3.9(13) & 3.6(13) & 5.8(13) & 4.3(13) \\
HCN         & 4.8(14) & 2.5(14) & 2.3(14) & 2.1(14) \\
CN          & 2.4(14) & 2.6(14) & 5.0(14) & 4.1(14) \\
CS          & 3.1(13) & 3.7(13) & 1.8(13) & 1.7(13) \\
C$_2$H      & 1.0(14) & 1.2(14) & 7.7(13) & 5.4(13) \\
H$_2$CO     & 4.4(12) & 3.7(12) & 8.2(12) & 2.9(12) \\
\textbf{N$_2$H$^+$}  & \textbf{1.9(11)} & \textbf{8.8(10)} & \textbf{1.5(12)} & \textbf{5.5(11)} \\
\textbf{OH}          & \textbf{9.0(14)} & \textbf{3.0(14)} & \textbf{6.3(15)} & \textbf{3.3(15)} \\
H$_2$O      & 4.9(16) & 2.5(16) & 3.1(16) & 1.4(16) \\
\textbf{CO$_2$}      & \textbf{2.2(15)} & \textbf{2.1(15)} & \textbf{5.9(16)} & \textbf{8.8(16)} \\
C$_2$H$_2$  & 3.0(13) & 3.1(13) & 2.4(13) & 1.6(13) \\ 
\textbf{CH$_3$OH}    & \textbf{4.1(10)} & \textbf{4.0(11)} & \textbf{7.9(10)} & \textbf{2.1(11)}
\enddata
\tablecomments{$a(b)$ means $a \times 10^{b}$}
\end{deluxetable}

\subsubsection{X-ray Ionization}
\label{xrayionization}

The model of \citet{nomura07} calculates an overall X-ray ionization rate at each point in the 
disk according to the theory of \citet{maloney96}.  
The rate at each point is calculated assuming a power-law fit for the X-ray absorption 
cross section dependent on X-ray energy and 
it is these approximate rates that are used in our chemical calculations in Paper I. 
In this work, we recalculate the X-ray ionization rate everywhere in the disk 
taking into account the X-ray energy spectrum, $F_\mathrm{X}(E)$, at each point and the explicit elemental 
composition of the gas \citep{glassgold97}.  
The overall X-ray ionization rate, $\zeta_\mathrm{XR}$, is given by summing over all elements,
\begin{equation}
\zeta_\mathrm{XR} = \sum_k \int_{E_k}^{E_\mathrm{max}} x_k \sigma_k(E)F_\mathrm{X}(E) 
\left[ \frac{E-E_k}{\Delta\epsilon}\right]\; \mathrm{d}E \; \mbox{s}^{-1},
\label{xrayrates1}
\end{equation}
where, for each element, $k$, $E_k$ is the ionization potential, $\sigma_k(E)$ is the cross section and 
$x_k$ is the fractional abundance with respect to H nuclei density.  
In this expression, the number of secondary ionizations per unit energy produced by 
primary photoelectrons, $N_\mathrm{sec}$, is given by the expression, $(E-E_k)/\Delta\epsilon$, 
where $\Delta\epsilon$ 
is the mean energy required to make an ion pair ($\approx$~37~eV).  
Only the number of secondary ionizations needs to be considered as this is generally much 
larger than 
the number of primary ionization events.  
Typically, each keV of secondary electron energy produces 
an average of 1000/37~$\approx$~27 ion pairs so that $N_\mathrm{sec}\gg N_\mathrm{pri}$.  
Note that X-rays interact only with atoms, regardless of whether an atom is bound within a molecule 
or free \citep{glassgold97}.  

We have also added the direct X-ray ionization of elements, the rate for which is given by
\begin{equation}
\zeta_k = \int_{E_k}^{E_\mathrm{max}} \sigma_k(E)F_\mathrm{X}(E)\;\mathrm{d}E \; \mbox{s}^{-1}.
\label{xrayrates2}
\end{equation}
using the ionization cross sections for each element, $k$, from \citet{verner93}.  

\subsection{Disk Ionization Fraction}
\label{ionizationfraction}

In addition to 
investigating the importance of photochemistry and X-ray chemistry in protoplanetary disks, 
we have determined the location and extent of potential dead zones where accretion may be
inhibited.  
Angular momentum transport in disks is thought to arise from turbulence
initiated by a weak-field magnetorotational instability or MRI \citep{balbus91}, hence, accretion may be
 inefficient in regions where the instability is suppressed.  
The turbulence generated by the instability can sustain a disordered magnetic field to which the gas 
is coupled.  The degree of the coupling, in turn, depends on the ionization fraction in the disk.  

Following \citet{gammie96}, we can define a magnetic Reynolds number, $Re_\mathrm{M}$, everywhere,
\begin{equation}
Re_\mathrm{M} = \frac{V_\mathrm{A}H}{\eta}, 
\end{equation}
where $V_\mathrm{A} \approx \alpha^{1/2}c_{s}$ is the Alfv\'{e}n speed, 
a function of $\alpha$, the scaling parameter for the viscosity ($\nu = \alpha c_s H$) from
the accretion disk model of \citet{shakura73} and $c_s$, the sound speed in the disk.  
$H$ is the disk scale height given by $H = c_{s}/\Omega$, where $\Omega$ is the Keplerian velocity at 
a particular radius, $R$, 
and $\eta$ is the magnetic resistivity which is related to the electron fraction, 
$\chi \equiv n_\mathrm{e}/n_\mathrm{H}$ by
\begin{equation}
\eta = (6.5 \times 10^{3}) \chi^{-1} \; \mathrm{cm}^{2} \mathrm{s}^{-1}, 
\end{equation}
(see \citet{gammie96} for further details).  
Accretion is likely suppressed in regions where the magnetic Reynolds number, $Re_\mathrm{M}$, falls 
below a critical value, $Re_\mathrm{M}^\mathrm{crit}$.  
 This parameter determines the degree to which the ionized gas is effectively coupled to the magnetic field.
Recent MHD simulations suggest $Re_\mathrm{M}^\mathrm{crit} \sim$ 100 \citep{sano02,ilgner06} which corresponds
roughly to an electron fraction, $\chi \sim$ 10$^{-12}$, although the exact value is dependent on the
disk model adopted.  
Using our chemical model results, we calculate the value of the magnetic Reynolds number, $Re_\mathrm{M}$,
everywhere and identify possible dead zones where $Re_\mathrm{M} \lesssim 100$.  

 \citet{chiang07} argue that a second criterion, taking into account the 
influence of ambipolar diffusion in suppressing the MRI, must be applied 
for protoplanetary disks. 
They define a dimensionless parameter, $Am$, which describes the degree to which 
neutral H$_2$ molecules (which make up the bulk of the gas), are coupled to the 
accreting plasma.  In order for neutral gas to be unstable, a H$_2$ molecule must collide with enough ions
within the e-folding time of the instability, $1/\Omega$, where $\Omega$ is 
the Keplerian velocity of the gas. 
\begin{equation}
Am = \frac{x_\mathrm{e} n \beta}{\Omega} > Am^\mathrm{crit}
\end{equation}
Here, $x_\mathrm{e}$ is the disk ionization fraction, $n$ is the number density of the neutral gas and 
$\beta$~$\approx$~2.0~$\times$~10$^{-9}$~cm$^{3}$~s$^{-1}$ is the 
rate coefficient for ion-neutral collisions.
Early MHD simulations by \citep{hawley98} suggested that sufficient turbulence and 
angular momentum transport is achieved when the `ambipolar diffusion parameter', 
$Am$, exceeds a critical value, $Am^\mathrm{crit}$~$\approx$~100.  
More recent simulations by \citet{bai11a} suggest that in very weakly ionized media where 
the `strong coupling' limit holds, 
such as protoplanetary disks, the critical value for $Am$ is $\approx$~1.  
The `strong coupling' limit is defined as when the ion inertia is 
negligible and the recombination time is much shorter than the orbital time \citep{bai11b}.
We adopt this latter criterion in our determination of the 
susceptibility of the disk to suppression of the MRI by ambipolar diffusion.
The results are reported in Section~\ref{results}. 

Note, the extent and location of a dead zone is also dependent on the treatment 
of gas-grain interactions and 
the size distribution of grains \citep{sano00,ilgner06}.  
\citet{sano00} calculated that, for a fixed gas-to-dust mass ratio, the dead zone shrinks as the 
grain size increases, 
assuming all grains have the same radius.  
As the grains increase in size (likely due to coagulation) the ion density increases since the 
total surface area of
grains decreases.  This subsequently decreases the recombination rate of gas-phase cations on 
grain surfaces.  
They also determine that in the midplane, grains are the dominant charged species (with charge $\pm e$).  

 It has also been postulated that gravitational grain settling (or sedimentation) 
towards the disk midplane influences the protoplanetary disk physical and chemical structure 
and thus, ionization fraction, the persistence of the MRI and the subsequent location and extent of dead zones 
\citep[see e.g.,][]{chiang01,dullemond04,dalessio06,nomura07,fogel11,vasyunin11}.
Since the dust is the dominant source of opacity in disks, grain settling allows deeper penetration 
of stellar and interstellar UV radiation potentially ionizing a larger 
proportion of the gaseous component of the disk.  
In particular, \citet{dalessio06} find that an absence of small grains in the upper disk layer,  
due to sedimentation, enhances the ionization in the disk surface and also decreases the 
temperature of the gas in the disk midplane since there is a decrease in the amount of radiation 
processed by grains and directed towards the midplane. 
In reality, disk ionization, turbulence by MRI and dust settling are coupled and thus,
ideally should be solved self-consistently \citep[see e.g.,][]{fromang06,ciesla07,turner10}.

In this work, for our gas-grain interactions, we assume a constant grain radius of 
0.1~$\mu$m and a fixed dust-grain
fractional abundance of 2.2~$\times$~10$^{-12}$.  
We assume that all grains are negatively charged and allow the recombination of cations 
on grain surfaces.  
This is valid assumption since negatively-charged grains dominate in regions where the 
 number density, $n$, is $\lesssim$~10$^{12}$~cm$^{-3}$ and this holds throughout most of our disk model.  
We intend to investigate the effects of adding neutral grains and a variable 
dust-grain size distribution caused by coagulation and settling in future models.

We have also neglected the thermal ionization of alkali metals, such as Na$^{+}$ and K$^+$, 
which becomes a significant source of ionization when the
gas temperature is greater than $\approx$~10$^3$~K \citep{fromang02,ilgner06}. 
In our disk model, we find in the midplane the gas temperature reaches values higher than this 
only within a radius of a
few tenths of an AU, nevertheless, it should looked at in future models since, 
due to this source of ionization, the gas may be magnetorotationally unstable 
close to the star.   

\section{RESULTS}
\label{results}

We calculate the chemical abundances in the disk as a function of radius, height 
and time.  
The results displayed in this section are those extracted at a time of 10$^6$~years, the 
typical age of visible T~Tauri stars with accompanying protoplanetary disks.  
Throughout this section, fractional abundance refers to the abundance of each species with respect 
to total particle number density.  
As in Paper I, we ran several different models with differing chemical ingredients in order 
to determine the influence of each chemical process and these are listed in Table~\ref{table3}.
Here, model UV-old, 
is our `fiducial' model in which we use the same method as in Paper I to calculate the 
photochemical rates and we use the X-ray ionization rates as calculated in \citet{nomura07}. 
We compare the results from model UV-old with those from models which include a recalculation of 
the photochemical rates only (model UV-new), the X-ray ionization rates only (model XR+UV-old) 
and both processes (model XR+UV-new).  
All models include freeze out, thermal desorption, cosmic-ray-induced desorption, photodesorption and 
grain-surface chemistry.  

\begin{deluxetable*}{lcccc}
\tablecaption{Chemical Models \label{table3}}
\tablewidth{0pt}
\tablehead{Chemical Process &\colhead{UV-old}&\colhead{UV-new}&\colhead{XR+UV-old}&\colhead{XR+UV-new}}
\startdata
Thermal desorption             & \checkmark  & \checkmark & \checkmark & \checkmark  \\
Cosmic-ray-induced desorption  & \checkmark  & \checkmark & \checkmark & \checkmark  \\
Photodesorption                & \checkmark  & \checkmark & \checkmark & \checkmark  \\
Grain-surface chemistry        & \checkmark  & \checkmark & \checkmark & \checkmark  \\
Photochemistry                 &             & \checkmark &            & \checkmark  \\
X-ray ionization               &             &            & \checkmark & \checkmark 
\enddata
\end{deluxetable*}

\subsection{Column Densities}
\label{columndensities}

\begin{figure*}
\centering
\subfigure{\includegraphics[width=0.32\textwidth]{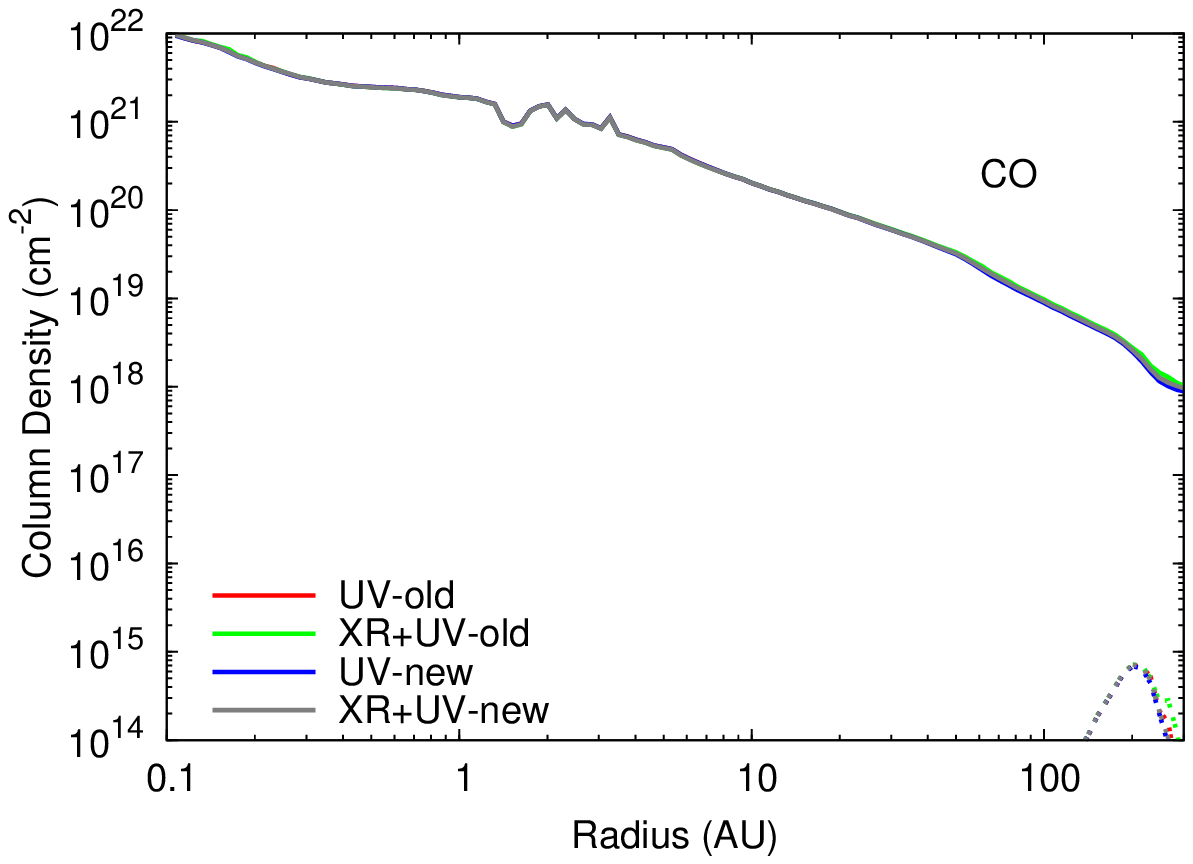}}
\subfigure{\includegraphics[width=0.32\textwidth]{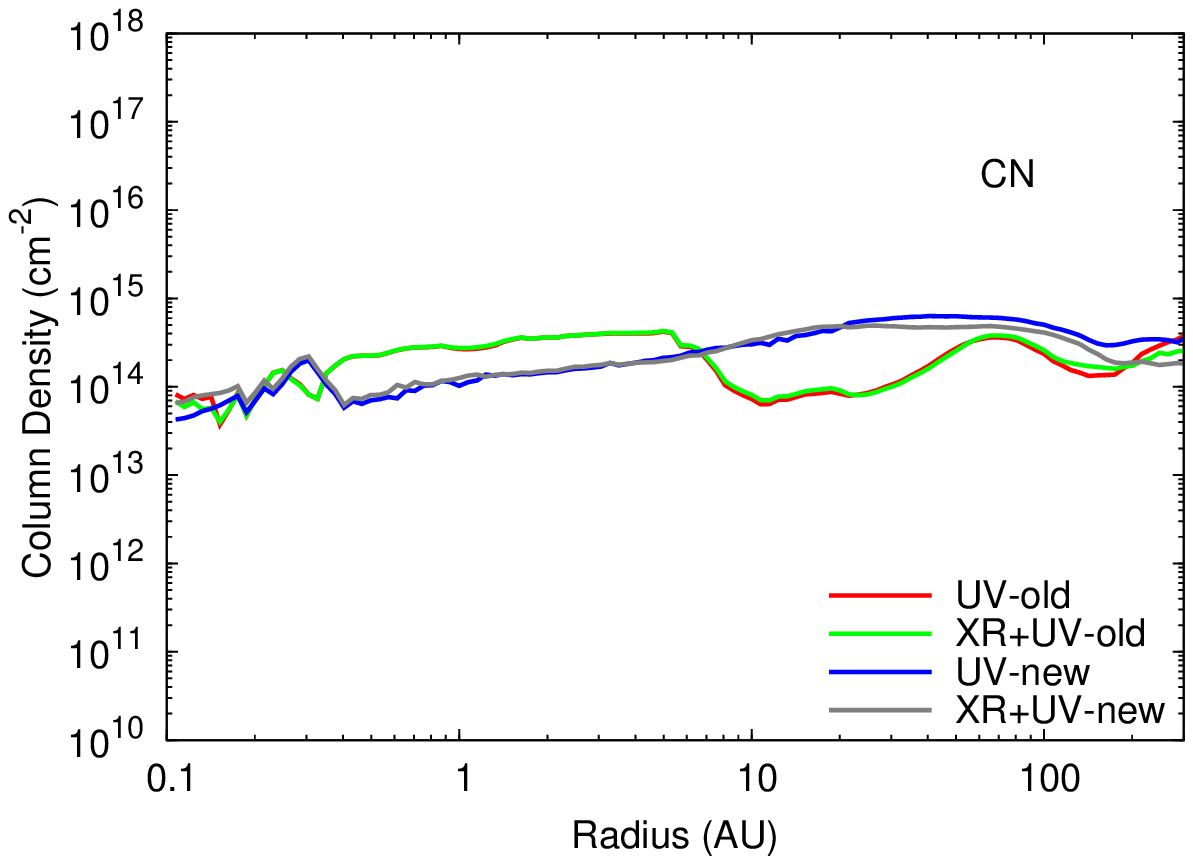}}
\subfigure{\includegraphics[width=0.32\textwidth]{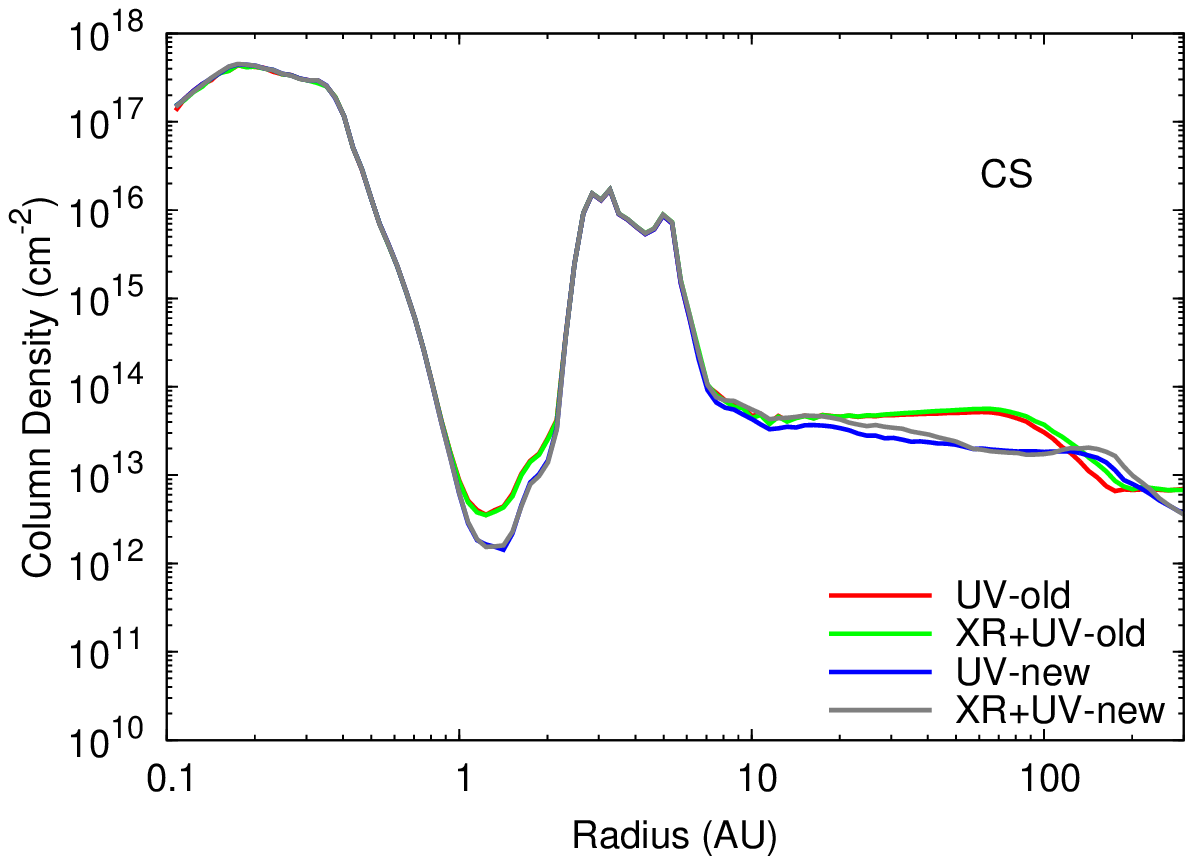}}
\subfigure{\includegraphics[width=0.32\textwidth]{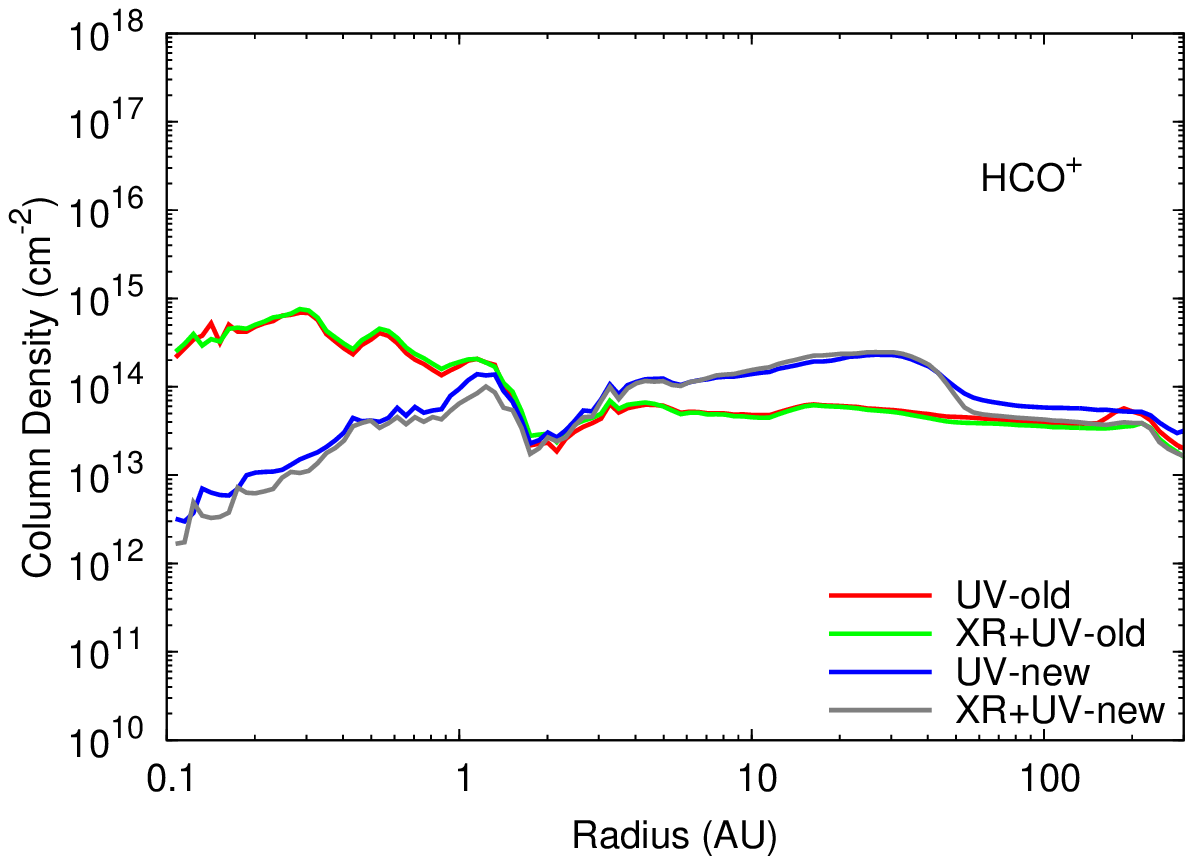}}
\subfigure{\includegraphics[width=0.32\textwidth]{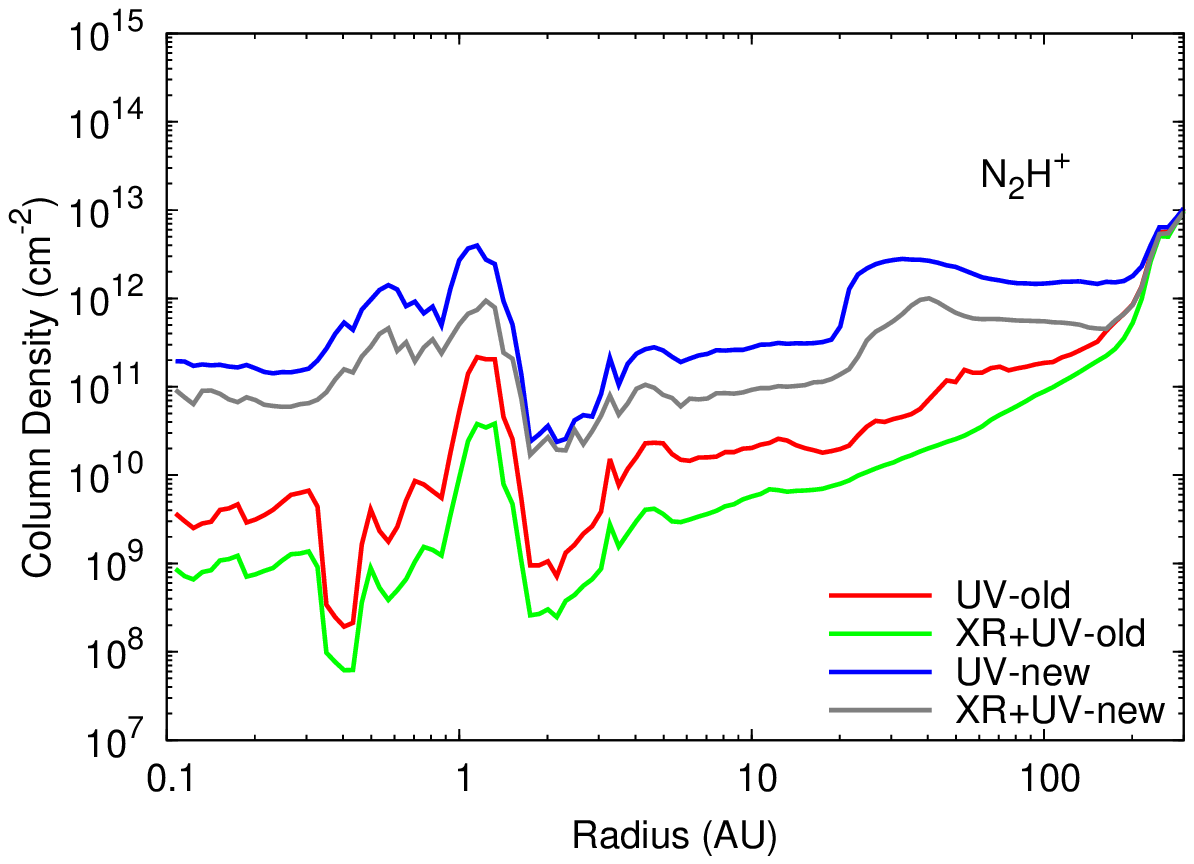}}
\subfigure{\includegraphics[width=0.32\textwidth]{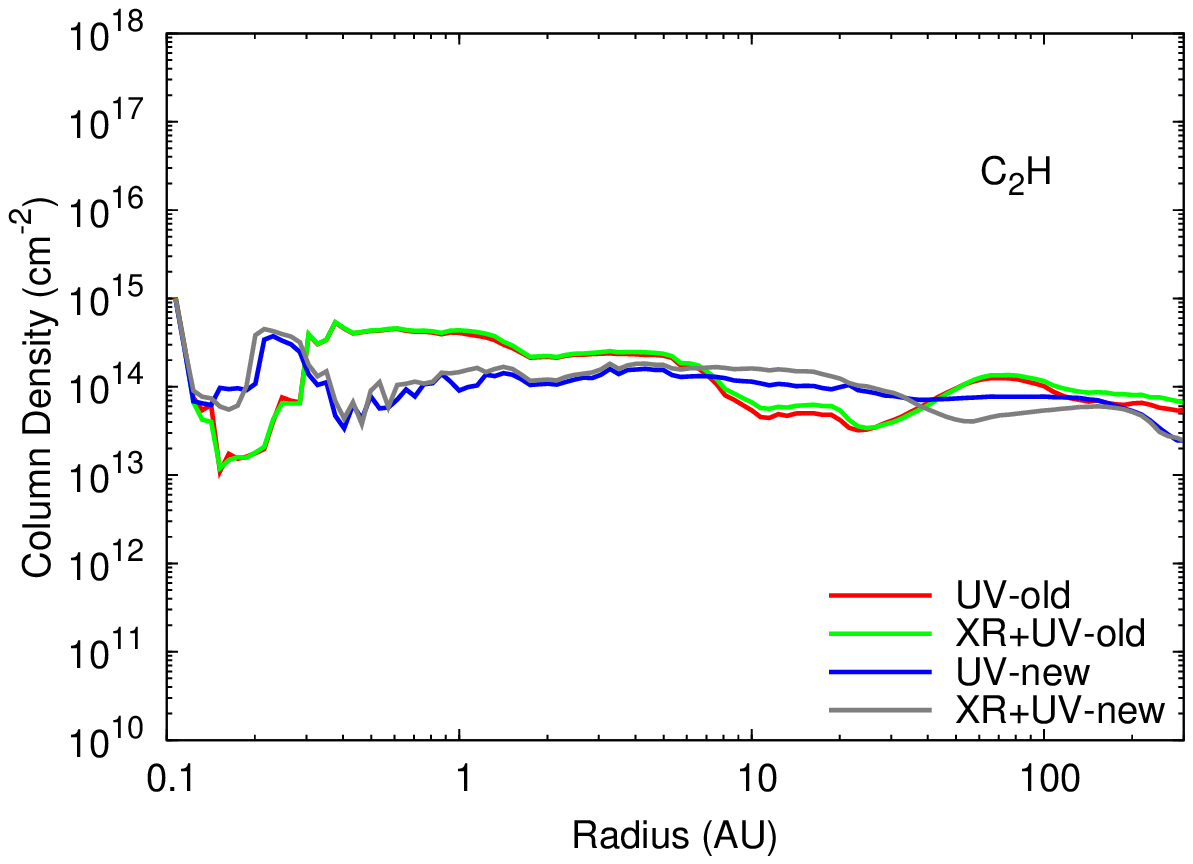}}
\subfigure{\includegraphics[width=0.32\textwidth]{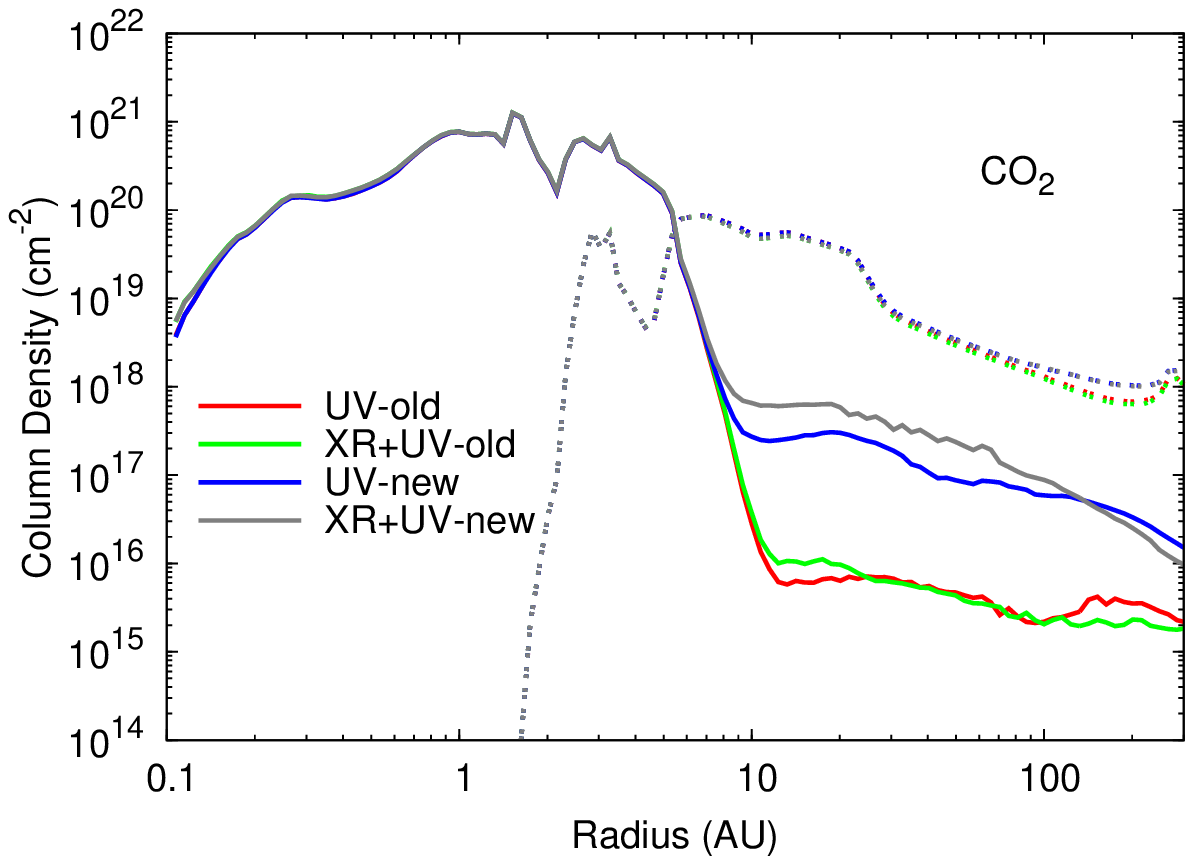}}
\subfigure{\includegraphics[width=0.32\textwidth]{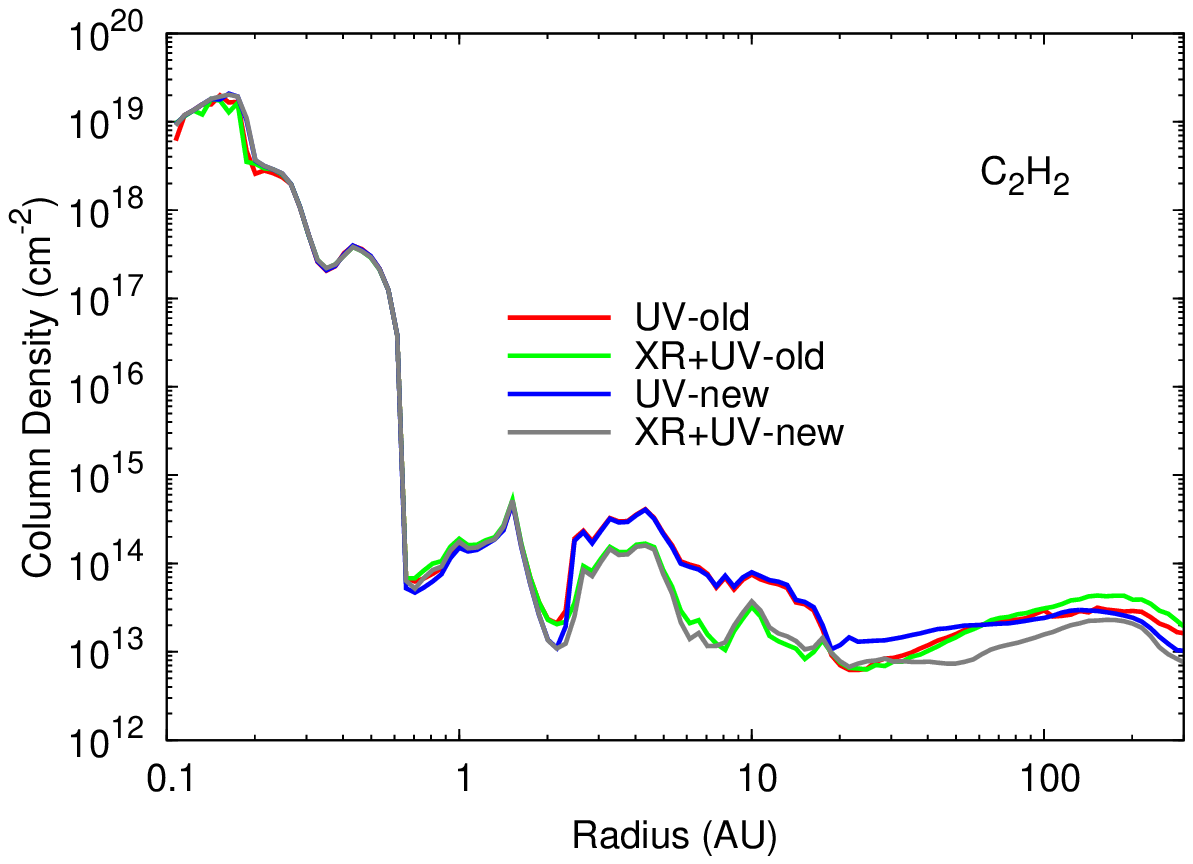}}
\subfigure{\includegraphics[width=0.32\textwidth]{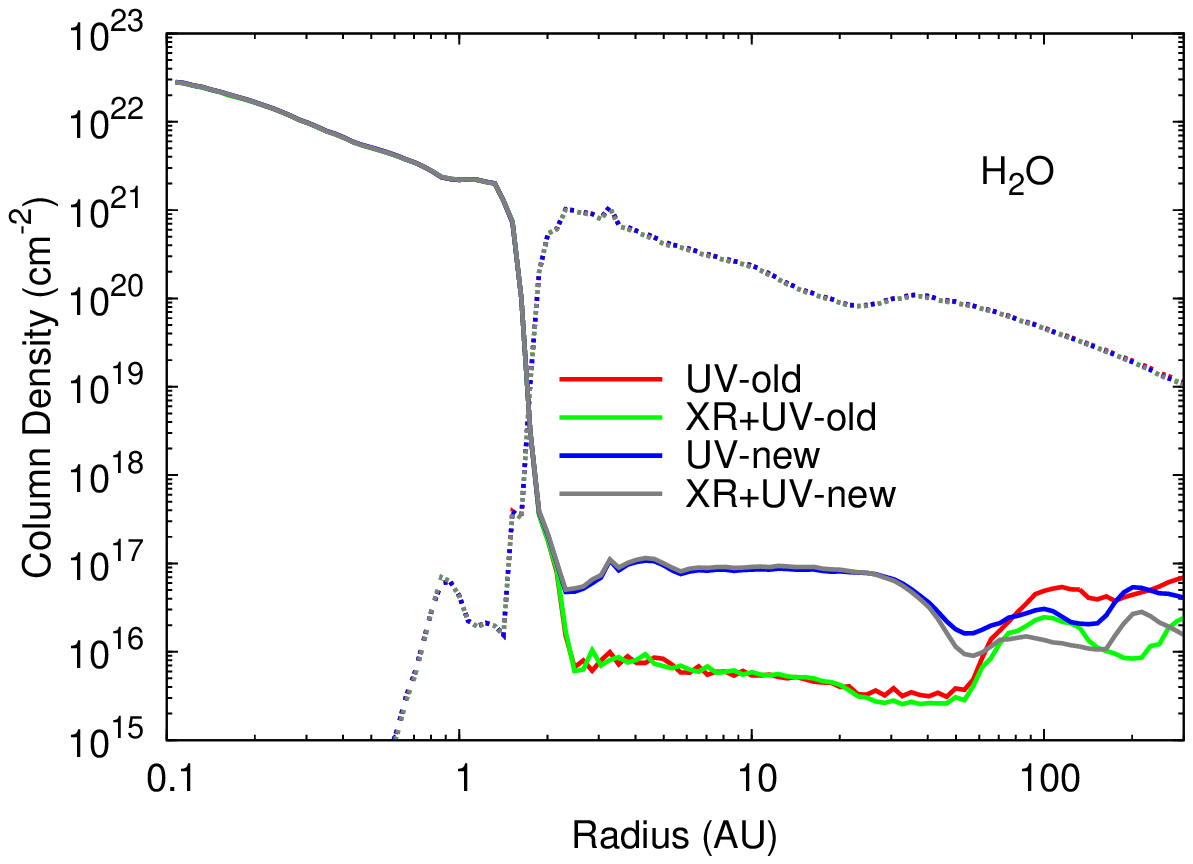}}
\subfigure{\includegraphics[width=0.32\textwidth]{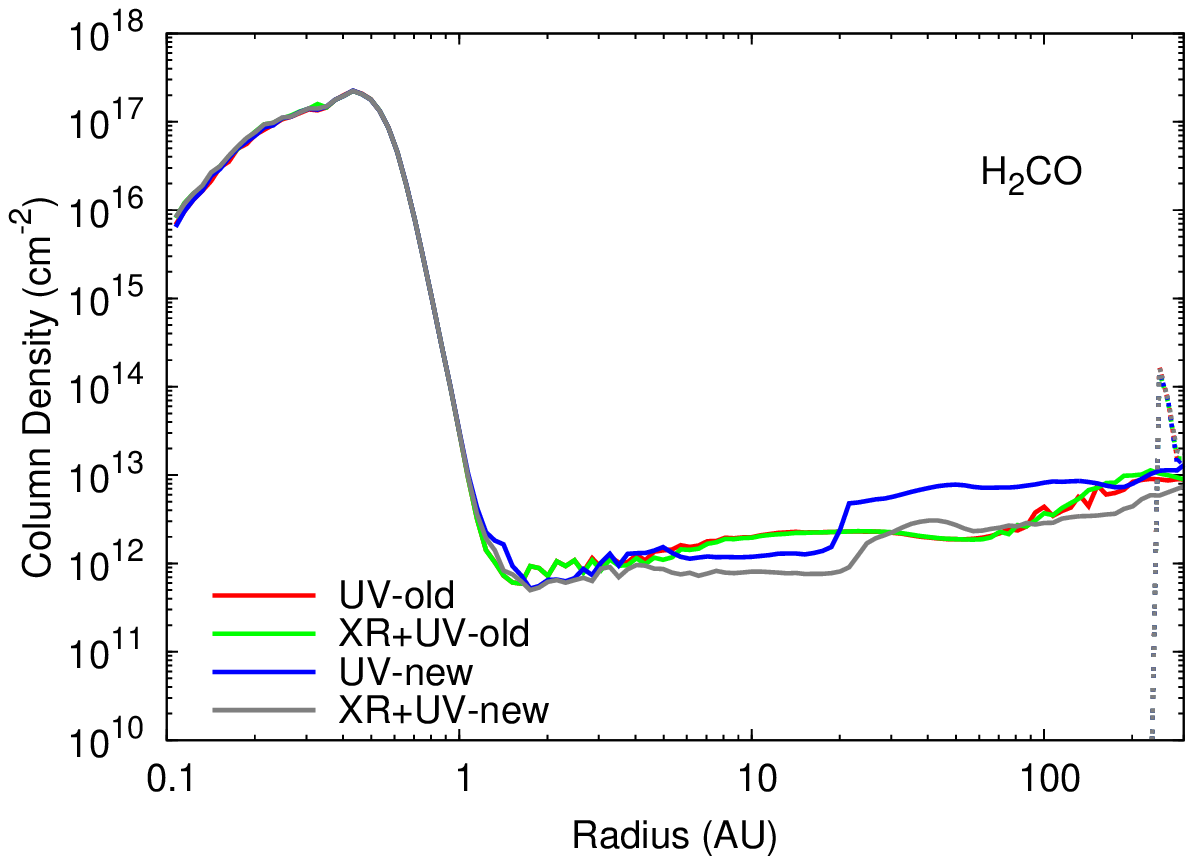}}
\subfigure{\includegraphics[width=0.32\textwidth]{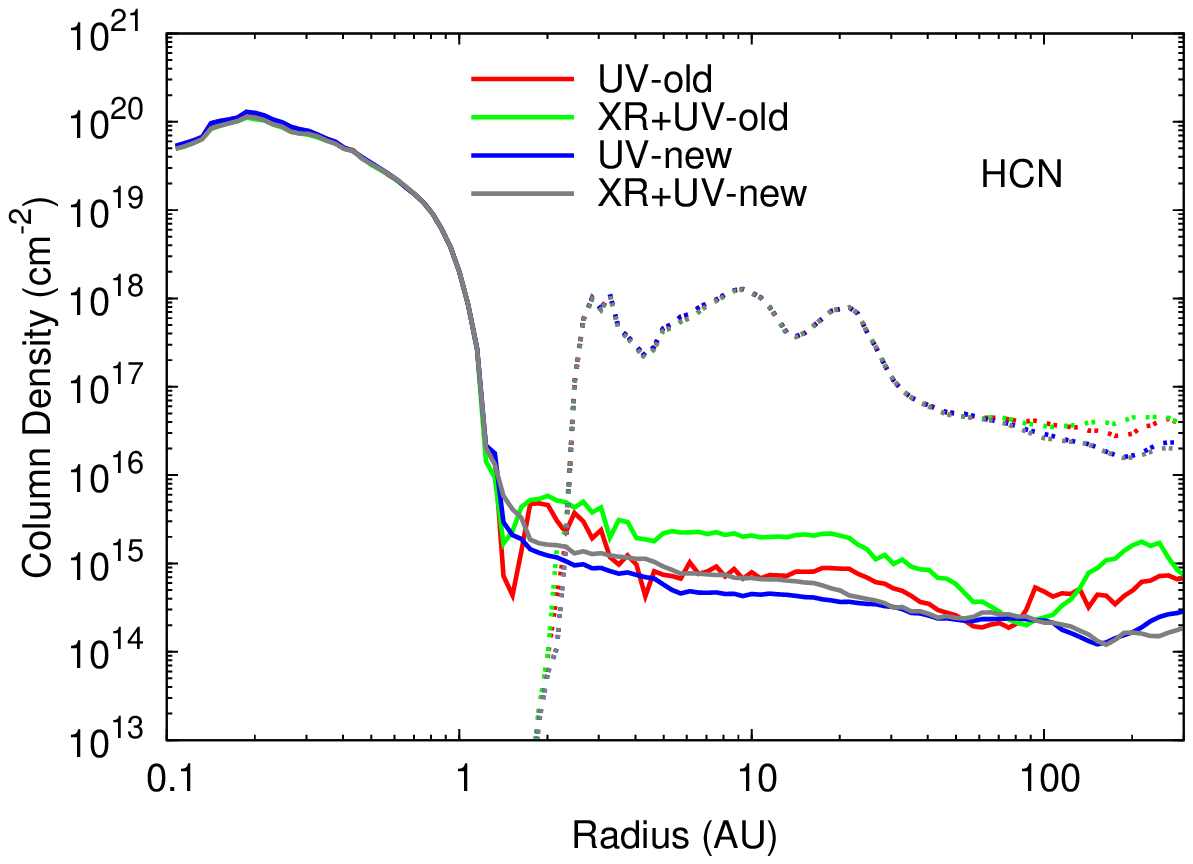}}
\subfigure{\includegraphics[width=0.32\textwidth]{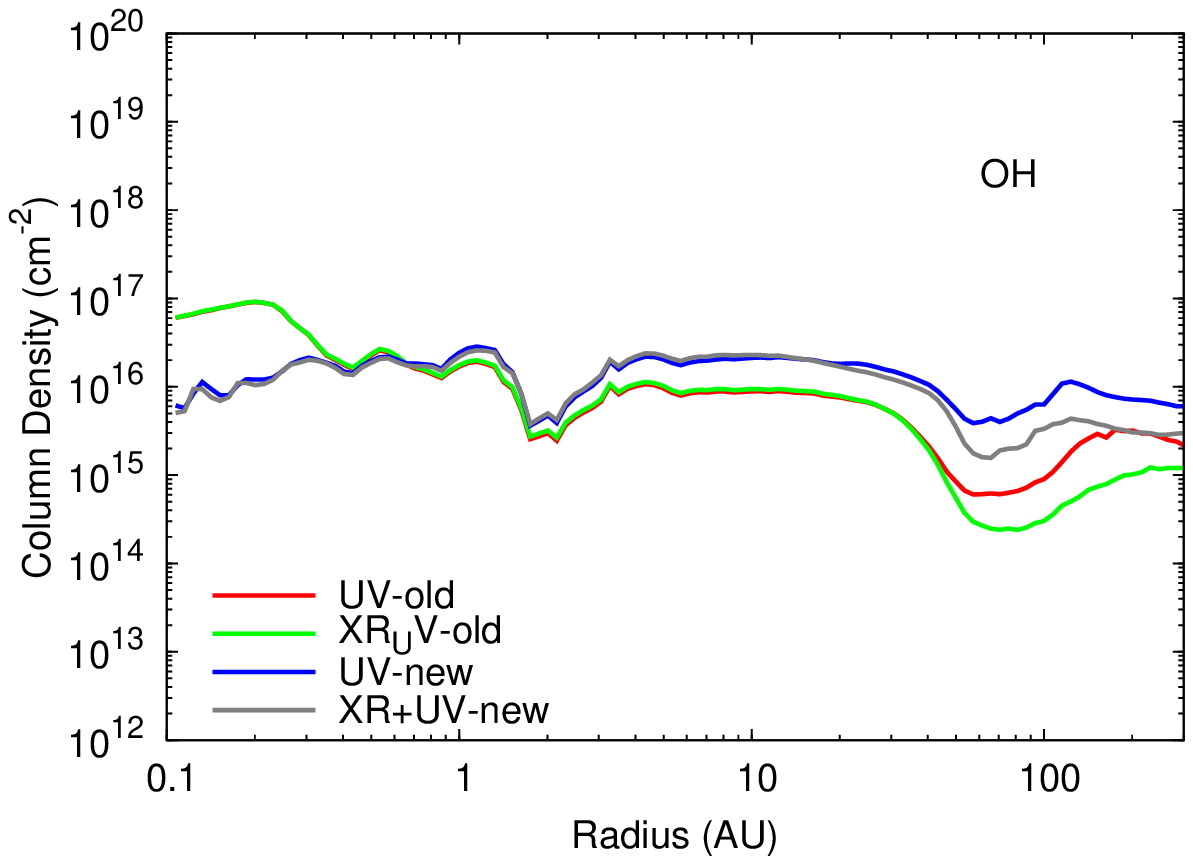}}
\subfigure{\includegraphics[width=0.32\textwidth]{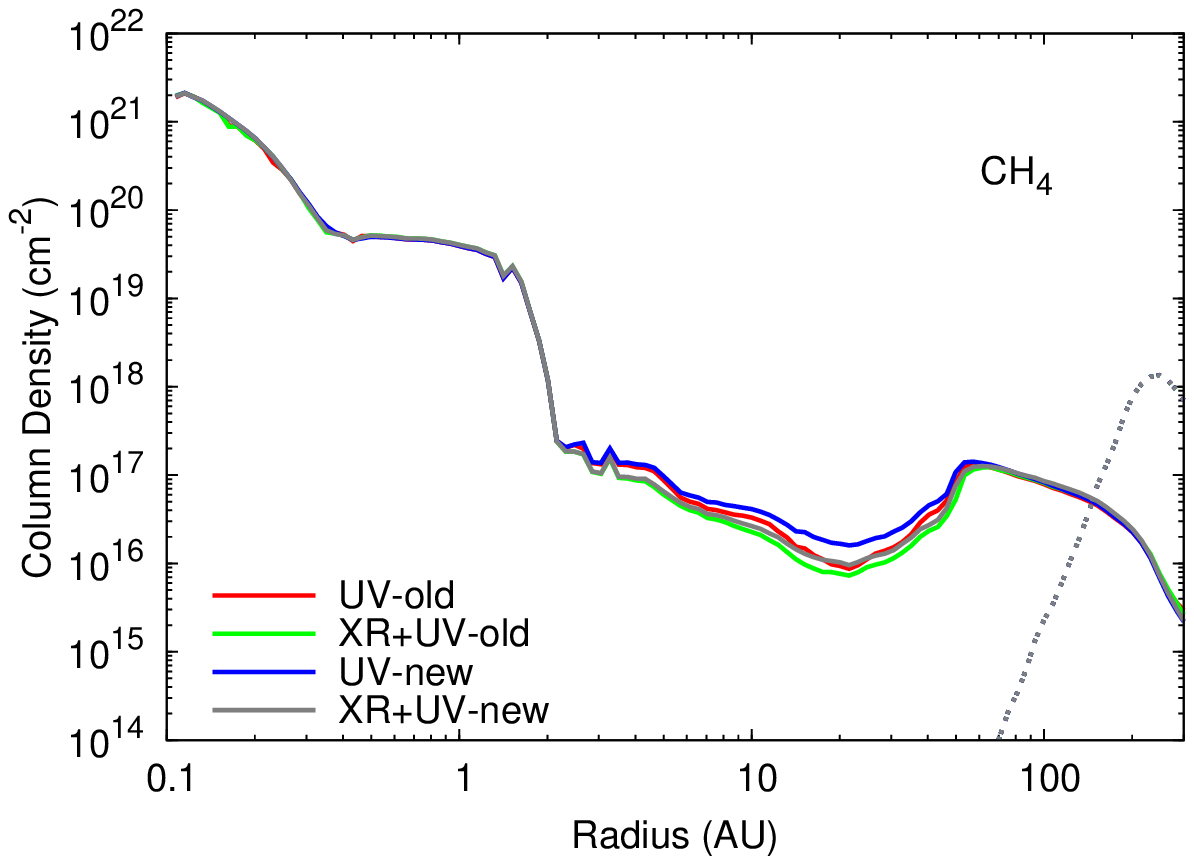}}
\subfigure{\includegraphics[width=0.32\textwidth]{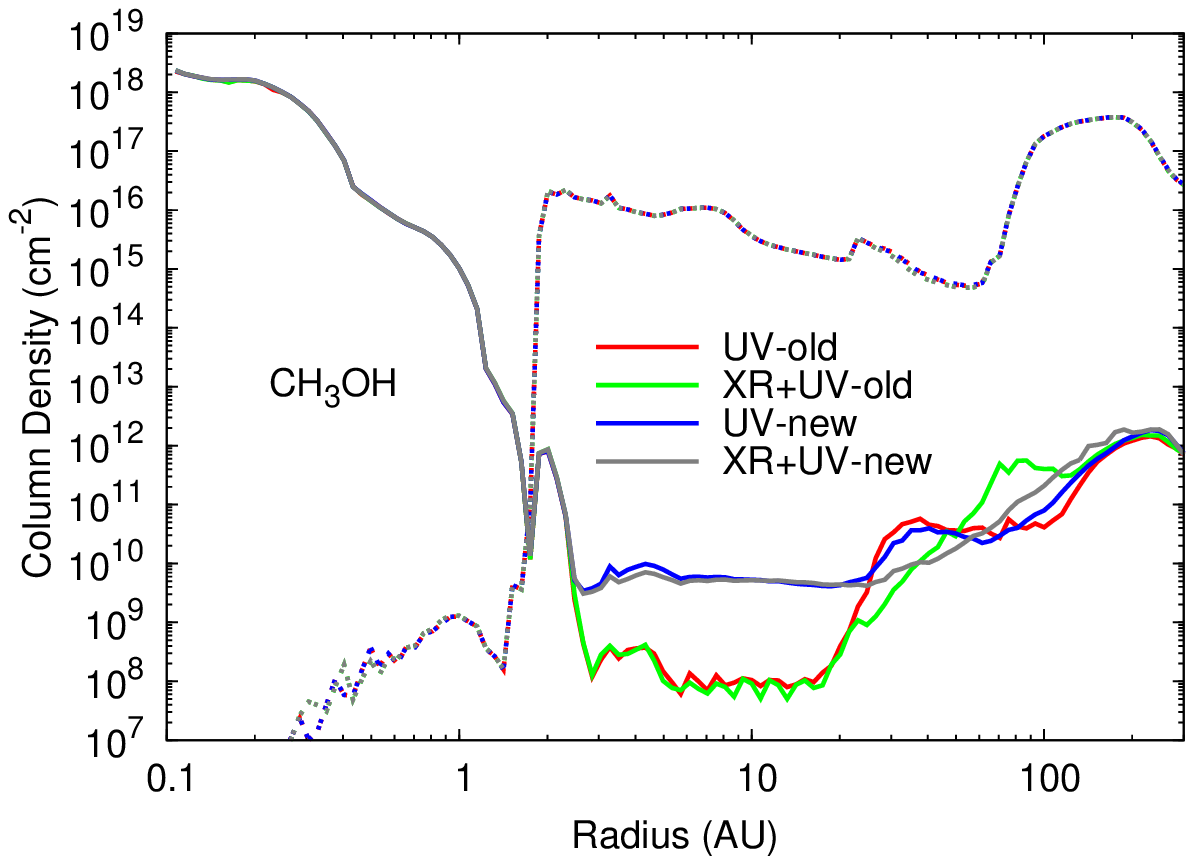}}
\subfigure{\includegraphics[width=0.32\textwidth]{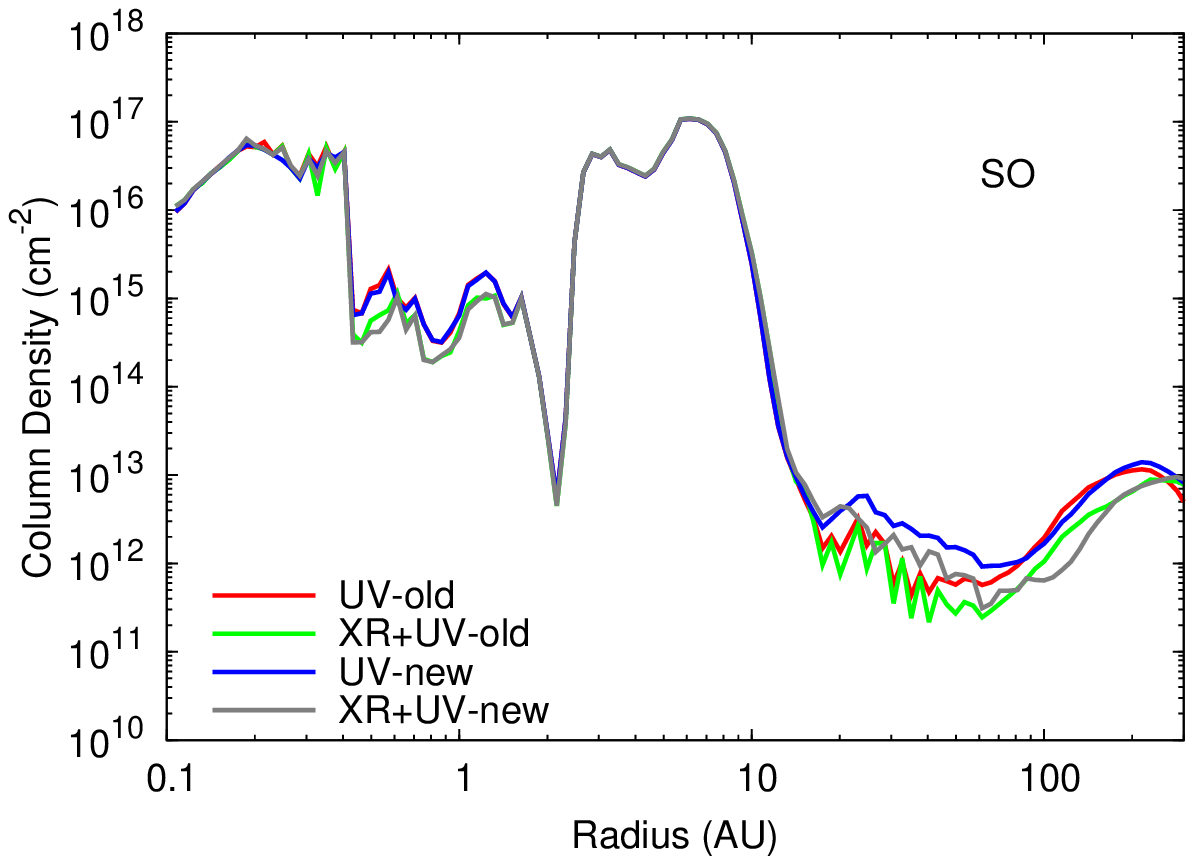}}
\caption{Column density (cm$^{-2}$) as a function of radius, $R$, for a range of molecules detected or searched for 
in protoplanetary disks.  The dashed lines represent grain-surface (ice) column densities.}
\label{figure3}
\end{figure*}

In Figure~\ref{figure3}, we present the radial column density 
(cm$^{-2}$) of molecules detected 
or searched for in protoplanetary disks at both (sub)mm and infrared wavelengths. 
The solid lines and dashed lines are the gas-phase and grain-surface column densities, respectively.  

Molecules whose column densities are relatively unaffected by the method employed to calculate the 
photoreaction rates and the X-ray ionization rate 
include CO, CN, CS, C$_2$H, C$_2$H$_2$, H$_2$CO, HCN, OH, CH$_{4}$ and SO.  
Over the radial extent of the disk, the 
column densities of the listed species vary, at most, by a factor of a few between chemical models.  
The gas-phase column densities of H$_2$O and CH$_3$OH are affected only beyond a radius of 
$\approx$~1~AU with that of CO$_2$ altered beyond a radius of $\approx$~10~AU.  
In all three cases we see a rise in the column density of each species when the photorates are recalculated. 
For the molecular ions, HCO$^{+}$ and N$_2$H$^+$, we see a different behaviour with 
the column density of the former species affected significantly within a radius of 
approximately 1~AU only.  
For the latter molecule, the column density is affected throughout the 
radial extent of the disk.  
For all molecules which possess an appreciable grain-surface column density, the values are 
relatively unaffected by the method used to compute the photochemical rates and the X-ray ionization rate.  

Overall, the recalculation of the photoreaction rates has a much bigger effect on the column densities than 
that of the X-ray ionization rate with models UV-old and XR+UV-old, on the whole, producing similar values and behaviour.  
This is also true for models UV-new and XR+UV-new.  
Exceptions to this include N$_2$H$^+$ (throughout the disk) and OH 
(in the outer disk beyond a radius of $\approx$~50~AU).  
A further general observation is that the recalculation of the photorates affects each molecule in a different
manner.  The spectrum-dependent photorates do not only directly affect the abundance and distribution of atoms 
and molecules which can undergo photoionization and photodissociation, they also indirectly 
affect the subsequent gas-phase chemistry, leading to enhancements/depletions of molecules 
which are not directly formed or destroyed via a photochemical route, e.g., HCO$^+$.  

We see an interesting structure in the column densities of the sulphur-bearing species, CS and SO.  
Both show a peak between $\approx$~2~AU and $\sim$~10~AU.  
In the outer disk, beyond $\sim$~10~AU, sulphur exists primarily in the disk midplane as H$_2$S ice on the 
grain mantle.  
Within 10~AU, the disk midplane is warm enough for H$_2$S to evaporate from the grain mantle replenishing the 
gas with sulphur which forms SO and CS.  
Within $\approx$~2~AU, SO$_2$ takes over from SO as the dominant gas-phase sulphur-bearing species.  
CS increases in abundance again within $\sim$~1~AU in the very warm, dense midplane.

The radial column densities, $N$, of a selection of molecules are listed in Table~\ref{table4} at radii of 
0.1, 1, 10 and 100~AU.  Molecules whose column densities are affected by more than one order of magnitude 
are highlighted in bold text.  
The general trend we see (with several exceptions discussed below) is a reduction in molecular column densities 
in the inner disk (within 1~AU) and an increase in the outer disk (beyond 1~AU) 
in models UV-new and XR+UV-new relative to model UV-old.  
At 0.1~AU, $N$(HCO$^+$) is reduced by two orders of magnitude with $N$(OH) reduced by one
order of magnitude, when the photochemistry is recalculated.  
The column density of N$_2$H$^+$ is affected by both X-ray ionization and photochemistry with
the former increasing the value relative to model UV-old and the latter reducing it.  
At 1~AU, only $N$(N$_2$H$^+$) is significantly affected where it is enhanced by almost two
orders of magnitude by the recalculation of the photochemistry 
and reduced by a factor of a few by the recalculation of the X-ray ionization rate.  
At 10~AU, the molecules affected are H$_2$O, CO$_2$, CH$_3$OH and N$_2$H$^+$.  
The column density of H$_2$O and CO$_2$ are both enhanced by over an order of magnitude in models UV-new 
and XR+UV-new.  
As at 1~AU, X-ray ionization reduces $N$(N$_2$H$^+$), whereas photochemistry
enhances it.  
It is a similar story for methanol, however, X-ray ionization has a much smaller effect than
photochemistry.  
At the final radius we consider here, 100~AU, again, N$_2$H$^+$ and CH$_3$OH are 
affected in a similar manner as at 10~AU.  
We also find the column density of OH increased significantly in model 
UV-new with that for CO$_2$ also enhanced in models UV-new and XR+UV-new.  
The only molecules for which X-ray ionization significantly affects the column density are 
N$_2$H$^+$ and CH$_3$OH.

The molecular column densities, although useful for tracing the general 
radial structure and identifying those molecules significantly affected by the inclusion or omission of each process, 
provide little information on the spatial distribution and abundance.  
This ultimately affects the strength of the line emission from the disk since this is influenced 
by the physical conditions in the region where each molecule is most abundant. 
In the following Section, we look more closely at the effects on the two-dimensional molecular structure
of the disk due to the recalculation of the photoreaction and X-ray ionization rates.  

\subsection{Photochemistry}
\label{photochemistryeffects}

\begin{figure*}
\centering
\subfigure{\includegraphics[width=0.4\textwidth]{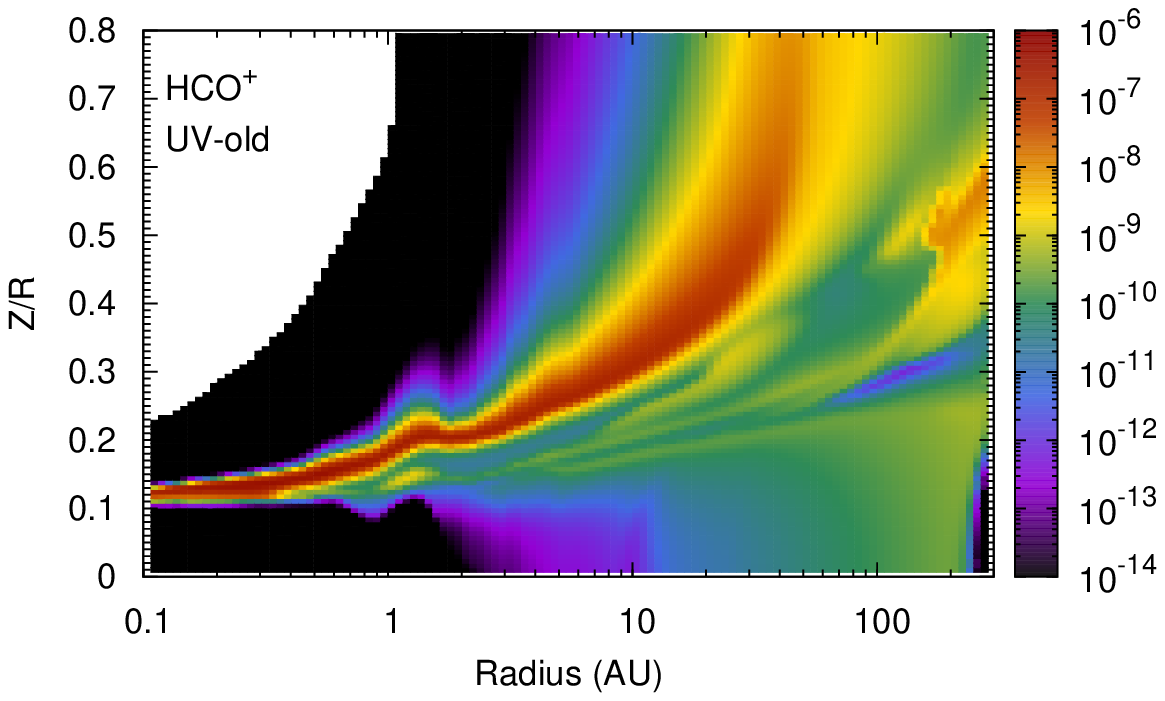}}
\subfigure{\includegraphics[width=0.4\textwidth]{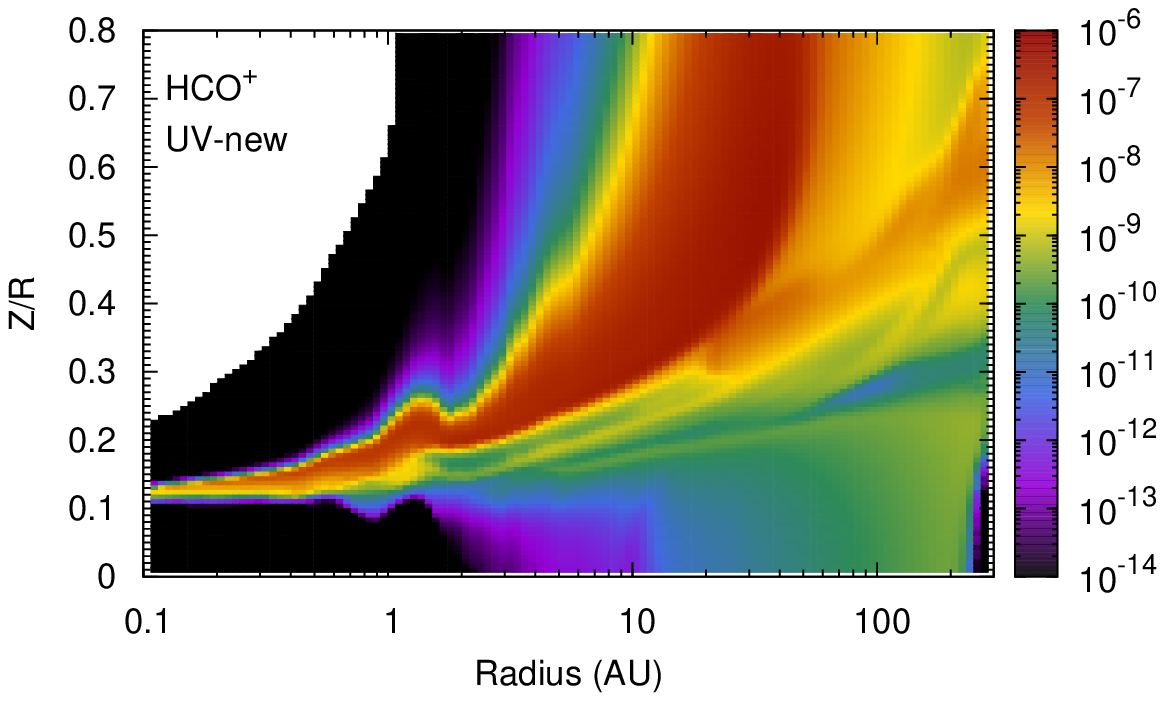}}
\subfigure{\includegraphics[width=0.4\textwidth]{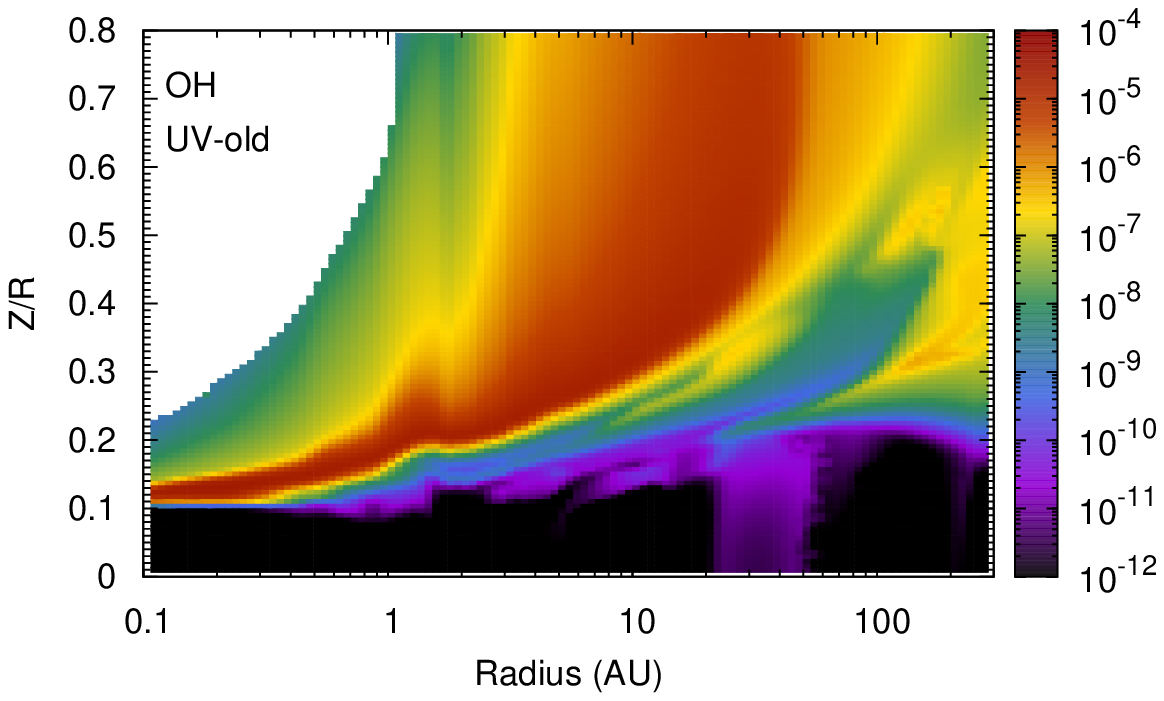}}
\subfigure{\includegraphics[width=0.4\textwidth]{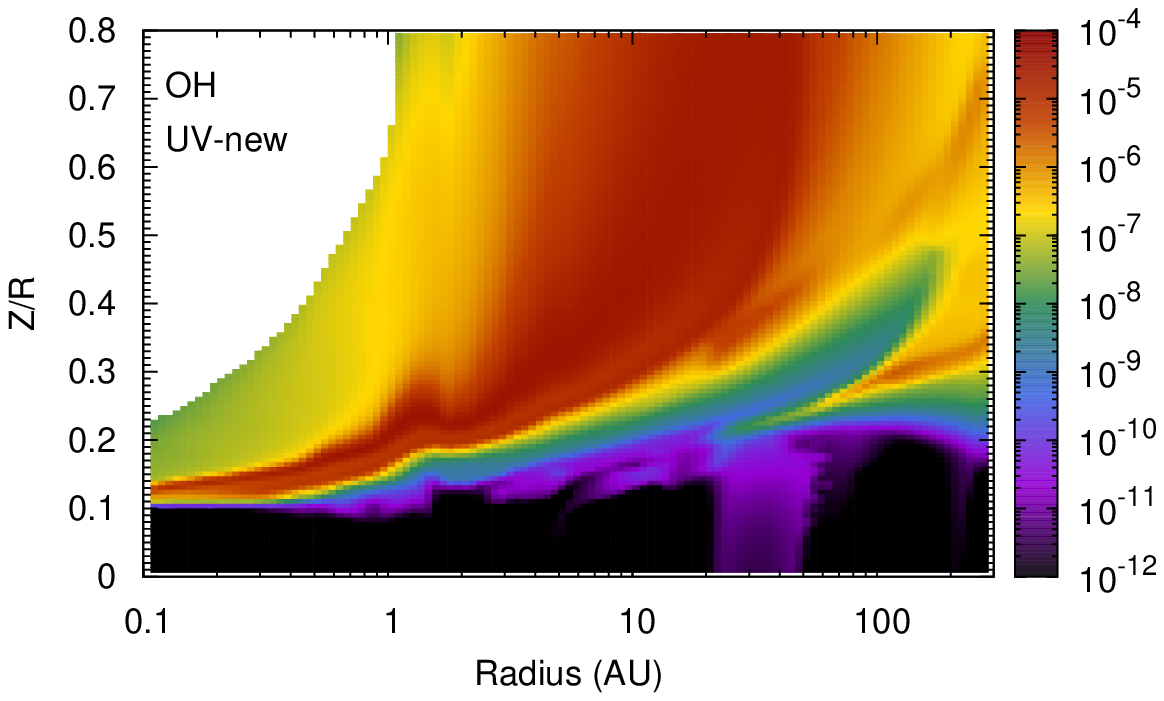}}
\subfigure{\includegraphics[width=0.4\textwidth]{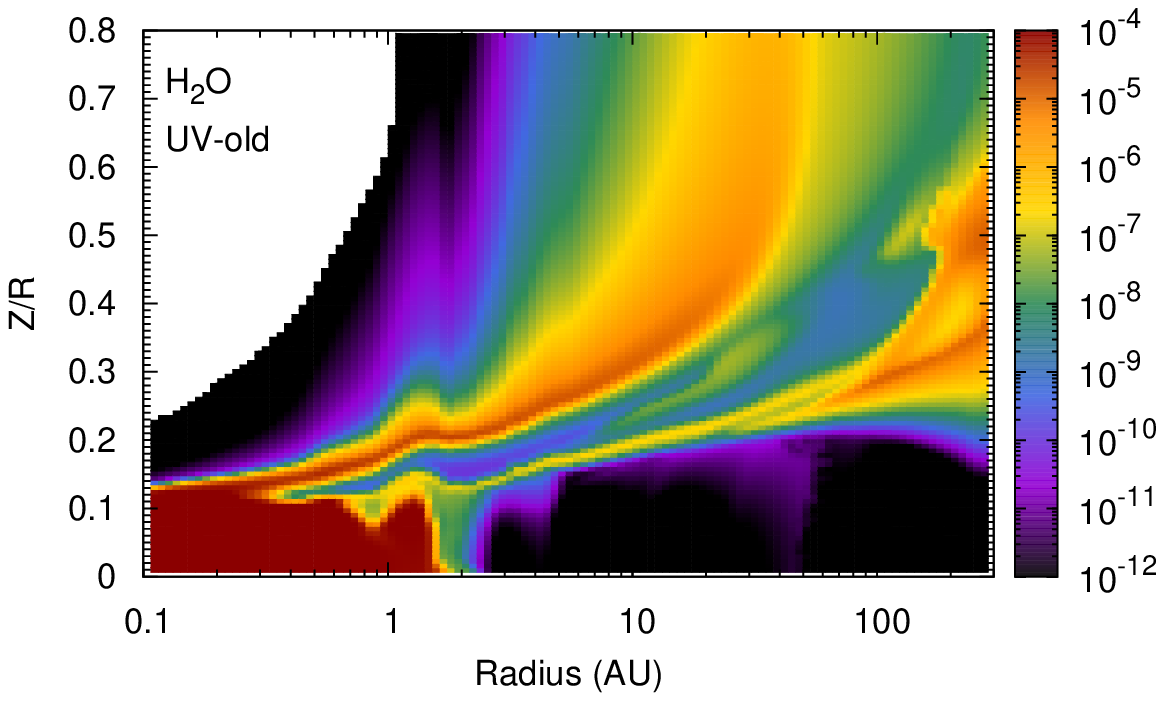}}
\subfigure{\includegraphics[width=0.4\textwidth]{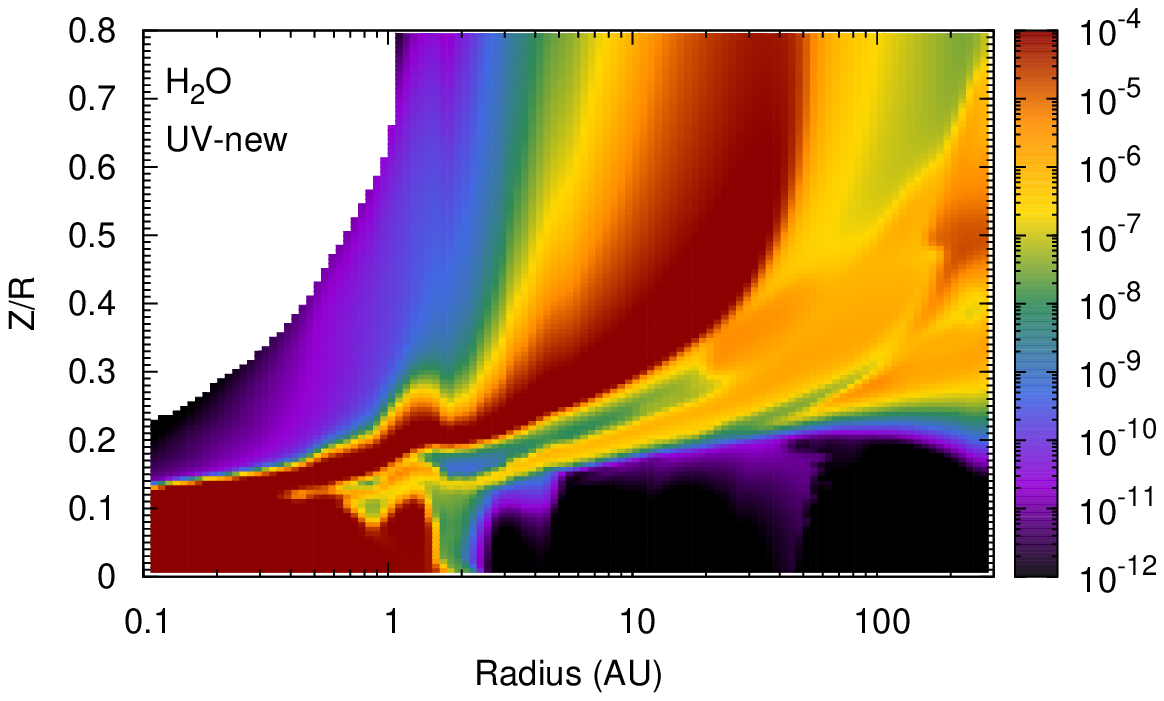}}
\subfigure{\includegraphics[width=0.4\textwidth]{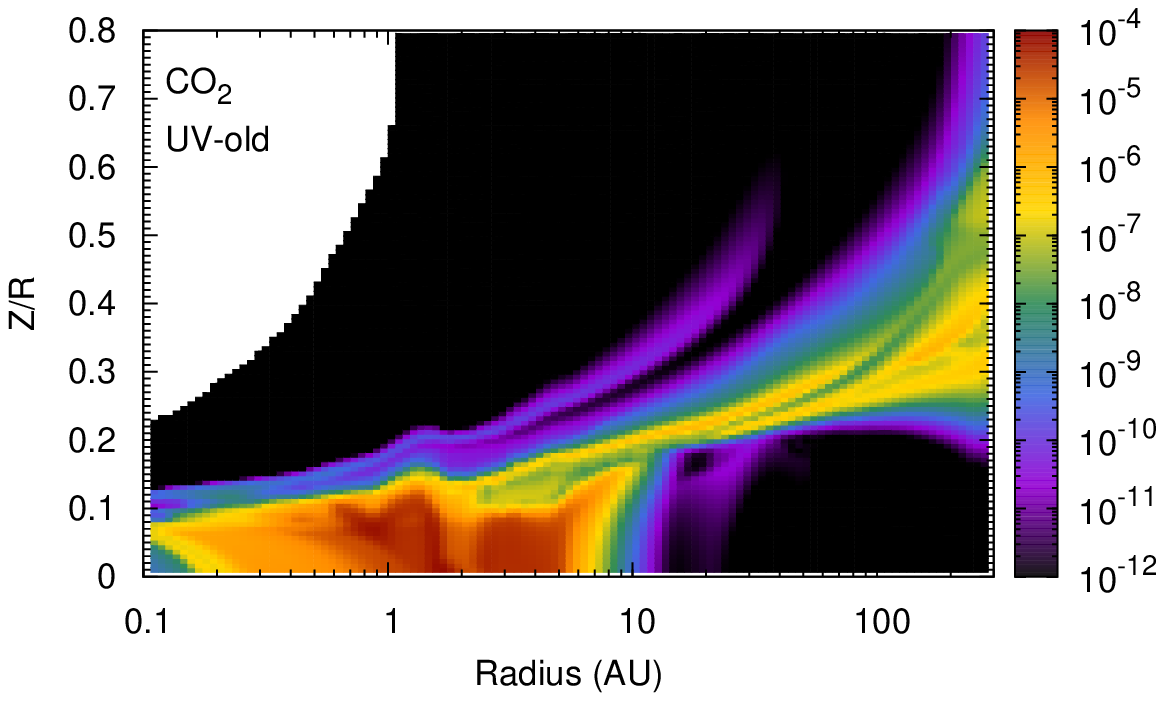}}
\subfigure{\includegraphics[width=0.4\textwidth]{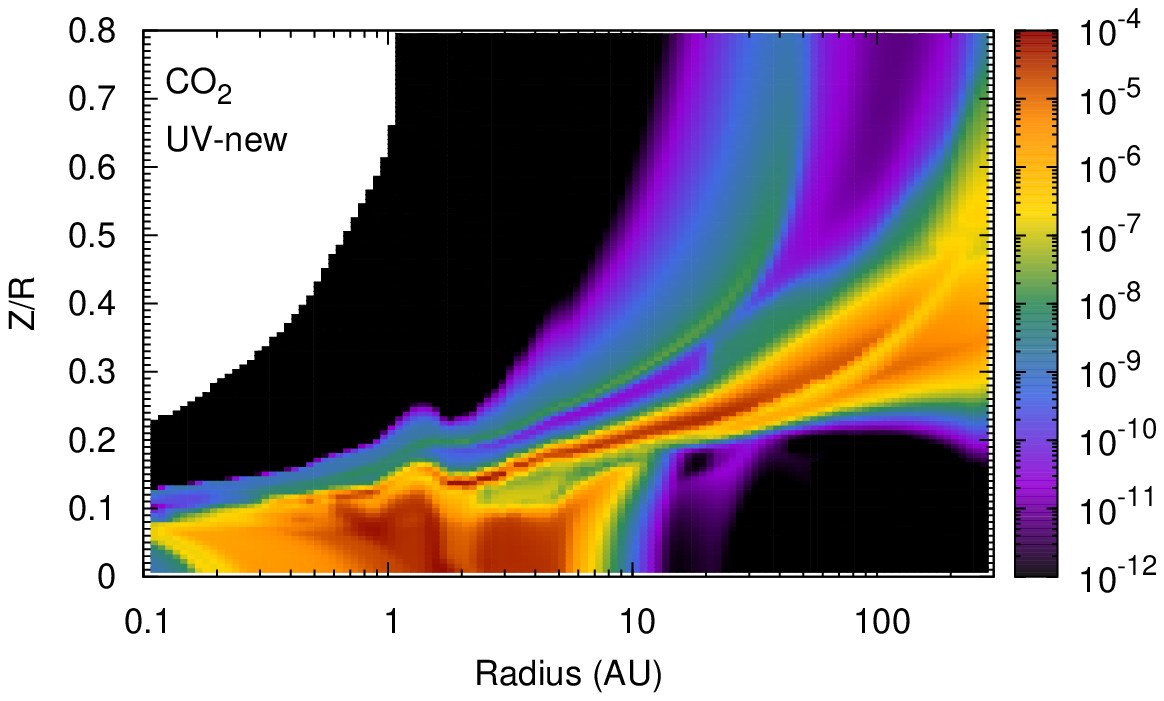}}
\subfigure{\includegraphics[width=0.4\textwidth]{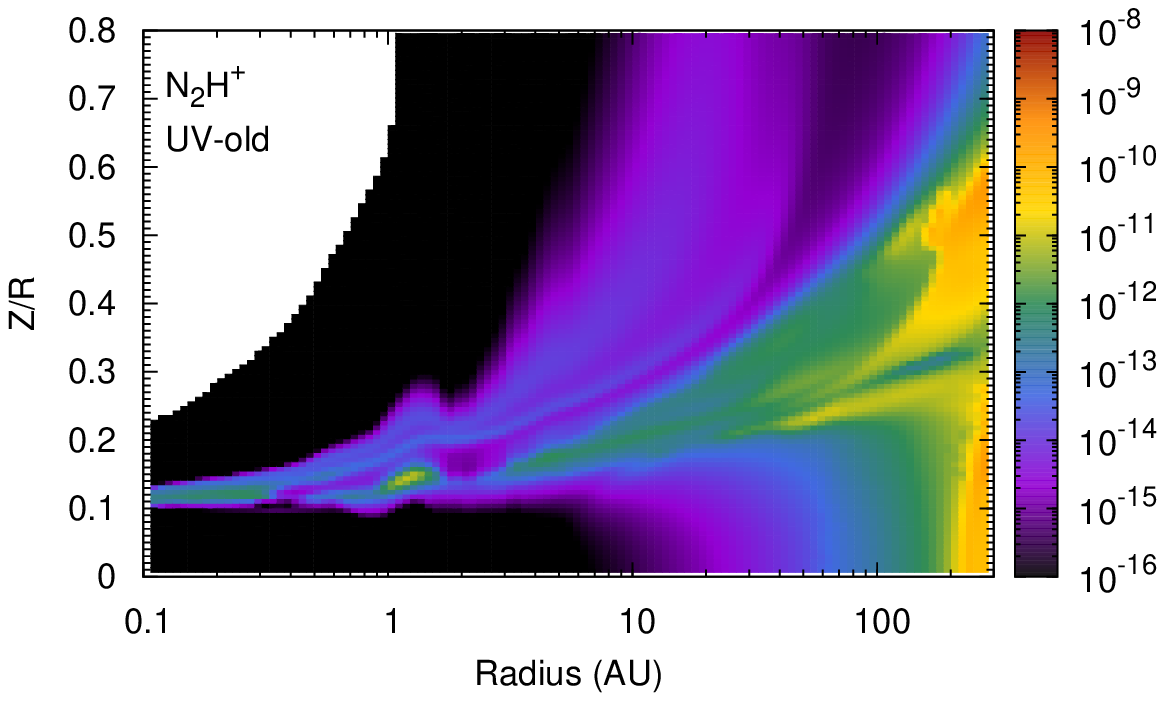}}
\subfigure{\includegraphics[width=0.4\textwidth]{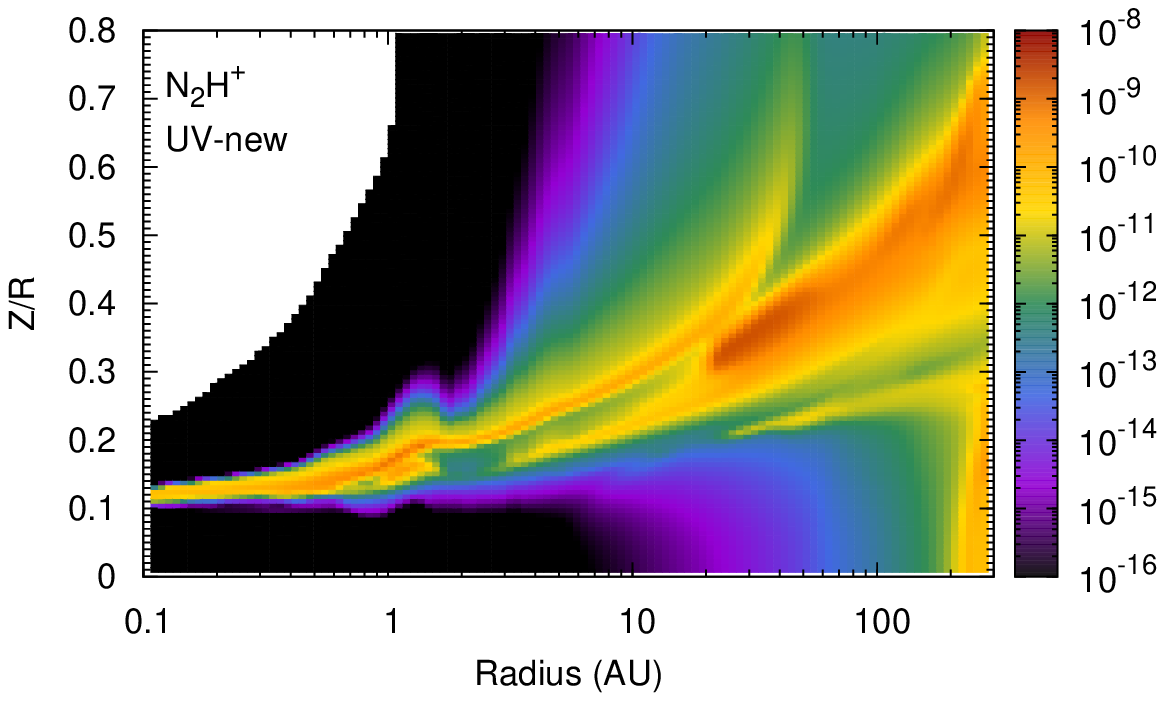}}
\caption{The fractional abundance of HCO$^+$, OH, H$_2$O, CO$_2$ and N$_2$H$^+$
as a function of disk radius, $R$, and height (scaled by the radius i.e., $Z/R$)
for models UV-old (left) and UV-new (right).}
\label{figure4}
\end{figure*}

In Figure~\ref{figure4} we display the fractional abundance of HCO$^+$, OH, H$_2$O, CO$_2$ 
and N$_2$H$^+$ as a function 
of disk radius and height (scaled by the radius) for models 
UV-old (left column) and UV-new (right column).  
These molecules are those we have identified as being most affected 
by the recalculation of the photorates.  

In the plots for HCO$^+$, it is clearly seen that within a radius of $\approx$~1~AU, the fractional abundance 
of HCO$^+$ in the `molecular layer' in this region, located 
at $Z/R$~=~0.1, is around two orders of magnitude lower in model UV-new than in
model UV-old.  Also clearly visible is the reason for the larger column density calculated beyond this radius, the
depth of the layer of HCO$^+$ in model UV-new is much larger than that in model UV-old although the maximum fractional 
abundance attained in both models is similar ($x$(HCO$^{+}$)~$\sim$~10$^{-6}$). 
The abundance of HCO$^+$ is controlled by ion-molecule chemistry and thus depends on the abundance of 
ionic and neutral precursors. An example of an ion-molecule gas-phase 
reaction which leads to the production of HCO$^+$ is
\begin{eqnarray*}
\mbox{CO} + \mbox{H}_3^+ & \longrightarrow & \mbox{HCO}^+ + \mbox{H}_2. 
\end{eqnarray*} 
In model UV-new, where we see an increase in HCO$^+$, we see a corresponding increase in CO and H$_3$$^+$.  
CO is directly influenced by the photochemistry since it can be photodissociated to 
produce C and O.  H$_3$$^+$, on the other hand, is primarily formed via the reaction of H$_2$ with 
H$_2$$^+$ and destroyed via electron recombination. 
The enhancement in HCO$^+$ in model UV-new 
also corresponds to where we see a slight decrease in electron abundance 
(see Figure~\ref{figure7}).  Relating this back to the photochemistry, this indicates 
an overestimation in this region in both the photodissociation of CO and photoionization (in general) 
in our fiducial model (model UV-old).  

The depth of the molecular layer where OH reaches its maximum fractional abundance 
($x$(OH)~$\sim$~10$^{-4}$) within 1~AU is smaller in model UV-old than in model UV-new, whereas, 
beyond this radius, the depth is larger.  
This accounts for the smaller column density in model UV-new at 1~AU and the larger at 10 and 100~AU.  
The distribution of OH is also somewhat different in model UV-new reaching an order of magnitude
higher fractional abundance in the disk surface throughout the radial extent.  
The abundance of OH is directly controlled by the photochemistry since 
OH is one of the products of the photodissociation of H$_2$O and can itself be photodissociated 
to form O and H.    
We see in the corresponding plots for water in Figure~\ref{figure4} that
its fractional abundance is also slightly higher in the disk surface in model UV-new relative to model UV-old.  
We also find that the abundance of free oxygen atoms decreases in this region, indicating in model UV-new, 
more atomic oxygen is locked up in oxygen-bearing molecules than in model UV-old.  

Looking at the distribution of water, we see a large enhancement in the fractional abundance 
in the molecular layer beyond a radius of $\approx$~1~AU, going from a value of $\sim$~10$^{-6}$ in 
model UV-old to $\sim$~10$^{-4}$ in model UV-new.  
Not only is the maximum fractional abundance reached much higher, the extent over which water exists with 
a value $\gtrsim$~10$^{-9}$ in model UV-new is also much larger.  
We conclude that the abundance of gas-phase water is very sensitive to the method 
employed to calculate the photorates.  

\begin{figure}
\includegraphics[width=0.5\textwidth]{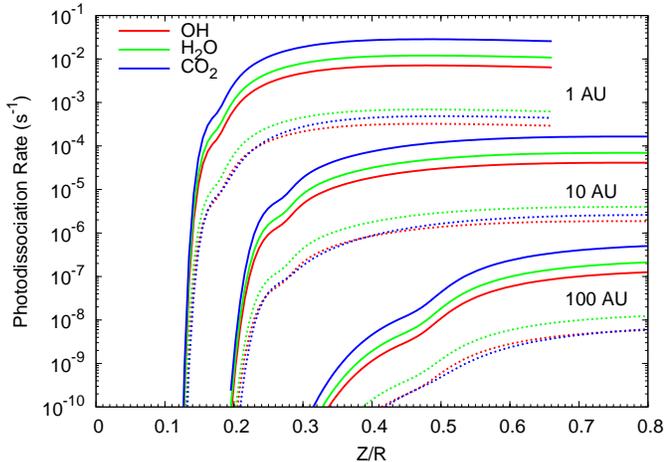}
\caption{The photodissociation rates of OH, H$_2$O and CO$_2$ as a function of 
disk height (scaled by the radius i.e., $Z/R$) at radii of 1~AU, 10~AU and 100~AU for models 
UV-old (solid lines) and UV-new (dashed lines).}
\label{figure5}
\end{figure}

Both OH and H$_2$O can be formed via neutral-neutral gas-phase reactions in warm regions of the disk where 
$T$~$\gtrsim$~200~K \citep{glassgold09}.
\begin{eqnarray*}
\mbox{H}_2 + \mbox{O} & \longrightarrow & \mbox{OH} + \mbox{H} \\
\mbox{H}_2 + \mbox{OH}& \longrightarrow & \mbox{H}_2\mbox{O} + \mbox{H}
\end{eqnarray*}
Since we see an increase in the
fractional abundance of H$_2$ in model UV-new relative to UV-old 
over the region where both OH and H$_2$O are increased, this
gas-phase production route for both species is more important in model UV-new than in model UV-old.  
Hence, the enhancement seen in the abundances of OH and H$_2$O in model UV-new is a 
combination of increased gas-phase production and decreased photodestruction.  
In Figure~\ref{figure5}, we display the photodissociation rates 
of OH and H$_2$O as a function of disk height (scaled by the radius) at radii of 1~AU, 10 ~AU and 100~AU.  
A decrease of around an order of magnitude in the photodissociation rates of both species 
is clearly seen in the upper disk layers accounting for the increase in abundance of both species in this region.  
The gross overestimation of the the photodissociation rates in model UV-old is due to the inclusion of 
Lyman-$\alpha$ photons in the calculation of the wavelength-integrated UV photon flux 
(see Section~\ref{photochemistry}).  
Also, note that the dissociation rates vary differently as a function of height in model 
UV-new versus UV-old 
since in the former model, the wavelength dependence of both the photo cross sections and UV field 
are included.

Gas-phase CO$_2$ is affected mainly 
in the outer disk beyond a radius of $\approx$~10~AU.   
We see that CO$_2$ in this region exists in a layer lower than that of water 
(due to its lower binding energy to dust grains) 
and is enhanced in the outer disk from a fractional abundance $\sim$~10$^{-6}$ 
in model UV-old to $\sim$~10$^{-4}$ in model UV-new.  
In model  UV-new, in this region, it possesses a comparable fractional abundance to the 
other main oxygen-bearing species, CO and O$_2$.  
As in the case for water, the extent over which $x$(CO$_2$) has a value $\gtrsim$~10$^{-9}$ is much larger 
in model UV-new than in model UV-old.   
We see an enhancement in CO$_2$ when the fractional abundance of ionic carbon begins to increase 
relative to atomiccarbon.  
Note, CO$_2$ is destroyed via reaction with C$^+$ to form CO$^+$ and CO.   
In model UV-new, the boundary where C$^+$/C~$\sim$~1 is much higher in the disk than in model UV-old 
(at a height of $\approx$~5~AU and $\approx$~2~AU, respectively, at a radius of 10~AU).
Thus in model UV-new, gas-phase CO$_2$ can evaporate from the grain surface and 
remain intact in this layer due to the lack of sufficient gas-phase destruction mechanisms.   
We display the photodissociation rate of CO$_2$ as a function of disk height 
(scaled by the radius) at radii of 1~AU, 10~AU and 100~AU in Figure~\ref{figure5}.  
As in the case for OH and H$_2$O, we see an decrease of around an order of magnitude in model 
UV-new compared with UV-old.  This accounts for the 
increase in the abundance of CO$_2$ in the upper disk layers in model UV-new.  
Note, in model UV-old, the dissociation rates for CO$_2$ are consistently larger 
than that of H$_2$O and OH, reflecting the the rates calculated for the 
unshielded interstellar medium (14, 5.9 and 3.5 $\times$~10$^{-10}$~s$^{-1}$, respectively).  
However, in model UV-new, the H$_2$O photodissociation has the largest rate at each 
radius, with OH also having a larger rate than CO$_2$ at 100~AU.  
This demonstrates how the relative photorates are sensitive to the shape of the radiation field 
due to the specific variation in the photo cross section of each species.  
 
When the photochemistry is recalculated, we see a dramatic increase in the region over which N$_2$H$^+$ 
reaches its maximum fractional abundance, now residing in a molecular layer which permeates the entire disk.  
The most dramatic increase is seen in the inner disk ($<$~100~AU) where $x$(N$_2$H$^+$) 
jumps from a maximum value of $\sim$~10$^{-12}$ in model UV-old to $\sim$~10$^{-9}$ in model UV-new.  
Similar to HCO$^+$, the abundance of N$_2$H$^+$ is controlled by ion-molecule chemistry, it is formed 
via the reaction of N$_2$ with H$_3$$^+$ and is destroyed
effectively by reaction with CO and electron recombination.  
Looking at Figure~\ref{figure7} which displays the electron fractional abundance for all four models as a function of
disk radius and height, we see in the
layer where N$_2$H$^+$ is significantly enhanced, there is a respective decrease in the electron 
fraction in model UV-new. 
N$_2$H$^+$ and HCO$^+$ show different behaviours since the abundance of N$_2$H$^+$ is more sensitive to the 
abundance of H$_3$$^+$ than HCO$^+$.  
HCO$^+$ has various alternative routes to formation other than the reaction of CO and H$_3$$^+$, e.g., C$^+$ + H$_2$O, 
whereas N$_2$H$^+$ forms mainly via the reaction of 
N$_2$ with H$_3$$^+$.  We discuss N$_2$H$^+$ chemistry further in Section~\ref{xrayionizationeffects}.

\begin{figure}
\subfigure{\includegraphics[width=0.5\textwidth]{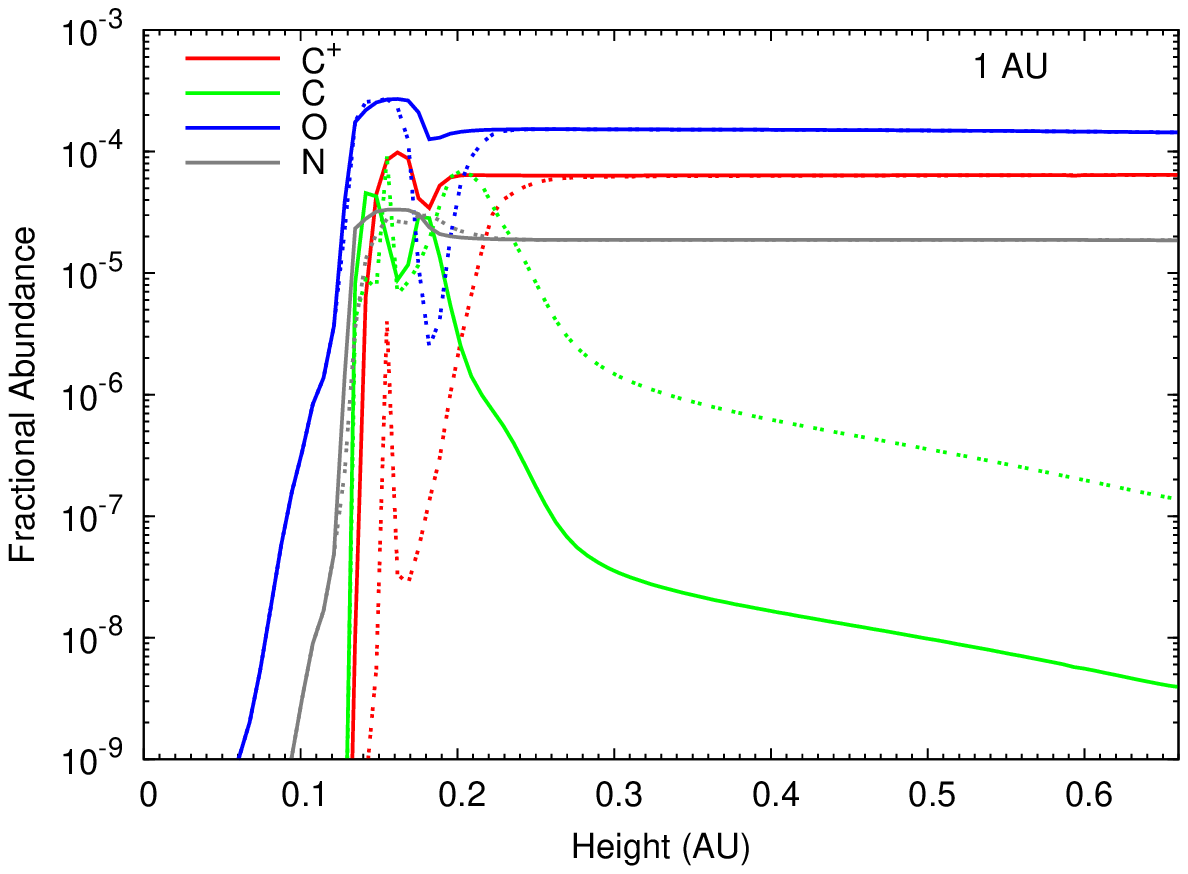}}
\subfigure{\includegraphics[width=0.5\textwidth]{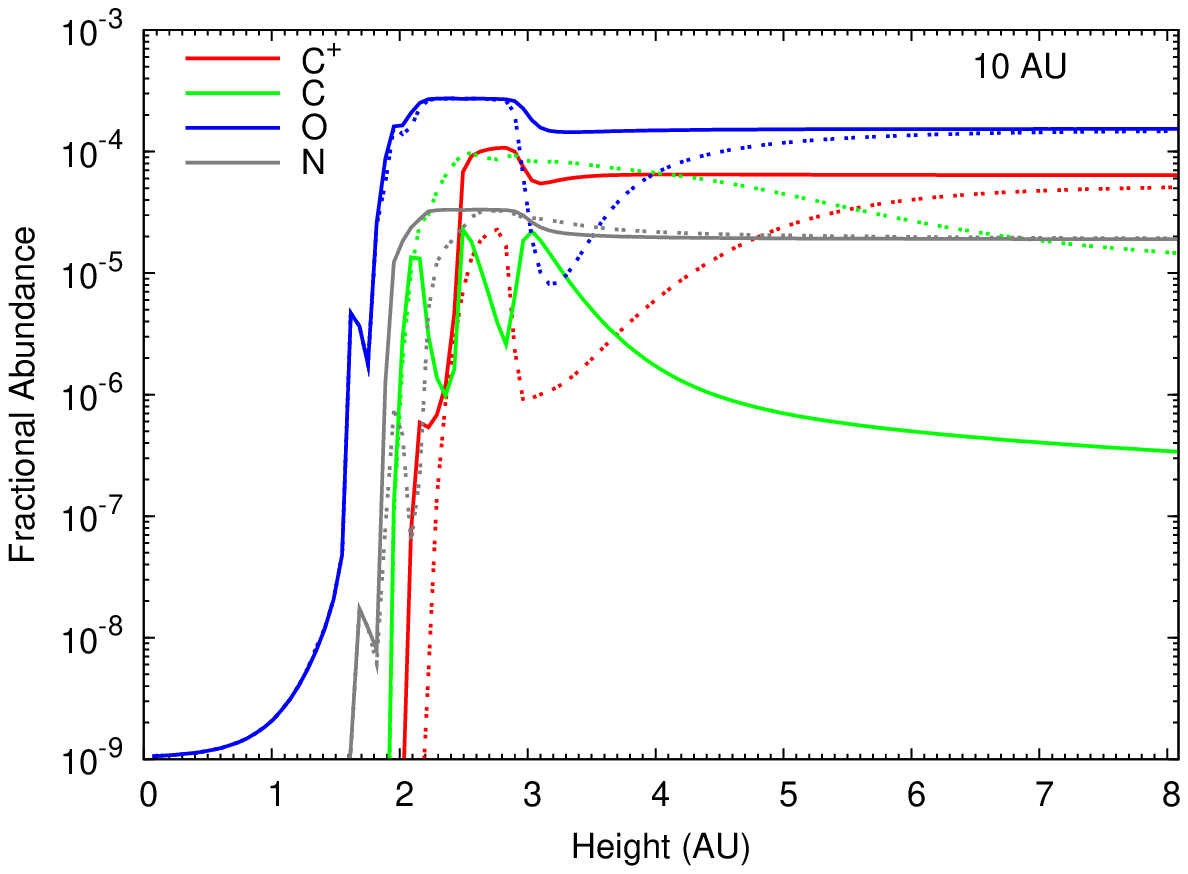}}
\caption{The fractional abundance of C$^+$, C, O and N   
as a function of height, $Z$, at radii of 1~AU (top) and 10~AU (bottom) 
for models UV-old (solid lines) and UV-new (dashed lines).}
\label{figure6}
\end{figure}

\begin{figure*}
\subfigure{\includegraphics[width=0.5\textwidth]{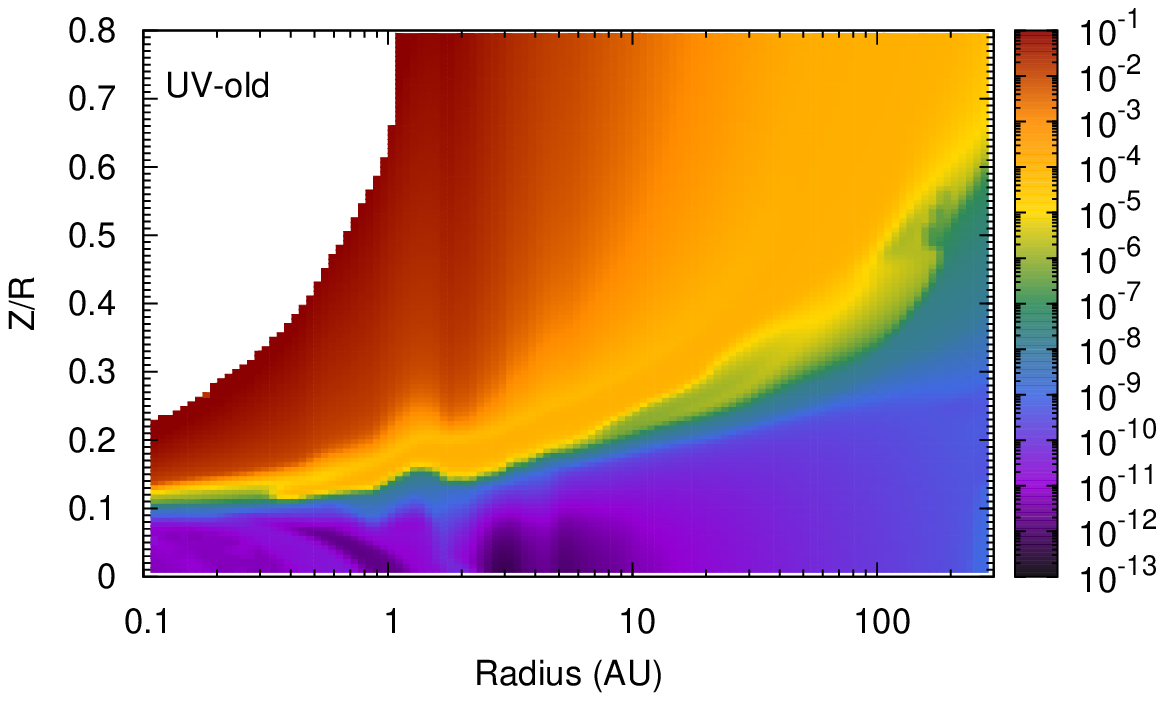}}
\subfigure{\includegraphics[width=0.5\textwidth]{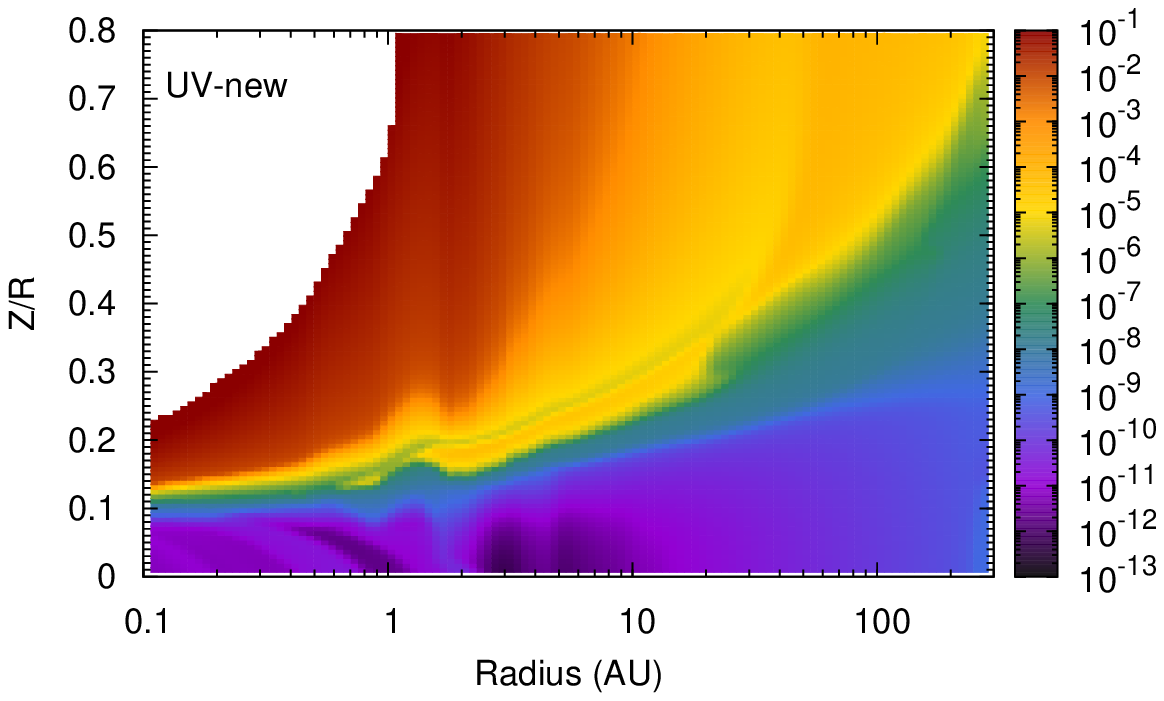}}
\subfigure{\includegraphics[width=0.5\textwidth]{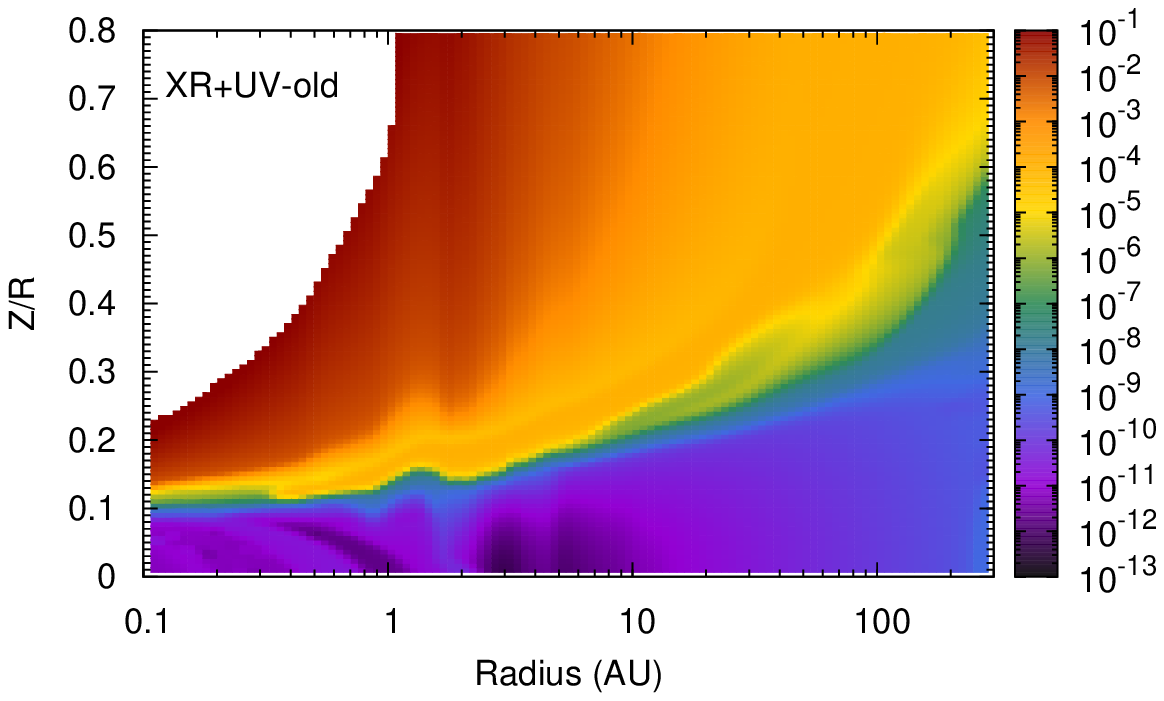}}
\subfigure{\includegraphics[width=0.5\textwidth]{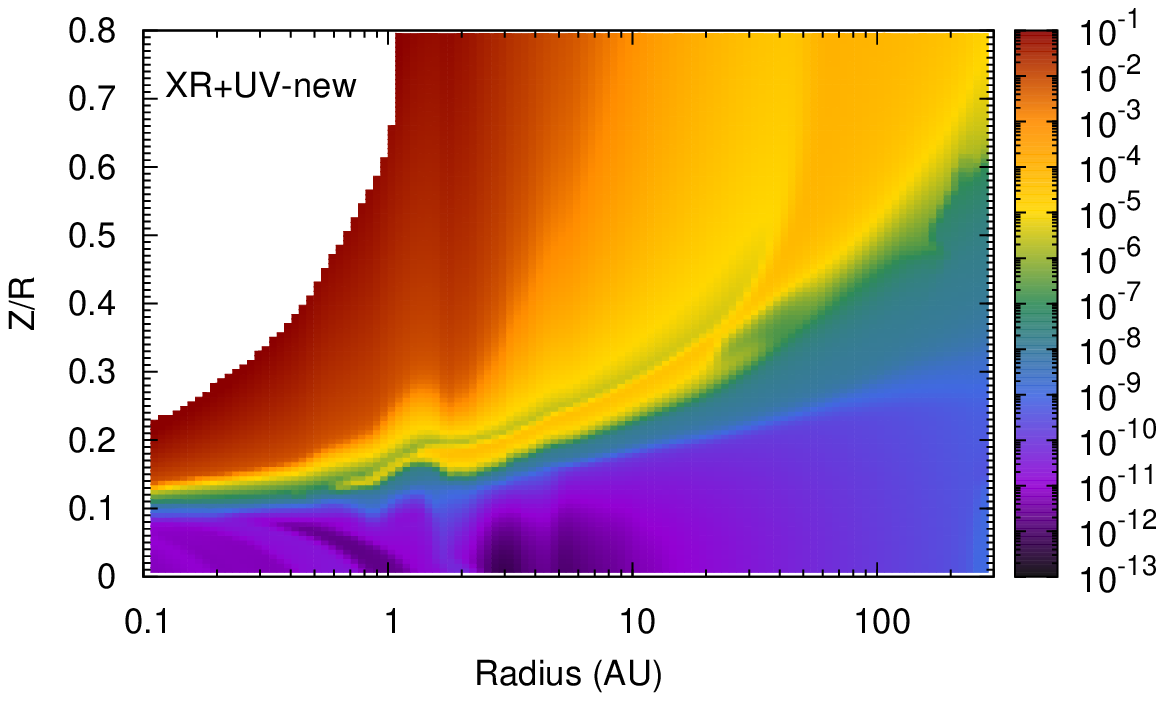}}
\caption{The fractional abundance of electrons as a function of disk radius, $R$, and height (scaled by the radius
i.e., $Z/R$)
for models UV-old (top left),  UV-new (top right), XR+UV-old (bottom left) and  XR+UV-new (bottom right).}
\label{figure7}
\end{figure*}

In Figure~\ref{figure6}, we present the fractional abundances of several atoms and ions at a radius 
of 1 (top panel) and 10~AU (bottom panel).  
We display the results for C$^+$ also since the ionization potential for carbon is much lower than that for
oxygen and nitrogen.  
In model UV-new, there is a significant depletion of free O atoms and C$^+$ ions in the region 
where we see an increase in the fractional abundance of OH, H$_2$O and the main carbon species, C.   
Note that the atomic carbon abundance in the disk surface is larger in model UV-new than in 
model UV-old, however, here
free carbon exists mainly in ionic form as C$^+$.  
In this region, we also see an enhancement in the fractional abundances of the molecular ions, HCO$^+$ and 
N$_2$H$^+$.  
Hence, in model UV-new, in the molecular layer where molecules reach their maximum fractional abundance, more 
oxygen atoms are contained within molecules than in model UV-old.  
Also, in this region, less carbon exists as C$^+$ so the ratio of C$^+$/C is generally 
lower in model UV-new than in model UV-old.  
This can be attributed to (1) the photorates in model UV-old being overestimated for particular molecules and atoms 
and (2) an adjustment of the gas-phase chemistry due to the enhancement/depletion of species 
directly affected by the radiation field.  

The layer in which there is an enhancement in the abundance of molecules also coincides with the boundary where
we cross from hydrogen in mainly molecular form (as H$_2$) to atomic form.  
In model UV-new, this transition is smoother and results in a slightly lower fractional abundance of H atoms over the
extent of the molecular layer.  
At high temperatures, the destruction of molecules by free hydrogen atoms can become an important process.  
The region where we find the large increase in OH and H$_2$O, coincides with a gas temperature of
$\gtrsim$~1000~K so that the reduction of free H atoms (which act to destroy molecules) 
in model UV-new is also influencing the gas-phase chemistry and subsequent abundances.  

\subsection{X-ray Ionization}
\label{xrayionizationeffects}

We find the only molecule significantly affected by X-ray ionization is N$_2$H$^+$.  
We see a decrease 
in both the maximum fractional abundance attained by N$_2$H$^+$ and a reduction in the spatial 
extent over which it exists with an appreciable abundance.  
Note, in model UV-old, the distribution of N$_2$H$^+$ is also different to that of any of the other molecules 
discussed thus far, existing mainly in the outer disk beyond a radius of $\approx$~100~AU 
and extending into the disk midplane (see Figure~\ref{figure4}).  
Since the abundance of N$_2$H$^+$ is controlled by ion-molecule chemistry, the relative abundances of precursor 
ions and neutral molecules will influence the amount of N$_2$H$^+$ which can exist.  
N$_2$H$^+$ is formed in cold, dense regions via reaction of H$_3$$^+$ with N$_2$, 
the latter of which is formed via radical-radical reactions e.g., N + NH $\rightarrow$ N$_2$ + H.  
The abundance of H$_3$$^+$ is controlled by the ionization of H$_2$ to form H$_2$$^+$.  
In reactions where molecular ions can form via proton donation, the ionization rate can have a large impact on
the resulting abundances reached, as seen here for the case of N$_2$H$^+$.  
Also, N$_2$H$^+$ is effectively destroyed by electron recombination so that the ionization fraction in the disk 
plays an important role (see Section~\ref{ionizationfractionresults}).  
However, we do not see this effect in the abundance and distribution of HCO$^+$ which also 
depends, albeit to a lesser extent (see later), on the abundance of H$_3$$^+$ and 
which appears independent of the treatment of X-ray ionization. 
HCO$^+$ resides in a different layer in the disk where the abundance and distribution of H$_3$$^+$ and electrons 
do not vary significantly between models UV-old and XR+UV-old.  
We do see corresponding differences in the abundance and distribution of H$_3$$^+$, in particular, between 
models UV-old and XR+UV-old in the region where N$_2$H$^+$ is most abundant.  
HCO$^+$ also has a plethora of different routes to formation, other than the protonation of 
CO by H$_3$$^+$.  

In our disk model, X-ray ionization rate dominates in the outer disk upper layers over cosmic-ray ionization.  
Our recalculation of the rates in this region leads to a decrease in N$_2$H$^+$ indicating 
an overestimation of the X-ray ionization rate in our original model.  

\subsection{Disk Ionization Fraction}
\label{ionizationfractionresults}

\begin{figure}
\subfigure{\includegraphics[width=0.5\textwidth]{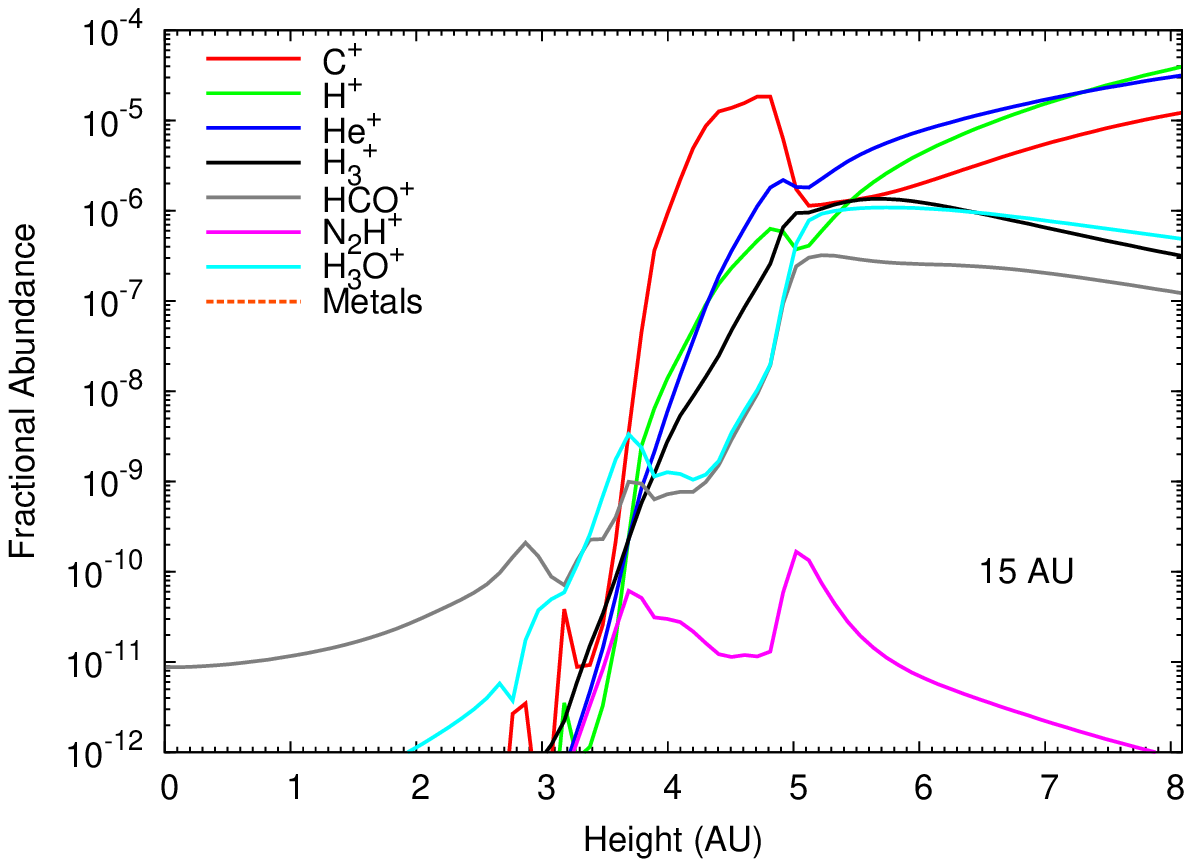}}
\subfigure{\includegraphics[width=0.5\textwidth]{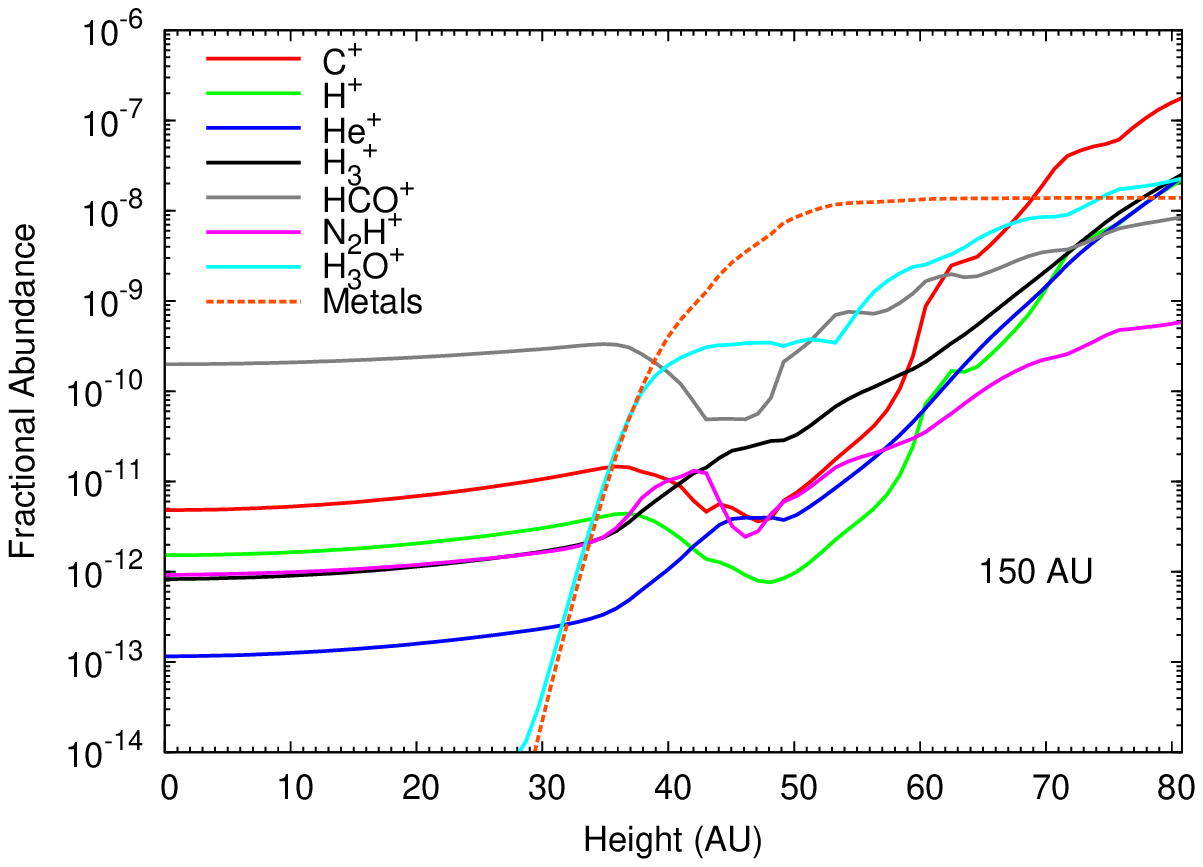}}
\caption{The fractional abundance of several important cations as a function of height, $Z$, 
at radii of 15~AU (top) and 150~AU (bottom) for model UV-new.}
\label{figure8}
\end{figure}

In Figure~\ref{figure7} we display the fractional abundance of electrons as a function of disk radius and height
for all models considered in this work and in Figure~\ref{figure8}, 
we present the fractional abundances of several abundant cations as a 
function of disk height at radii of 15~AU (top) and 150~AU (bottom).   

At first glance, there appears little difference in the distribution of electrons between the different models.  
The ionization fraction in the disk ranges from close to 1, in the disk surface nearest the star, 
to less than 10$^{-12}$ in the disk midplane between a radius of 1~AU and 10~AU.  
Comparing the plots for models UV-old and XR+UV-old, we see only minor differences; 
in the boundary layer where the
ionization gradient is highest and coinciding with the aforementioned molecular layer, and in the 
 midplane layer where the electron fraction reaches values $<$~10$^{-11}$ which is slightly larger in depth in  
model XR+UV-old.  

Comparing models UV-old and UV-new we see there is more of an effect in the boundary layer 
described above 
which extends into
the outer disk, in fact, we see that at the outermost radius, $\approx$~300~AU, the ionization 
fraction in the disk
surface layer ($Z/R$~$\approx$~0.6 - 0.7) in model UV-new is orders of magnitude lower than that 
in model UV-old.  
This is coincident with where we see an increase in the abundance of N$_2$H$^+$.
We also note the slight increase in the electron abundance in the disk surface in model UV-new 
compared with model UV-old.  
This is unsurprising, as we would expect the stellar radiation field to impact on the 
chemistry in the disk
surface, in particular. Comparing the graphs for models UV-new and XR+UV-new, we see the 
effects of the X-ray ionization 
in the disk surface,
which negates the decrease in the electron abundance to some extent.  

Looking at Figure~\ref{figure8}, in the disk surface at 15~AU, the most abundant cations 
are H$^+$ and He$^+$ whereas at 150~AU, the most abundant is C$^+$.  
In the disk midplane, at all radii $\gtrsim$~1~AU, HCO$^+$ dominates the cation fraction.  
Within $\sim$~1~AU, metallic ions such as Fe$^+$ and Na$^+$ also contribute to the 
cationic abundance in the midplane forming efficiently via charge exchange 
(at the expense of molecular cations such as HCO$^+$).  
In the molecular layer, the picture is more complex with several molecular and atomic ions contributing to 
the cation fraction including H$_3$$^+$, H$_3$O$^+$, HCO$^+$ and C$^+$.  
In this region, at 150~AU, metallic ions with an appreciable fractional abundance 
($\sim$~10$^{-9}$ to 10$^{-8}$) such as Fe$^+$, Mg$^+$ and Na$^+$ also contribute to the cation fraction.  
N$_2$H$^+$ only reaches significant abundances in the the outer disk ($\gtrsim$~100~AU) caused, to a degree, 
by the depletion of gas-phase CO due to freeze out onto grain surfaces in this region.

\begin{figure*}
\subfigure{\includegraphics[width=0.5\textwidth]{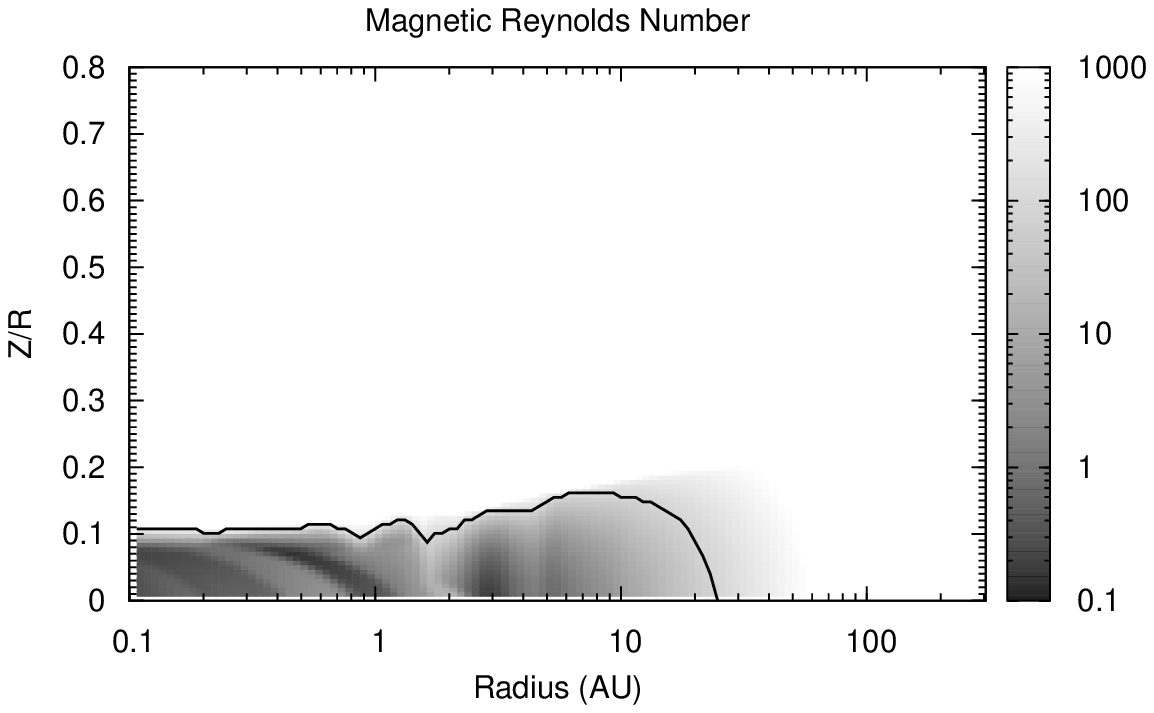}}
\subfigure{\includegraphics[width=0.5\textwidth]{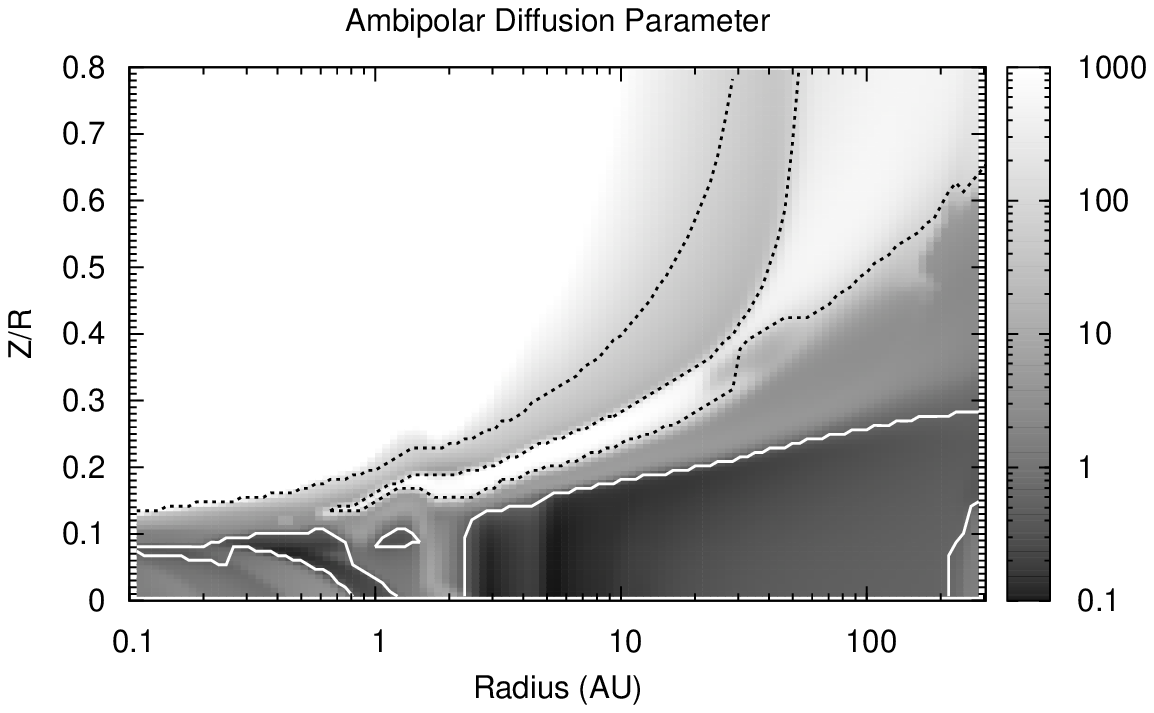}}
\caption{The magnetic Reynolds number, $Re_\mathrm{M}$ (left),  
and ambipolar diffusion parameter, $Am$ (right), 
as a function of disk radius, $R$, and height (scaled by the radius
i.e., $Z/R$) for model XR+UV-new.  
The black contours on both plots represent the boundary where $Re_\mathrm{M}$ and $Am$~=~100, whereas, 
the white contours on the right-hand plot represent the boundary where $Am$~=1.}
\label{figure9}
\end{figure*}

Following the theory outlined in Section~\ref{ionizationfraction}, we calculated the magnetic Reynolds 
number, $Re_\mathrm{M}$, and the ambipolar diffusion parameter, $Am$, 
as a function of disk radius and height and our results are displayed in   
Figure~\ref{figure9}.  
We plot the results for model XR+UV-new only, since we find both the photochemistry and X-ray ionization 
have little impact on the values of $Re_\mathrm{M}$ and $Am$ in the inner disk midplane.  

Simulations by \citet{sano02} suggest that a suitable value for $Re_\mathrm{M}^\mathrm{crit}$ is $\approx$~100.  
At values lower than this, accretion in the disk is likely inhibited due to suppression of the  
MRI by Ohmic dissipation.  
Looking at the left panel of Figure~\ref{figure9} and adopting this criterion, 
we see that there is a probable dead zone where the MRI is susceptible to Ohmic dissipation 
in the midplane extending from the
innermost radius, 0.1~AU, to $\approx$~20~AU.  
The depth of this dead zone varies from $Z/R$ = 0.1 at 0.1~AU to $Z/R$~$\approx$~0.15 at 10~AU.    
We find little difference in the extent of the dead zone between the various chemical models.

Similarly, the simulations of \citet{bai11a} suggest that ambipolar diffusion can inhibit 
mass and angular momentum transport in regions in the disk where the ambipolar 
diffusion parameter, $Am$, exceeds a critical value, $Am^\mathrm{crit}$~$\sim$~1.  
In the right-hand panel of Figure~\ref{figure9}, the white contours 
represent the boundary where $Am$~=~1 whereas the black contours represent $Am$~=~100.   
We see ambipolar diffusion has a bigger effect creating a much larger dead zone  
in the disk midplane, ranging from just beyond 2~AU out to around 200~AU. 
If we adopt the earlier criterion of \citet{hawley98}, 
i.e. $Am^\mathrm{crit}$~=~100 (black contours), we find that 
the MRI is suppressed throughout most of the disk mass. 

\begin{figure}
\includegraphics[width=0.5\textwidth]{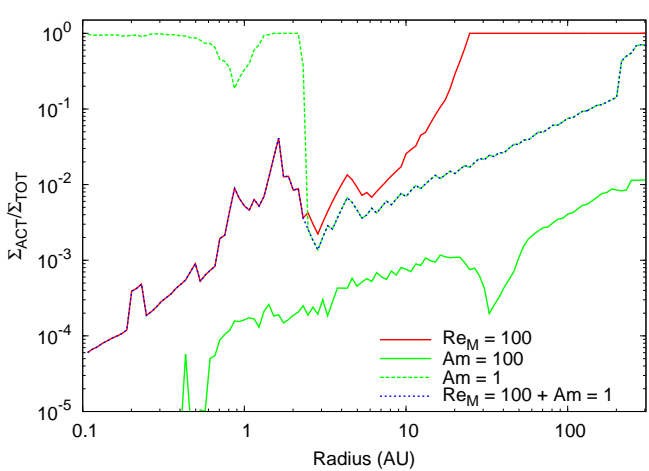}
\caption{The ratio of the active column density to the total column density, $\Sigma_\mathrm{ACT}/\Sigma_\mathrm{TOT}$, 
as a function of radius, $R$, for four criterion: $Re_\mathrm{M}$~$>$~100 (solid red line), 
 $Am$~$>$~100 (solid green line), $Am$~$>$~1 (dashed green line), and $Re_\mathrm{M}$~$>$~100 and $Am$~$>$~1.}
\label{figure10}
\end{figure}

\citet{perez-becker11} argue that, in order for efficient accretion in protoplanetary disks, 
both criterion for $Re_\mathrm{M}$ and $Am$ must be met.  
Our calculations suggest, in our disk model, most of the midplane is inactive and 
so accretion can occur via the surface layers only, a similar result to that found by \citet{gammie96}.  
As a measure of the effect on the mass accretion rate of the disk due to the presence of dead zones, 
in Figure~\ref{figure10} we show the ratio of the `active' column density 
to the total column density, $\Sigma_\mathrm{ACT}/\Sigma_\mathrm{TOT}$, as a function of disk radius, $R$,   
since this ratio estimates the efficiency of accretion.
We plot $\Sigma_\mathrm{ACT}$ for four criteria: $Re_\mathrm{M}$~$>$~100 (solid red line), $Am$~$>$~100 
(solid green line), 
$Am$~$>$~1 (dashed green line), and $Re_\mathrm{M}$~$>$~100 and $Am$~$>$~1 (dotted blue line).  
For the latter criterion, we see that accretion efficiency ranges from $\lesssim$~1~\% in the inner 
disk ($<$~10~AU) to around 70~\% in the outer disk ($>$~200~AU). 
Although most of the outer disk is unaffected by Ohmic dissipation (solid red line), it is susceptible to 
ambipolar diffusion (dashed green line).  
Applying the more stringent condition of $Am$~$>$~100 (solid green line), we find that most of 
the disk is constrained to an accretion rate $\lesssim$~1\% that which dictates the disk structure. 
The structure seen in the active column density is due to the corresponding structure 
seen in the spatial distribution of $Re_\mathrm{M}$ and $Am$ (displayed in Figure~\ref{figure9}) 
which are related to the fractional and absolute electron density, respectively.  

Compared with Ohmic dissipation, ambipolar diffusion suppresses the MRI in lower density regions \citep[see e.g., Figure~1 in][]{kunz04}.  
Our results are consistent with this since we see the dead zone extending into the 
lower density regions in our disk i.e., to larger radii in the disk midplane and towards the disk surface, 
when the ambipolar diffusion criterion is included.   

The suppression of accretion in specific areas 
leading to a build up of material flowing through the disk
midplane, will change the physical nature of the disk. 
This, in turn, influences the chemical composition 
and thus the penetration of UV photons, X-ray photons and cosmic-ray particles.   
The changing physical and chemical conditions may act to increase $Re_\mathrm{M}$ and $Am$ 
to the extent that they exceed their critical values, hence, the presence of dead zones may be time dependent.  
Young stars and protostellar objects often undergo transient periods of enhancements in luminosity 
often ascribed to spells where the accretion disk possesses an increased mass flow rate going from 
$\lesssim$~10~$^{-7}$~$M_\odot$~yr$^{-1}$ to $\sim$~10~$^{-4}$~$M_\odot$~yr$^{-1}$ for periods of $\sim$~100~yr.  
An episodic build up and release of material in the disk due to the development and subsequent dissipation 
of an accompanying dead zone may be responsible 
\citep[see e.g.,][]{herbig77,hartmann96,calvet00,armitage01,vorobyov06}.  

We note here that the steady disk model is a first step in our 
attempt to approach reality.  
A more sophisticated model, in which we treat the charge balance more realistically 
(including detailed balance of grain charge and PAHs) 
and in which we calculate the dust evolution in conjunction with 
the chemical evolution, may go some way to addressing the apparent discrepancy between
the observed and calculated accretion rates in the inner disk.  
We also note here, the disk mass and surface density profile of our model 
is consistent with observed thermal dust emission 
from the outer regions of protoplanetary disks and observed accretion rates 
are mainly derived from optical and UV spectral lines originating from close to the stellar 
surface \citep[see e.g.,][]{calvet00}. 

The suppression of turbulence (and thus angular momentum transport) has various further 
consequences on the disk physical structure and resulting chemistry.  
A significant source of heating in the disk midplane close to the star is 
viscous dissipation due to accretion.  
This can elevate the temperature in this region to $\sim$~1000~K which has implications on the disk chemical 
structure and resulting line emission.  
Since this is typically the densest region of the disk, a lot of material is contained within a 
few AU of the star and 
thus can have a significant contribution to the disk-integrated line profiles 
\citep[see e.g.,][]{carr08}.  
If accretion is suppressed in this region, this heating cannot occur and so will affect 
the chemistry and resulting line emission from the inner disk. 
It is possible that there are alternative sources of ionization in the very hot, dense 
inner regions of disks, such as thermal ionization, which allow the 
gas to become magnetorotationally unstable very close to the star \citep[see e.g.,][]{umebayashi88}. 

We have investigated the effect of turbulent mixing on disk chemical structure \citep{heinzeller11} 
and find that in the `planet-forming' region, within approximately 10~AU and where the bulk of the infrared 
line emission originates, turbulent mixing acts to mix material in the vertical direction, 
dredging material from the disk midplane towards the surface.  
In fact, we find that our models with turbulent mixing included agree best with current infrared data from 
disks.  
In dead zones, the disk flow is essentially laminar and so turbulent mixing within the dead zone itself 
may be suppressed.
This could act to confine a proportion of infra-red active molecules to the cold, dense midplane.     

Finally, the only negative charge carriers we 
consider here are electrons and simple negative ions such as 
H$^-$ and CN$^-$.  
Dust grains and polycyclic aromatic hydrocarbons (PAHs) are also able to capture electrons to form 
negatively charged grains and PAHs.  
As discussed previously in Section~\ref{ionizationfraction}, dust grains may be the 
dominant charge carriers in the disk midplane.  
Effective accretion depends on the degree of coupling of charged particles to the magnetic field.  
If PAHs and dust grains are the dominant charge carriers, 
the relatively large, heavy PAHs and dust grains will increase drag, 
further opposing accretion.  
The addition of charged grains is a modification we intend to investigate in future models.  

\subsection{Self and Mutual Shielding of H$_2$ and CO}
\label{selfshielding}

The photodissociation of H$_2$ and CO is dominated by line absorption rather 
than continuum absorption \citep[see e.g.,][]{lee96}, hence, in any atmosphere where there is a significant 
column density of foreground H$_2$ or CO, there will be a degree of self shielding against 
photodissociation dependent on the amount of intervening material.  
Also, there is an overlap in wavelength ranges over which both molecules absorb leading to 
a degree of mutual shielding due to the removal of UV flux by the photodissociation of 
foreground H$_2$.  

To date, significant investigation into the self- and mutual shielding of H$_2$ and CO 
in photon-dominated regions (PDRs) has been conducted leading to a variety of 
approximate methods for the computation of self-shielding factors in plane-parallel media 
\citep[see e.g.,][]{draine96,lee96,visser09}.  
In protoplanetary disks, detailed two-dimensional radiative transfer is required since the 
radiation field has two sources, the central star and the interstellar medium.  
Also, the addition of self-shielding requires that the abundances of H$_2$ and CO are known 
everywhere in the disk at each time step in the calculation simultaneously and this 
is computationally impossible in a spatially high-resolution model such as this.  
A compromise often employed in disks is to assume the plane-parallel approximation in the vertical 
direction and calculate the chemical structure from the disk surface downwards.  
In this way, the column density of H$_2$ and CO from the disk surface to the point of 
interest can be estimated by extracting out the abundances in the upper layers 
at a particular time (e.g., 10$^6$~years).  
The self-shielding factors calculated for plane-parallel media are then employed in the calculation of the 
photodissociation rates.  
However, in the upper disk, where photodestruction is most significant, the direct stellar 
radiation field is the dominant component so that the column density of material from the star to the 
point of interest becomes the important parameter opposed to the column density from the disk surface 
(see Figure~4 in \citet{nomura05}). 
Also, many self-shielding factors are computed assuming irradiation by the interstellar field only.  
As we have already discussed, the radiation field in protoplanetary disks will have a very different 
shape and strength to that found in irradiated interstellar clouds.    
The computation of such factors also depends on the assumed gas temperature and density since these 
parameters affect the line shapes and thus absorption cross sections of H$_2$ and CO.  
In summary, the treatment of self- and mutual shielding in protoplanetary disks is a complex problem and 
the use of existing approximations to model the self-shielding in disks is problematic as the application
of shielding factors computed for irradiated interstellar clouds is debatable.   

In the generation of our disk physical model, which originates from \citet{nomura05}, the self-shielding 
of H$_2$ is included in the calculation \citep{federman79}.  
Here, although we use the calculated H and H$_2$ abundances from \citet{nomura05} as initial conditions 
in our chemical model, the abundances of both species are 
adjusted to an extent when the full, and vastly more complex, time-dependent chemistry is computed.  

\begin{figure}
\includegraphics[width=0.5\textwidth]{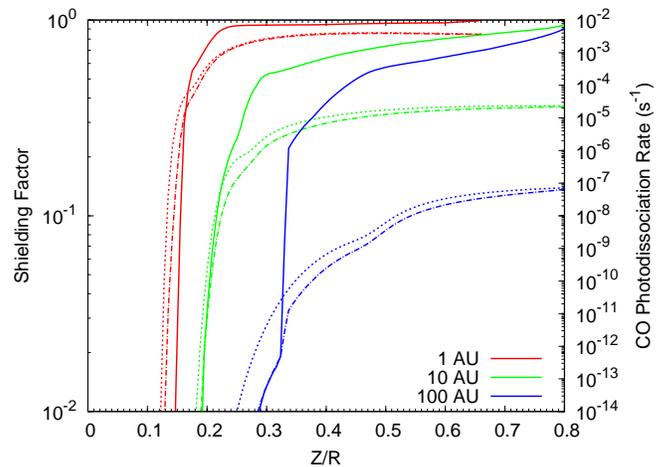}
\caption{CO shielding factors (solid lines) and photodissociation rates (s$^{-1}$) with (dot-dashed lines) 
and without (dotted lines) the shielding factors implemented as functions 
of disk height (scaled by the radius i.e., $Z/R$) at radii 
of 1~AU, 10~AU and 100~AU.}
\label{figure11}
\end{figure}

In order to quantify the effect that self- and mutual shielding of CO may have in our model, 
we adopt the method described in \citet{lee96} to compute the photodissociation rate of CO as a function of 
disk height at a radii of 1~AU, 10~AU and 100~AU assuming the usual plane-parallel approximation.  
The results of this calculation are shown in Figure~\ref{figure11}.  
The shielding factors only become significant deep in the atmosphere where the radiation field decreases 
significantly in strength.  
We conclude, the addition of self-shielding of CO using approximation methods has a minor effect on the 
overall disk structure when compared with the effects of the recalculation of the photorates taking into 
consideration the two-dimensional radiation field in the disk.  
We expect the inclusion of self- and mutual shielding in the radial and vertical directions will lead to an increase
in the abundances of H$_2$ and CO in the outer disk, in particular, since the radial column densities here
will be larger than the vertical and the radial stellar radiation field is dominant.  
We intend to look more closely at the complex issue of self- and mutual shielding of H$_2$ and CO in 
disks in future work.  

\subsection{On the Effects of Lyman-$\alpha$ Radiation}
\label{lyman-alpha}

In Section~\ref{photochemistry} we discussed how we included Lyman-$\alpha$ radiation in the 
calculation of the wavelength-integrated flux yet neglected it in the determination of the 
wavelength-dependent UV spectrum throughout the disk.  
This was due to the difficulty in treating Lyman-$\alpha$ scattering in protoplanetary disks, an issue only 
recently addressed in work by \citet{bethell11}.  
Here, we consider the effects of the addition of Lyman-$\alpha$ on the calculated photodissociation 
rates of several molecules.  

We calculate the photodissociation rates using the stellar spectrum displayed in the left-hand 
panel of Figure~\ref{figure1} which is the \emph{unshielded} UV photon flux at the disk 
surface at a radius of 1~AU.  
Note, this spectrum is different to that used to calculate the photorates displayed in 
Figure~\ref{figure6}.  
In our full disk model, we assume that the stellar radiation has travelled through, 
and thus been attenuated by, 
material along the line of sight between the star and the disk surface.  

We calculate the unshielded photorates  
with and without Lyman-$\alpha$, using the calculated Lyman-$\alpha$ 
cross sections from Table~1 in \citet{vandishoeck06}. 
Our results are presented in Table~\ref{table5}. 
In our determination of the photorate at the Lyman-$\alpha$ wavelength, 
we assume the cross section is constant across the line, 
determine the rate due to Lyman-$\alpha$ only, and add that to 
the rate calculated by integrating over the background UV field (see Equations~\ref{photorates1} and 
\ref{photorates2}).  
The integrated photon flux over the Lyman-$\alpha$ line is 8.4~$\times$~10$^{14}$~photons~cm$^{-2}$~s$^{-1}$.  

Note that the photodissociation cross section of most molecules at the Lyman-$\alpha$ wavelength is generally 
lower than at other wavelengths, hence,   
we find that the inclusion of Lyman-$\alpha$ increases the photodissociation rates by no more than a factor 
of 2.52 (as determined for HCN).  
In general, the enhancement is less than a factor of 2 which is well within the uncertainties 
in both the UV spectrum and the photo cross sections.  
Even though the UV field is weaker at wavelengths other than the Lyman-$\alpha$ wavelength
(see Figure~\ref{figure1}) the overall contribution of Lyman-$\alpha$ photons to the photodissociation 
rates is around the same order of magnitude as the contribution by the background UV photons, 
i.e., we do not see the order of magnitude difference in the photorates when the Lyman-$\alpha$ 
photon flux is included in the integrated UV field and 
the interstellar rates scaled by this flux (see Figure~\ref{figure5}).  
This result is in agreement with the work of \citet{fogel11} who find that the inclusion of Lyman-$\alpha$ 
has an effect on the chemistry \emph{only} when grain settling is included in their model.  

\begin{deluxetable*}{lccccc}
\tablecaption{Unshielded Photodissociation Rates at 1~AU \label{table5}}
\tablewidth{0pt}
\tablehead{\colhead{Species} & \colhead{$\sigma_\mathrm{Ly}$\tablenotemark{1}} & \colhead{$k_\mathrm{Ly}$} 
& \colhead{$k_\mathrm{UV}$} & \colhead{$k_\mathrm{Ly}$ + $k_\mathrm{UV}$} & 
\colhead{($k_\mathrm{Ly}$+$k_\mathrm{UV}$)/$k_\mathrm{UV}$}}
\startdata
CO$_2$        & 6.1~(-20) & 5.12~(-05) & 9.69~(-03) & 9.74~(-03) & 1.01 \\
H$_2$O        & 1.2~(-17) & 1.01~(-02) & 1.41~(-02) & 2.42~(-02) & 1.72 \\
HCN           & 3.0~(-17) & 2.52~(-02) & 1.78~(-02) & 4.30~(-02) & 2.52 \\
OH            & 1.8~(-18) & 1.51~(-03) & 6.33~(-03) & 7.84~(-03) & 1.51 \\
C$_2$H$_2$    & 4.0~(-17) & 3.36~(-02) & 5.16~(-02) & 8.52~(-02) & 1.65 \\
CH$_4$        & 1.8~(-17) & 1.51~(-02) & 1.44~(-02) & 2.95~(-02) & 1.51 \\
NH$_3$        & 1.0~(-17) & 8.40~(-03) & 2.56~(-02) & 3.40~(-02) & 1.33
\enddata
\tablerefs{(1) \citet{vandishoeck06}}
\tablecomments{$a(b)$ means $a \times 10^{b}$}
\end{deluxetable*}

\subsection{Comparison with Other Models}
\label{comparisonmods}

A full two-dimensional calculation of the UV radiative transfer in disks was 
achieved by \citet{vanzadelhoff03}.  
They coupled the UV radiation field and photochemistry and
applied their chemical model to the established protoplanetary disk model of \citet{dalessio99}.  
In this earlier work, they calculate the photorates adopting difference shapes for the stellar UV field and 
come to a similar conclusion as ourselves i.e., the abundances of molecules are sensitive 
to the shape of the adopted stellar UV spectrum.  
They also find at a radius of 105~AU in their model, that the photodissociation rates 
of C$_2$H and H$_2$CO are generally underestimated when the UV stellar spectrum 
is approximated by a scaling of the interstellar radiation field.  

Early investigations into the calculation of the X-ray ionization in disks 
include the work of \citet{aikawa99,aikawa01}.  
In \citet{aikawa99}, they adopt a power-law approximation for the X-ray absorption cross-section 
similar to that used in Paper I \citep{maloney96}.  
In \citet{aikawa01} they perform a calculation similar to that adopted here, 
where the explicit elemental composition and absorption cross-section are taken into account as is the 
direct X-ray ionization of elements.  
Although their underlying disk physical model is analytical in nature and different to ours in that the 
X-rays are included in purely a chemical sense i.e., X-ray heating of the disk is not taken into account, 
they too find little difference between results using either method.  
More recent models based on the work of \citet{aikawa01} \citep[e.g.,][]{woods09}, have adopted this method 
of calculating the X-ray ionization rates.  

\citet{aikawa06} include spectrum-dependent photochemistry looking at the effect of grain growth on 
disk chemical structure although they neglect the effects of X-rays.    
The physical model used in their work is based on the model of 
\citet{nomura05} where a full two-dimensional 
calculation of the radiative transfer allowed the determination of the UV spectrum 
(including UV excess emission from the star) everywhere in the disk.
We currently have in preparation a paper where we look at the effects of grain growth in a protoplanetary 
disk in which X-rays are included (Walsh et al. 2011, in preparation).    

The disk model of \citet{woitke09}, `ProDiMo', has recently been updated with 
spectrum-dependent photochemistry \citep{kamp10} and X-ray ionization \citep{aresu11} with the 
methods used and cross-sectional data adopted similar to that used here.  
The recent work by \citet{vasyunin11} now includes wavelength-dependent photochemistry, updating their 
previous model \citep{semenov05,vasyunin08} to investigate the effects of 
grain-settling on the disk chemistry.  
Also, \citet{fogel11} have included photochemistry investigating the effect of Lyman-$\alpha$ 
radiation on the disk chemical structure  
They find, in models in which dust settling is 
included, those species with significant photodissociation 
cross sections around the Lyman-$\alpha$ line are particularly affected, generally 
leading to a reduction in column density for species such as HCN, NH$_3$ and CH$_4$.  
This effect is significantly less in models where they assume the dust is well-mixed with the gas, an assumption 
we adopt in our work.

\subsection{Comparison with Observation}
\label{comparisonobs}

The observation of molecular line emission from protoplanetary disks remains a challenge due to 
the small angular size of these objects and the limitations of existing (sub)mm observing facilities.  
Ideally, to directly compare our results with observations, full radiative transfer should be 
performed and disk-averaged line profiles and intensities calculated to compare with those measured 
by telescopes.  

Our model is not specific to any particular source and we have adopted the star-disk parameters 
of a typical T~Tauri star \citep{hartmann96}.  
However, since we adopt the X-ray spectrum and UV excess observed from the T~Tauri star, TW Hya, 
we compare our calculated column densities with those derived from molecular line 
observations (single dish and interferometric) of this source.    
TW Hya is a well-studied, relatively old ($\sim$~10~Myr), nearby (51~$\pm$~4~pc), 
almost face-on ($\lesssim$~10\textdegree), 
isolated classical T~Tauri star \citep[see e.g.,][]{kastner97,webb99,krist00,mamajek05}.  
Observations of continuum dust emission from TW Hya suggest it has an cleared inner hole, 
possibly caused by a forming planet, and a 
truncated outer disk with radius $\approx$~200~AU \citep{calvet02,setiawan08, akeson11}.  

Rotational line emission at (sub)mm wavelengths 
from several molecules has been observed in TW Hya including 
CO, CN, HCO$^+$, HCN, 
as well as the isotopologues, $^{13}$CO, H$^{13}$CO$^+$, DCO$^+$ and DCN 
\citep{kastner97,vanzadelhoff01,vandishoeck03,wilner03,qi04,thi04,qi06,qi08}.  
Molecular line emission has been spatially mapped by \citet{qi04,qi06,qi08} using the 
Submillimeter Array (SMA) and by \citet{wilner03} using the Australia Telescope 
Compact Array (ATCA).  
More recently, \citet{hogerheijde11} report the detection of the ground-state 
emission lines of both spin isomers of H$_2$O 
in TW Hya observed using the Heterodyne Instrument for the Far-Infrared 
(HIFI) on the Herschel Space Observatory. 
In addition to (sub)mm data, 
\citet{najita10} present detections of OH, CO$_2$, HCO$^+$, 
and possibly CH$_3$, at mid-infrared wavelengths 
using the Infrared Spectrograph (IRS) on board the Spitzer Space Telescope.  

\citet{thi04} and \citet{qi08} each use a method to derive the column density of each 
molecule using their observed line profiles.    
\citet{thi04} assume the emitting region has a typical density and temperature, 
constrained by line ratios in a single molecule, and assuming the energy 
levels are thermalised (i.e. they assume local thermodynamic equilibrium or LTE).  
The optical depth can be estimated using line ratios of the same transition 
in two isotopologues assuming the same excitation temperature.  
They list their beam-averaged column densities for TW Hya assuming a disk radius 
of 165~AU and an excitation temperature of 25~K.  
We reproduce their calculated values here in Table~\ref{table6}.
\citet{qi08} constrain their estimated column densities between 10~AU and 100~AU by 
assuming a simple power-law distribution for the surface density of the 
disk and fitting the 
vertical extent and abundance of each molecule to match the observed line profiles 
using a $\chi^2$ method.  
Again, we list their fitted values at a radius of $\approx$~100~AU here in Table~\ref{table6}.  

Comparing our calculated column densities at a radius of 100~AU (listed in Table~\ref{table4}) 
with those derived from observation, 
we see our value for CO is more than two orders of magnitude larger than 
the value from \citet{thi04}.  
Comparing the values for HCO$^+$ and HCN, we see good agreement between the column densities from 
both authors, however, our calculated column densities for both species 
are around one order of magnitude larger than the observed values.  
Looking at CN, we see the same behaviour with our model predicting a value between 
2 to 5 $\times$~10$^{14}$~cm$^{2}$ compared with the observed value of 
6.6~$\times$~10$^{13}$~cm$^{2}$.  

Unfortunately, \citet{hogerheijde11}, do not report their results in terms of 
column density, however, they do interpret their results using the 
detailed disk model of \citet{woitke09} calculating a total 
water vapour mass of 7.3~$\times$~10$^{24}$~g in a disk with total mass 
1.9~$\times$~10$^{-2}$~$M_\odot$.  
This corresponds to a disk-averaged 
water vapour fractional abundance of $\approx$~2~$\times$~10$^{-8}$.  
Taking our modelled column densities of H$_2$ and H$_2$O at 100~AU 
for model XR+UV-new, we find a 
column-averaged fractional abundance of 7~$\times$~10$^{-8}$ which compares
well with the disk-averaged value.  
\citet{hogerheijde11} report that their observations are optically thin hence the 
line emission is tracing the entire column density of water throughout the vertical 
extent of the disk.  
Also, \citet{najita10} do not convert their line observations to physical quantities 
such as column density or fractional abundance since they plan to 
disuss their results for OH, HCO$^+$ and CO$_2$ in more detail in a future publication.  
They do mention their apparent anomalous result for TW Hya which is their non-detection of 
H$_2$O, C$_2$H$_2$ and HCN line emission, in contrast to 
results for most classical T~Tauri stars observed using IRS \citep{carr11,salyk11}.  
One explanation they give is that TW Hya is unusual in that is apparently at 
a more advanced stage than its accretion rate suggests i.e., TW Hya is not 
an archetypical classical T~Tauri star.      

Why is our model overestimating the column densities, especially those observed 
with ground-based facilities?  One reason is that 
our disk is not meant to be specific to TW Hya even though we adopt the X-ray and UV 
spectrum of this source in our model.  
Another reason is that our column densities involve integration over the full 
vertical extent of the disk from the lower 
disk surface to the upper disk surface.  
In reality, the disk is likely to be optically thick, 
for $^{12}$CO and H$^{12}$CO$^+$ emission particularly, and so the  
observed line emission is only tracing a fraction of the total column density.  
A third reason is that approximations for the disk chemical and physical 
structure are used for generating observed column densities rather than a detailed model 
which allows for vertical and radial structure in the disk density and temperature and 
resulting molecular abundances.  

As stated previously, in order to determine how accurate our model is, one must 
perform full radiative transfer and compare our modelled line profiles with those 
observed.

\begin{deluxetable}{lc}
\tablecaption{TW Hya Observed Column Densities\label{table6}}
\tablewidth{0pt}
\tablehead{\colhead{Species} & \colhead{N (cm$^{-2}$)}}
\startdata
\cutinhead{Beam-averaged values from \citet{thi04}}
CO        &  3.2~(16) \\
HCO$^+$   &  4.4~(12) \\
CN        &  6.6~(13) \\
HCN       &  9.2~(12) \\
\cutinhead{Fitted values at 100~AU from \citet{qi08}}
HCO$^+$   &  5~(12) \\
HCN       &  2~(13) \\
\enddata
\end{deluxetable}

\section{SUMMARY}
\label{summary}

We have investigated the effects of photochemistry and X-ray ionization on the chemical structure of 
a protoplanetary disk surrounding a typical T~Tauri star.  
We used a high-resolution complex physical model which takes into account irradiation by both
the central star and the interstellar medium.   
We compiled a comprehensive chemical network including a large gas-phase reaction set extracted from the 
UMIST Database for Astrochemistry (Rate06), gas-grain interactions (accretion, thermal desorption, 
cosmic-ray-induced desorption and photodesorption) and a grain-surface network.  

In previous work, we presented results from a model in which we approximated the photorates in the disk by
scaling those in Rate06 (which assume the interstellar radiation field) 
by the wavelength-integrated UV flux at each point.  
The X-ray ionization rate everywhere was also approximated using a power law to describe the energy-dependent
cross section for ionization.  
Here, we recalculated both the photorates and the X-ray ionization rate, in the first case taking into account
the unique UV wavelength spectrum at each point, and in the second, the unique X-ray energy spectrum and the
explicit elemental composition of our gas.  
We also added the direct X-ray ionization of elements.  

We found that the recalculation of the photochemistry has a much larger effect on the disk chemical structure
than that for the X-ray ionization.  
Concentrating on those molecules which have been observed or searched for in disks at 
either (sub)mm or
infrared wavelengths, we find that the species most sensitive to the photochemistry are OH, HCO$^+$, N$_2$H$^+$, 
H$_2$O, CO$_2$ and methanol.  
We also find the radicals, CN and C$_2$H, are also affected although not to the same extent 
as those listed above.  
The recalculation of the photochemistry affects each molecule in a different manner indicating the inherent
non-linearity of the chemistry. 
Molecules affected throughout the disk extent include HCO$^+$, N$_2$H$^+$ and OH i.e.,
predominantly molecular ions and radicals.  
Water and methanol are mainly affected beyond a radius of around 1~AU, whereas CO$_2$ is altered beyond
approximately 10~AU.  
The differences in the behaviour of the saturated molecules can be attributed to the location of the
molecular layer in each case, with methanol and water residing in a layer higher in the disk than CO$_2$ which 
in turn is linked to the desorption temperature of both molecules, with water more strongly bound to the 
grain surfaces than carbon dioxide (see Table~\ref{table2}).  

In general, we find the change in the photochemistry leads to a depletion of molecules in the inner disk (within
1~AU) and an enhancement in the outer disk (beyond 1~AU).  
We conclude that, in the outer disk, our original approximation was overestimating the 
photorates, whereas, in the inner disk, where the stellar irradiation is strongest, 
we were underestimating the photorates. 
We should note that the photochemistry also indirectly affects the abundances of molecules via gas-phase chemistry.  
In models where we recalculate the photorates, we find differences in the location of the H/H$_2$ and C$^+$/C 
transition regions with both ratios generally decreasing in the molecular layer 
in models where we recalculate the photochemistry.  
Neutral molecules such as water and CO$_2$ are effectively destroyed by reaction with 
C$^+$ and in regions of high temperature by reaction with H atoms. 
Also, the molecular ions, HCO$^+$ and N$_2$H$^+$ are efficiently 
destroyed via dissociative electron recombination.  

Concerning the X-ray ionization rate, we find the only molecule especially affected is N$_2$H$^+$.  Residing in a
different region of the disk to those molecules discussed thus far, we see the effect of X-ray ionization in the
outer disk, in particular, where the abundance and distribution of N$_2$H$^+$ is decreased.  
Since the chemistry of N$_2$H$^+$ is linked closely to the ionization of molecular
hydrogen in the disk, this is unsurprising, however, one conclusion we can draw is that observations of 
N$_2$H$^+$ emission from disks in conjunction with observations of those molecules unaffected by X-rays 
e.g., CO, could give us information on the effects of X-ray ionization in the outer regions of protoplanetary disks 
around classical T~Tauri stars.  

The ionization fraction in the disk is affected only marginally by the recalculation of the photochemical and
X-ray ionization rates.  We find photochemistry affects the electron abundance in the disk surface whereas
X-rays affect the disk midplane and outer disk region, since X-rays are able to penetrate the disk more 
effectively than UV photons.  

A calculation of the magnetic Reynolds number and ambipolar diffusion parameter, 
everywhere in our disk allowed us to determine the location of a
possible significant dead zone in the disk midplane 
extending to $\approx$~200~AU where accretion will likely be suppressed.  
This has implications on both the physical and chemical structure since a significant source of 
heating in the disk midplane close to the star is viscous dissipation due to accretion flow.  
Our calculations suggest the accretion rate in the outer regions of our disk model 
is only $\approx$~70~\% that used to determine the disk physical structure. 
In the inner disk, this value falls to $\approx$~1~\%, however, we have 
neglected alternative sources of ionization which may dominate in the inner disk 
e.g., the thermal ionization of alkaki models, which should be considered in future 
models.

We conclude that a wavelength-dependent treatment of the photochemistry in protoplanetary disks is necessary 
since it significantly affects the chemical structure.   
We also determine that existing approximations of the X-ray ionization rate in such objects are sufficient,
affecting only the abundance and distribution of N$_2$H$^+$.  
In order to directly compare our model results with observation and to determine the effects of the various 
chemical processes investigated in our work on observable line emission we have computed the radiative transfer 
and modelled molecular line emission from the disk both at the resolution of existing facilities and at the 
expected resolution of ALMA.   
We report the results from these calculations in a subsequent publication (C.\ Walsh et al. 2011, in preparation).

\acknowledgments

We thank an anonymous referee for his or her comments which greatly 
improved the content and scope of the manuscript.  
C.\ Walsh acknowledges DEL for a studentship and JSPS for the award of a short-term 
fellowship to conduct research in Japan.  H.\ Nomura acknowledges the JGC-S Scholarship 
Foundation, the Grant-in-Aid for Scientific Research 21740137 and the 
Global COE Program ``The Next Generation of Physics, Spun from Universality and 
Emergence'' from MEXT, Japan.  
Astrophysics at QUB is supported by a grant from the STFC.

\end{document}